\documentclass[11pt]{article}

\newcommand{\be}{\begin{equation}}
\newcommand{\ee}{\end{equation}}
\newcommand{\bea}{\begin{eqnarray}}
\newcommand{\eea}{\end{eqnarray}}
\newcommand{\bg}{\begin{gather}}

\newcommand{\bseq}{\begin{subequations}}
\newcommand{\eseq}{\end{subequations}}

\renewcommand{\ln}{\mathop{\rm ln}\nolimits}

\def\be{\begin{eqnarray}}
\def\ee{\end{eqnarray}}
\def\lb{\label}

\usepackage{amsmath,amssymb}
\usepackage{array}
\usepackage{graphicx}
\usepackage{graphicx}
\usepackage{wrapfig}
\usepackage{color}
\usepackage{tensor}
\usepackage{hyperref}
\usepackage[sort,compress]{cite}
\usepackage{tikz}
\usepackage[T1]{fontenc}
\usepackage{multirow}
\usepackage{longtable}
\usetikzlibrary{decorations.pathmorphing, decorations.pathreplacing, decorations.shapes} 

\numberwithin{equation}{section}

\renewcommand{\d}{\mathrm{d}}

\DeclareMathOperator{\arsinh}{arsinh}

\begin{document}
\title{\textbf{
Exploring the landscape of black hole mimickers}}
\vspace{0.5cm}
\author{ \textbf{ Sergey N. Solodukhin   and   Vagif Tagiev }} %\copyright

\date{}
\maketitle
\begin{center}
    \emph{Institut Denis Poisson UMR-CNRS 7013,
  Universit\'e de Tours,}\\
  \emph{Parc de Grandmont, 37200 Tours, France} 

\end{center}

\vspace{0.5cm}
%  \texttt{ sergey.solodukhin@univ-tours.fr}

%{\vspace{-11cm}
%\begin{flushright}
%preprint
%\end{flushright}
%\vspace{10cm}
%}

%\hfill{\tt IUB-TH/***}\\\mbox{} \\
%@twocolumnfalse\endcsname

%\maketitle \thispagestyle{empty}% \vspace*{.5cm}

%\vspace{0.2mm}

\begin{abstract}
We identify a general class of spacetime metrics that mimic the properties of black holes without possessing a true event horizon. These metrics are constrained by the requirements of being singularity-free and geodesically complete. Specifically, we study metrics that do not possess $Z_2$ symmetry and may deviate slightly or significantly from the symmetric case. Focusing on scalar perturbations propagating on such backgrounds, we analyze the resulting effective radial potentials and their dependence on different corners of the mimicker landscape. We further investigate the corresponding quasinormal modes and explore their characteristic features. Finally, we survey the landscape for potential observational signatures, including shadow properties and the possible presence or absence of echo effects.
%\noindent {PACS: 04.70Dy, 04.60.Kz, 11.25.Hf }}
\end{abstract}
\rule{7.7 cm}{.5 pt}\\
\noindent ~~~ {\footnotesize  sergey.solodukhin@univ-tours.fr\\
                              vagif.tagiev@univ-tours.fr }

\newpage
\tableofcontents
\pagebreak

\section{Introduction}

One of the central predictions of general relativity is the existence of black holes, first emerging as a consequence of the Schwarzschild metric \cite{Schwarzschild:1916uq}. For many years, however, the physical reality of such objects was regarded with skepticism. With the steady advancement of observational techniques, accumulating indirect evidence \cite{Narayan:2005ie}  has progressively strengthened the case that black holes are indeed astrophysical objects.

Over the past decade, two landmark observations have revolutionized the study of compact gravitational objects.

\noindent (i) The first is the detection of gravitational waves from black hole mergers by the LIGO and Virgo collaborations \cite{LIGOScientific:2016sjg, LIGOScientific:2016aoc}. The observed signal consists of three distinct phases: the inspiral phase, accurately described by quasi-Newtonian dynamics within the post-Newtonian framework \cite{Blanchet:2013haa}, \cite{Buonanno:1998gg} and found to be in excellent agreement with observations; the merger phase, a fully relativistic regime dominated by the strong-field dynamics of general relativity and requiring numerical simulations \cite{Blanchet:2009sd, Campanelli:2005dd}; and finally, the ringdown phase, during which the remnant object relaxes to equilibrium while emitting gravitational radiation. This last stage is of particular importance, as it is governed by the object's quasinormal modes (QNMs) \cite{Vishveshwara:1970zz, Nollert:1999ji, Dreyer:2003bv, Berti:2009kk, Konoplya:2011qq}, which depend solely on its intrinsic parameters: mass, charge, and angular momentum. Measuring the ringdown phase and extracting the corresponding QNMs provides a powerful means of testing the no-hair theorems \cite{Berti:2015itd, Konoplya:2016pmh, Berti:2005ys}.

\noindent (ii) The second breakthrough came in 2019, when the Event Horizon Telescope (EHT) collaboration achieved the first direct image of the shadow cast by the compact object at the center of M87 \cite{EventHorizonTelescope:2019ggy}, followed by the shadow of Sagittarius A*, the supermassive object at the center of our own Galaxy \cite{EventHorizonTelescope:2022xqj}. Observations of shadows enable tests of the geodesic structure of the surrounding spacetime.

However, both of these observational channels are largely insensitive to the geometry in the immediate vicinity of the event horizon \cite{Cardoso:2008bp, Konoplya:2017wot}. In both cases, the measurable features are primarily determined by the region near the photon sphere \cite{Abramowicz:2002vt}, where the effective potential attains its maximum and where signals or geodesics are scattered. As a result, these observations cannot directly probe the deeper regions of spacetime. This limitation naturally raises a fundamental question: are the observed compact objects truly black holes? Do they indeed possess event horizons and central singularities, as predicted by general relativity?

This question motivates the study of horizonless alternatives to black holes, such as gravastars \cite{Mazur:2001fv}, boson stars \cite{Schunck:2003kk}, \cite{Herdeiro:2021lwl}, ultracompact stars \cite{Abramowicz:1997qk}, wormholes \cite{Morris:1988tu},\cite{Lemos:2008cv}, \cite{Damour:2007ap}, fuzzballs \cite{Mathur:2005zp}, and others \cite{Cardoso:2019rvt}, often referred to as black hole mimickers or Exotic Compact Objects (ECO). However, the absence of a horizon in these models leads to certain problems, such as superradiant instabilities \cite{Cardoso:2014sna}, \cite{Maggio:2017ivp}, \cite{Pani:2008bzt} or the requirement of exotic matter that violates standard energy conditions \cite{Morris:1988tu} in order to sustain them.

The spacetimes of compact objects may contain unstable photon spheres \cite{Claudel:2000yi, Teo:2020sey}, which give rise to the appearance of shadows. Null geodesics with specific impact parameters can become effectively ``trapped'' on these spheres \cite{Perlick:2021aok, Vagnozzi:2022moj}, preventing them from reaching distant observers. Similarly, timelike geodesics with smaller impact parameters are unable to escape. Consequently, the existence of a shadow does not constitute definitive proof that a compact object is a black hole. Indeed, shadows may arise for a variety of compact configurations, including black holes \cite{Perlick:2021aok}, wormholes \cite{Tsukamoto:2024pid, Shaikh:2018kfv}, and compact stars \cite{Wang:2020jek, Rosa:2022tfv, Herdeiro:2021lwl}. However, unlike the Schwarzschild or Kerr black holes, other exotic compact objects (ECOs) may exhibit additional structures in their effective potentials. These may include an extra unstable photon sphere\footnote{Generically there may also appear a stable light ring,
which, as was discussed earlier \cite{Cunha:2017qtt}, \cite{Keir:2014oka}, \cite{Xavier:2024iwr}, \cite{Benomio:2024lev}, may potentially lead to
non-linear instability. The issue of instability is interesting. However, it goes beyond the
scope of the present paper and will not be discussed here.}.
% , for instance, it was shown in \cite{Cunha:2017qtt} that ultracompact objects formed as a result of classical gravitational collapse of matter must possess an additional photon sphere, which is stable and may lead to nonlinear instability \cite{Keir:2014oka}. However, this statement does not apply to wormholes because of their nontrivial topology. It was shown in \cite{Xavier:2024iwr} that, like black holes, wormholes must have at least one photon sphere. 
Additional unstable photon spheres have been found in certain wormhole \cite{Tsukamoto:2024pid} or hairy black hole models \cite{Guo:2022ghl, Wang:2025hzu}, or an infinite potential barrier at the center, as in the case of gravastars \cite{Sakai:2014pga}.
Such features can lead to a range of distinct observational effects. One striking example is the appearance of a second shadow in wormhole geometries possessing two asymmetric photon spheres \cite{Wielgus:2020uqz, Wang:2020emr, Guerrero:2022qkh, Guerrero:2021pxt}. Other potential signatures emerge when the object is surrounded by an accretion disk, in which case the resulting luminosity profile may reveal further deviations from the classical black hole predictions.

Another key observational signature is the ringdown stage of compact objects. This stage is entirely governed by the object's  QNMs, while information about the initial perturbation is almost completely lost, remaining only in the overall amplitude and width of the emitted signal. Unlike classical black holes, whose effective potentials possess a single peak and no additional structures near the horizon, ECO models introduce additional features on the horizon scales. These may include an extra potential peak (as in wormholes \cite{Damour:2007ap}, \cite{Bueno:2017hyj} and hairy black holes \cite{Guo:2021enm, Guo:2022umh}), effective reflecting walls at the centers of stars (such as ultracompact stars or gravastars \cite{Cardoso:2014sna}), or discontinuities associated with thin-shell structures (in certain ultracompact star models \cite{Shen:2025yiy}).

These additional structures give rise to the phenomenon of gravitational-wave echoes \cite{Cardoso:2016oxy, Cardoso:2016rao, Hui:2019aox, Abedi:2016hgu}, a sequence of repeated, time-delayed bursts following the primary ringdown signal. The primary burst is governed by the black hole-like QNMs \cite{Cardoso:2016oxy, Hui:2019aox} and is therefore almost indistinguishable from the ringdown of a true black hole merger. However, the true QNMs of the system, defined as the poles of the Green's function associated with the full potential, are distinct: they do not include the black hole modes, and they correspond to long-lived oscillations with small imaginary parts.

The class of metrics that resemble a classical black hole in the exterior region but exhibit small or large deviations from the Schwarzschild geometry in the interior is vast. It is therefore natural to speak of a landscape of metrics describing black hole mimickers. To meaningfully constrain this landscape, certain physical conditions must be imposed, most notably, the absence of curvature singularities and the geodesic completeness of the spacetime. In the present work, we examine the restrictions that follow from these requirements.

We then explore different corners of this landscape by introducing a set of representative examples, which we refer to as test metrics. Each of these metrics exhibits a characteristic behavior in the inner asymptotic region, effectively replacing the interior of a classical black hole with a modified geometric structure. The test metrics considered here represent various generalizations of the $Z_2$-symmetric Damour-Solodukhin (DS) wormhole metric \cite{Damour:2007ap}. The most direct generalization is to relax the $Z_2$ symmetry. This symmetry breaking can be small, as in our test metric I, or large, as in metrics II, III, and IV, leading to a profound modification of the spacetime structure in the inner region.

We subsequently confront these metrics with observational constraints on the shadow size derived from the recent EHT observations. This analysis allows us to bound the deformation parameters that quantify the deviations from the classical black hole geometry. Finally, we investigate the corresponding waveforms and QNMs, focusing on how these observables depart from those of the Schwarzschild black hole.

It should be noted that wormhole geometries generally require exotic matter to support them. In the present work, we do not specify a concrete formation mechanism for such wormholes; instead, we consider representative examples of metrics characterized by a particular asymptotic behavior in the inner region. Nevertheless, wormhole solutions can also arise in certain alternative theories of gravity \cite{Harko:2013yb, Moraes:2017dbs} or within higher-dimensional general relativity \cite{Svitek:2016nvm}, where the presence of exotic stress-energy tensors is not necessary. Moreover, within the framework of general relativity, wormholes may be supported by quantum effects, such as the Casimir energy \cite{Garattini:2019ivd}, or through backreaction from quantum conformal fields associated with the conformal anomaly \cite{Berthiere:2017tms}.

Further insight comes from two-dimensional models \cite{Potaux:2021yan, Potaux:2022uxa, Potaux:2023fwm}, which demonstrate that when the quantum backreaction is taken into account, the classical event horizon may be replaced by a wormhole throat.

 The paper is organized as follows.
In Sec.~\ref{sec:General class of mimicker metrics}, we analyze the functions describing static, spherically symmetric metrics, subject to the conditions of regularity and geodesic completeness. We then introduce four test metrics that satisfy these requirements.
In Sec.~\ref{sec:Shadow and constraints to the parameters}, we examine the behavior of null geodesics in these geometries and derive parameter constraints from the observed shadow sizes of M87$^{\ast}$ \cite{EventHorizonTelescope:2019ggy} and Sgr~A$^{\ast}$ \cite{EventHorizonTelescope:2022xqj}.
Section~\ref{sec:Quasinormal Modes} is devoted to the study of the ringdown phase of scalar perturbations in these spacetimes and the computation of their corresponding QNMs. We also describe the numerical methods employed in these calculations and provide a detailed discussion of the ringdown behavior for the symmetric DS wormhole.
 
\section{General class of  mimicker metrics} 
 \label{sec:General class of  mimicker metrics}

 The spacetime metrics that describe black hole mimickers can be regarded as small or large deformations of the classical black hole metric, namely, the Schwarzschild metric. One of the basic assumptions in this work is that the mimickers are described by static, spherically symmetric, and asymptotically flat metrics in the physical region. Although spherical symmetry is expected to be preserved even at the quantum level, the assumption of staticity is less evident. It is conceivable that the full quantum gravitational equations do not admit static solutions with horizons. In the present study, however, we restrict our attention to static metrics in order to maintain a close connection with the classical case and to simplify the analysis.
 
 Thus, we consider a general class of  metrics of the following form:
 \be
 ds^2=-g(\rho)dt^2+d\rho^2+r^2(\rho)(d\theta^2+\sin^2\theta d\phi^2)
 \lb{1}
 \ee
 It contains two functions $g(\rho)$ and $r(\rho)$, the possible form of which will be further constrained.
 Since our  starting point is the classical black hole we present below the form of the metric functions in this case,
 \be
 g_{\text{sch}}(r)=1-\frac{2M}{r}\, , \  \   \pm\rho=\int^r \frac{1}{\sqrt{g_{\text{sch}}(r)}}dr=\sqrt{r(r-2M)}+M\ln\frac{r+\sqrt{r-2M}}{r-\sqrt{r-2M}}
 \lb{2}
 \ee
 We use units in which $G=1$ and, moreover,  in our numerical computations  below in the paper we will often  set the mass parameter $M=1$.

 \begin{figure}[h!]
    \begin{minipage}[h!]{0.49\linewidth}
    \center{\includegraphics[width=8cm,height=5cm]{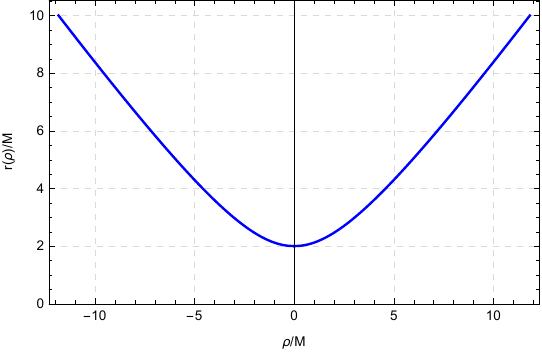}}\\ 
    \end{minipage}
    \hfill
    \begin{minipage}[h!]{0.49\linewidth}
    \center{\includegraphics[width=8cm,height=5cm]{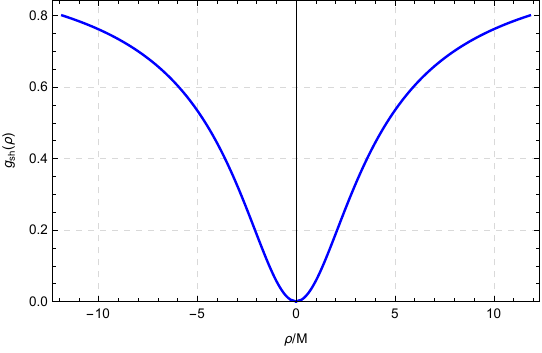}}\\
    \end{minipage}
    
    \caption{\footnotesize  Metric functions $r(\rho)$ and $g(\rho)$ as functions of the radial coordinate $\rho$.}
    \label{PlotMetricSCH}
\end{figure}
 
In the metric of the classical black hole (\ref{1})-(\ref{2}) the surface $\rho=0$ corresponds to the position of the black hole horizon, $r=2M$.
The positive and negative values of the radial coordinate $\rho$ correspond to the same values of $r\geq 2M$ so that one simply has a double-fold cover
 of this region. This is a $Z_2$ configuration. In the modified metrics that we consider below the interpretation of the regions $\rho>0$ and $\rho<0$ differs significantly.  The region $\rho>0$ will be referred as the {\it outer} or {\it physical} region while $\rho<0$ corresponds to the
 {\it  inner} region. 
 
 At asymptotic infinity in the physical region $\rho\rightarrow+\infty$,   one has that $r(\rho)=\rho -M\ln\rho+O(1)$ and $g_{\text{sch}}(\rho)=1-2M/\rho+O(\rho^{-2}\ln\rho)$.
 Near the horizon, $r=2M$ ($\rho=0$),  one has that 
 \be
 r(\rho)=2M+\frac{\rho^2}{8M}+O(\rho^4)\, , \  \   g_{\text{sch}}(\rho)=\frac{\rho^2}{16M^2}+O(\rho^4)\, .
 \lb{3}
 \ee
 The metric can be formally extended to negative values of $\rho$ so that the entire metric is $Z_2$ symmetric.
 We observe that the two functions $r(\rho)$ and  $g(\rho)$ develop a minimum at $\rho=0$, in which function $g(\rho)$ also vanishes.
 This is the horizon. Its regularity requires that the two-dimensional sphere at $\rho=0$  be a minimal surface, implying that
 $r'(\rho)=0$ at $\rho=0$.
 
Generalizing the general metric (\ref{1}) to a class of mimickers we assume that the global spacetime corresponds to all possible values of $\rho$,
 $-\infty<\rho<+\infty$. The behavior of the metric is quite different in the outer region $\rho>0$  and in the inner region $\rho<0$.
 Constraining the possible form of the metric we shall make the following assumptions:
 
 \vspace{0.5cm}
 
\noindent i) In the outer region, where $\rho>0$, the functions $g(\rho)$ and $r(\rho)$ are given by their classical form plus some small corrections.

\vspace{0.5cm}

\noindent ii) Function $g(\rho)$  is non-vanishing for any finite value of $\rho$. At $\rho=0$ (where the classical horizon used to stay), its value is small,
$g(\rho=0)\ll 1$, and it is determined by the small parameters present in the deformation of the classical metric.

\vspace{0.5cm}

\noindent iii) In the inner region both functions $g(\rho)$ and $r(\rho)$ may have either small or large deviations from the classical $Z_2$ configuration.

\vspace{0.5cm}

\noindent iv)   The global spacetime is geodesically complete both for null geodesics and time-like geodesics.

 \vspace{0.5cm}

\noindent v) The global spacetime is free from the curvature singularities.

\vspace{0.5cm}
 
\noindent  The two last conditions iv) and v) are the most restrictive. So that let us discuss them in more detail.

 \bigskip
 
 \noindent{\bf Geodesic completeness.} The null geodesics in the metric (\ref{1}) are described by the equation
 \be
 \frac{d\rho}{d\lambda}=\frac{E}{\sqrt{g(\rho})}
 \lb{4}
 \ee
if the geodesic is radial, i.e.  the respective impact parameter is zero (and the angular momentum $L=0$), and by the equation
 \be
 \frac{d\rho}{d\lambda}=\frac{L}{\sqrt{g(\rho})}\sqrt{\frac{E^2}{L^2}-V_{\text{null}}(\rho)}\, , \  V_{\text{null}}(\rho)=\frac{g(\rho)}{r^2(\rho)}
 \label{5}
 \ee
 The time-like geodesics, for any value of the angular momentum $L$, are described by the equation
 \be
  \frac{d\rho}{d\tau}=\frac{1}{\sqrt{g(\rho})}\sqrt{E^2-V_{\text{time}}(\rho)}\, , \  \ V_{\text{time}}(\rho)=g(\rho)(1+\frac{L^2}{r^2(\rho)})
  \lb{6}
  \ee
 We assume that in the asymptotic inner region, $\rho\rightarrow -\infty$, the asymptotic behavior is as follows
 \be
 g(\rho)\sim g_0(-\rho)^{\kappa_1}\, , \  \  \  r(\rho)\sim r_0(-\rho)^{\kappa_2}\, .
 \lb{7}
 \ee
 As it is seen from equation (\ref{4}) the completeness of the radial null geodesics imposes constraint only on function $g(\rho)$.
 Indeed, we find that
 \be
 \Delta \lambda =\frac{1}{E}\int^{-\infty} d\rho \sqrt{g(\rho}) =\infty  \, \  \Rightarrow \,  \  \kappa_1\geq - 2
 \lb{8}
 \ee
 This condition admits both positive and negative values of $\kappa_1$.
 For positive values of $\kappa_1$ function $g(\rho)$ grows to infinity in the asymptotic regime in the inner region, while for 
negative values of $\kappa_1$ function $g(\rho)$ decreases asymptotically  to zero in the inner region.
 
 \bigskip
 
 \noindent{\bf The radial potential.}  The condition of completeness of geodesics with the non-vanishing angular momentum $L$ does not impose extra
 constraints on parameters $\kappa_1$ and $\kappa_2$. However, the character of these geodesics can be quite different depending on values of
 $(\kappa_1, \kappa_2)$. Indeed, the effective radial potential for null-geodesics $V_{\text{null}}(\rho)$ (\ref{5}) asymptotically, in the inner region, behaves as
 $V_{\text{null}}(\rho)\sim(- \rho)^{\kappa_1-2\kappa_2}$. There are several cases to consider:
 
 \medskip
 
\noindent $A1$:  $\kappa_1-2\kappa_2<0$  the potential tends to vanish asymptotically.
 
 \medskip
 
 \noindent $A2$: $\kappa_1=2\kappa_2$  the potential approaches a constant.
 
 \medskip
 
 \noindent $A3$: $\kappa_1-2\kappa_2>0$  the potential diverges in the asymptotic region.
 
 \medskip
 
 Thus, in the inner region a non-radial null geodesic reaches (for the infinite values of the affine parameter, provided condition (\ref{8}) is satisfied) the asymptotic infinity in the case $A1$,  
 it may reach the asymptotic infinity for sufficiently small impact parameter $b=L/E$ in the case $A2$ and it will never reach the infinity for any values of $E/L$
 in the case $A3$. The latter case is quite interesting and deserves a separate analysis that will be done later in the paper.

 For the time-like geodesics the analysis is similar, although,  it involves more particular cases:
 
 \medskip
 
 \noindent $B1$:  $\kappa_1>0$  no time-like geodesic with any value of $L$  can reach infinity in the inner region.
 
  \medskip
  
  \noindent  $B2$: $ \kappa_1\leq 0$ and $\kappa_1-2\kappa_2>0$ only radial geodesic ($L=0)$ can reach the asymptotic region.
  
  \medskip
  
   \noindent $B3$: $\kappa_1<0$ and $\kappa_1=2\kappa_2$  geodesics with sufficiently large ratio  $E/L$ may reach the infinity.
   
   \medskip  
  
  \noindent $B4$:  $\kappa_1<0$ and $\kappa_1-2\kappa_2<0$  any geodesic reaches infinity for any values of $E$ and $L$.
 
 \medskip
 
 In resume, for certain values of parameters $\kappa_1$ and $\kappa_2$ we find an interesting behavior when the geodesics falling into the mimicker eventually come back. Such a mimicker then, for sufficiently long observation times,  is a smaller dark object than a black hole or even not dark at all. This of course depends on the travel time
 for the particle falling into the inner region that needed to come back. Such a travel time is supposed to be large so that the growing brightness of the
 initially dark object would be really slow. Later in the paper we analyse this and other aspects of the geodesics  in more detail for the test metrics to be present below.
 
  \bigskip
 
 \noindent{\bf Absence of curvature singularities.} The other important condition we impose on the black hole mimickers is their regularity. The curvature singularity  if it exists may appear either in the Ricci scalar $R$ or in the Riemann tensor.  For the metric (\ref{1}) we find
 \be
 R=-\frac{g''}{g}+\frac{1}{2}\left(\frac{g'}{g}\right)^2-\frac{2r'g'}{rg}-\frac{4r''}{r}+2\left(\frac{1-r'^2}{r^2}\right)
 \lb{9}
 \ee
 for the Ricci scalar and
 \be
 R_{\alpha\beta\mu\nu}R^{\alpha\beta\mu\nu}=2\frac{g''}{g}-\left(\frac{g'}{g}\right)^2+8\left(\frac{r'}{r}\right)^2+2\left(\frac{r'g'}{rg}\right)^2+4\left(\frac{1}{r^2}-\frac{r'^2}{r^2}\right)^2
 \lb{10}
 \ee
 for the square of the Riemann tensor.
 
 For the asymptotic metric (\ref{7}) we then find that
 \be
 &&R=\frac{2}{r_0\rho^{2\kappa_2}}+O(\frac{1}{\rho^2}) < \infty \Rightarrow \kappa_2\geq 0 \nonumber \\
 &&R_{\alpha\beta\mu\nu}R^{\alpha\beta\mu\nu}=\frac{4}{r_0^4\rho^{4\kappa_2}}-\frac{8\kappa_2^2}{\rho^{2\kappa_2+2}}+O(\frac{1}{\rho^2})< \infty \Rightarrow \kappa_2\geq 0 
 \lb{11}
 \ee
 Thus, the condition of regularity imposes constraint on the possible values of $\kappa_2\geq 0$. This constraint should be combined with
 the constraint (\ref{8}),
  \be
  \kappa_1\geq -2\, , \  \   \kappa_2\geq 0 \, .
  \lb{12}
  \ee
  These are the conditions to define a class of metrics that are geodesically complete and that are free from curvature singularities.

 \bigskip

 \noindent{\bf Travel time in the throat.}  For null geodesics, the travel time, in the clock of a distant observer,  between two points along the radial direction is defined by the  integral 
 \be
\Delta t =\int_{\rho_1}^{\rho_2} \dfrac{d\rho}{\sqrt{g(\rho)}} \, .
\lb{x1}
 \ee
 We assume that one point is lying in the outer region and the other is in the inner region. Between the two points there lies a region of very small values of $g(\rho)$.
 Clearly, the integral over this region is dominating in (\ref{x1}): the light spends most of its travel time in the throat, the region where the metric function has a minimum.
 If metric function $g(\rho)$ has a minimum at some point $\rho_{\text{min}}$ between these two points 
then the integral can be estimated by expansion around it
\be
\Delta t \sim \int_{\rho_1}^{\rho_2}\dfrac{d\rho}{\sqrt{g(\rho_{\text{min}}) + \dfrac{1}{2}g''(\rho_{\text{min}})(\rho-\rho_{\text{min}})^2}}
\ee
It will be assumed that the minimum $g(\rho_{\text{min}})$, if it exists, is small which allows us to estimate  the integral as
\be
\Delta t \sim \sqrt{\dfrac{2}{g''(\rho_{\text{min}})}}\ln\dfrac{1}{g(\rho_{\text{min}})}\, \, .
\label{travel time}
\ee
So that the travel time in the radial direction is set by the minimal non-zero value of the metric function $g(\rho_{\text{min}})$ in the throat.
 
 For a non-radial null geodesic characterized by the angular momentum $L$ the travel time as measured by a distant observer is
 \be
 \Delta t=\frac{E}{L}\int_{\rho_1}^{\rho_2}\frac{d\rho}{\sqrt{g(\rho)}\sqrt{\frac{E^2}{L^2}-V_{\text {nul}}(\rho)}}\, .
 \lb{y1}
 \ee
 It is infinite if the light ray follows the local maximum  of the potential, $E/L=\text{max}(V_{\text {nul}}(\rho))$.  
 Such a divergence is also present in the classical black hole case and is therefore not a distinctive feature of the mimicker geometry. 
 
 \bigskip
 
 \noindent{\bf Singular but geodesically complete metrics.}
  The condition of geodesic completeness does not restrict $\kappa_2$, so spacetimes with $\kappa_2<0$ are in principle possible. As follows from the analysis of curvature invariants, such spacetimes would have a singularity at asymptotic infinity in the inner region. However, since this singularity can be reached only at infinite values of the affine parameter, these configurations may still be physically acceptable. We, however, do not consider this possibility in the present paper.
 
 \bigskip
 
 \noindent{\bf The compact radial coordinate.}  In our analysis we find it useful,  what the radial coordinate is concerned,  to use a compact coordinate $x$ that changes in the
 range $-1\leq x\leq 1$ and is defined as
 \be
 x=\sqrt{1-\frac{2M}{r}}\, , \ \ g(x)=x^2\, .
 \label{compactcoord}
 \ee
 The transition from the radial coordinate $\rho$ to $x$  is given by relation
 \be
 \rho(x)=\frac{2Mx}{1-x^2}+M\ln \frac{1+x}{1-x}\, .
 \label{rhoincompactcoord}
 \ee

 \section{ Test metrics} 
 
 In order to explore  the different corners in the landscape of the black hole mimickers in more detail we, in this paper, consider
 several test metrics.  We stress that these metrics are not solutions to any gravitational equations we know. Their only purpose in this paper is to serve
 as certain examples of what may be expected for the mimicker metrics of certain type.
 
 In order to simplify the analysis we assume that in all these test metrics the function $r(\rho)$ is given by its classical expression
 that has minimum at $\rho=0$ so that it asymptotically  grows as $r(\rho)\sim \pm \rho$ when $\rho\rightarrow \pm \infty$. Thus, we set $\kappa_2=1$
 in all our examples. The general form (\ref{1}) of the metric is assumed in all examples.
 
 \bigskip
 
 \subsection{Metric I:  the non-symmetric wormhole}
 \label{sec:metric1desc} 
 
 First, we consider a metric in which the metric function approaches a constant value in the asymptotic inner region.
 This metric is a generalization of the DS wormhole metric
 \cite{Damour:2007ap}. In the DS metric the function $g(\rho)$ develops a minimum at exactly the same value of $\rho=0$
 where $r(\rho$) has its minimum. This is the symmetric case.  The global wormhole spacetime is $Z_2$ symmetric.
 In the minimum $g(\rho=0)=b^2>0$ is non-zero.   The respective radial potential is $Z_2$ symmetric and has two identical peaks symmetrically located with respect to $\rho=0$.
 Here we want to generalize this to a non-symmetric case while keeping same asymptotic behavior of the metric. So that the metric is no more $Z_2$ symmetric.
 The respective test metric is then given by the metric function
 \be
 g_{\text{I}}(\rho)=\left(\sqrt{g_{\text{sch}}(\rho)}+a\right)^2+b^2
 \lb{13}
 \ee
 where $g_{\text{sch}}(\rho)$ is the function in the classical black hole metric (\ref{2})-(\ref{3}).
 The parameter $a$ is then responsible for the violation of $Z_2$-symmetry. 
 Function (\ref{13}) asymptotically  approaches $g_{\text{I}}(\rho)\rightarrow  (1\pm a)^2+b^2$  where the plus sign  stands for the outer region and the minus sign  is for the inner region.
 It is required to have $g_{00}=-1$ in the  asymptotic outer infinity. This could be achieved 
by rescaling appropriately the function (\ref{13}). This, however, makes the entire expression rather complicated. Therefore, we prefer to redefine
the time coordinate $t$ to time $t_\infty$ measured by an observer at outer infinity, $t_\infty=\sqrt{(1+a)^2+b^2}\,t$. Where applicable  it is assumed that this redefinition is used.

\begin{figure}[h!]
    \begin{minipage}[h!]{0.49\linewidth}
    \center{\includegraphics[width=8cm,height=5cm]{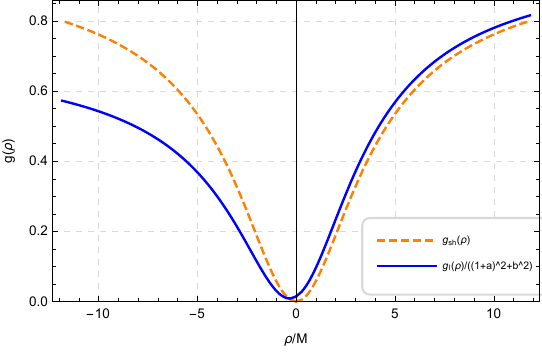}}\\ 
    \end{minipage}
    \hfill
    \begin{minipage}[h!]{0.49\linewidth}
    \center{\includegraphics[width=8cm,height=5cm]{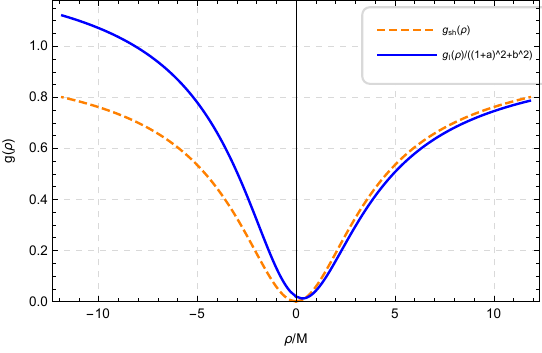}}\\
    \end{minipage}
    \caption{\footnotesize The metric function $g_{\text{I}}(\rho)$ for two sets of parameters compared with the function $g_{\text{sch}}(\rho)$ \eqref{2}. \textbf{The left plot} corresponds to the parameters $b = 0.1$ and $a = 0.08$. \textbf{The right plot} corresponds to the parameters $b=0.1$ and $a=-0.08$.}
    \label{PlotMetric1}
\end{figure}
 
 In terms of the compact radial coordinate $x$ function $(\ref{13})$ has a  simple expression
 \be
 g_{\text{I}}(x)=(x+a)^2+b^2\, .
 \lb{14}
 \ee
 
 It takes a minimal value at $x_{\text{min}}=-a$ that corresponds to $r_{\text{min}}=\frac{2M}{1-a^2}$. The respective value of coordinate $\rho$ is 
 \be
 \rho_{\text{min}}=-\frac{2Ma}{1-a^2}+M\ln\frac{1-a}{1+a}\, .
 \lb{15}
 \ee 
 We note that in the non-symmetric case (\ref{13}) the position $(\rho=0$) of the minimum of function $r(\rho)$ does not
 coincide with the position $(\rho=\rho_{\text{min}})$ of the minimum of function $g(\rho)$.
 The size of the wormhole throat, or, equivalently, the light travel time, can be estimated by the equation \eqref{travel time},
 \be
 \Delta t\sim \dfrac{4M}{(1-a^2)^2}\ln{\dfrac{1}{b^2}}\, .
 \label{traveltime1}
 \ee
 Note, that its order of magnitude, as in the symmetric case, is  determined by the parameter $b$.

 \bigskip
 
 \subsection{Metric II: infinite tube wormhole}

 The second metric that we consider is the one in which the metric function 
 is almost identical to the DS metric in the outer region while
 in the inner region the metric function $g(\rho)$ is monotonically decreasing to zero in the asymptotic inner region.
 This combined behavior is achieved for the following metric function, 
 \label{sec:metric2disc}
 \be
 g_{\text{II}}(\rho)=\left(g_{\text{sch}}(\rho)+b^2\right)\Delta(\rho)\, , \  \   \Delta(\rho)=\frac{1}{2}\left(1+\frac{\rho}{\sqrt{\rho^2+\rho_0^2}}\right)\, .
 \lb{metricII}
 \ee
 In this case we have two small parameters: $b$ and $\rho_0>0$. In the limiting case when both parameters vanish the metric function  (\ref{metricII}) becomes the one for the
 classical black hole.
 
\begin{figure}[h!]
    \centering
    \resizebox{1\linewidth}{!}{ % Масштабируем всю вставку
        \begin{minipage}{\linewidth}
           % \begin{minipage}{0.49\linewidth}
           \begin{minipage}{0.43\linewidth}
                \includegraphics[width=\linewidth]{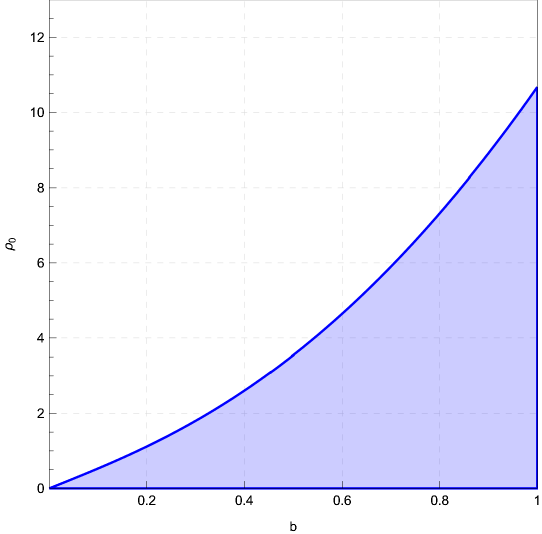}
            \end{minipage}
            \begin{minipage}{0.45\linewidth}
                \begin{minipage}{\linewidth}
                    \includegraphics[width=\linewidth]{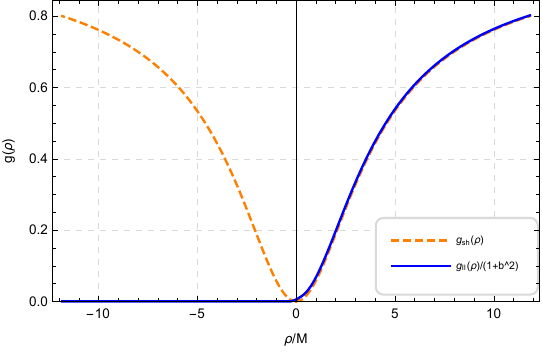}                
                \end{minipage}
                \vspace{0.5em}
                \begin{minipage}{\linewidth}
                    \includegraphics[width=\linewidth]{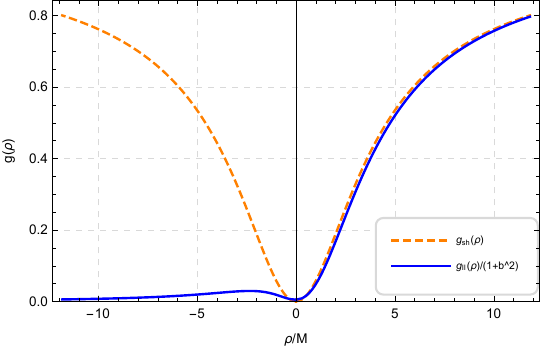}               
                \end{minipage}
            \end{minipage}
        \end{minipage}
    }
    \caption{\footnotesize \textbf{Left:}  region  in space of parameters  $b$ and $\rho_0>0$ for which the metric function $g_{\text{II}}(\rho)$ increases monotonically. \textbf{Right:}  the metric function $g_{\text{II}}(\rho)$ for two different values of parameters and the comparisons with the Schwarzschild metric function $g_{\text{sch}}(\rho)$. \textbf{The top plot} corresponds to the monotonic case with parameters $b=0.1$ and $\rho_=0.01$, \textbf{the bottom plot} corresponds to a non-monotonic case with parameters $b=0.1$ and $\rho_0=2$.}
    \label{PlotMetric2}
\end{figure}

 Function $\Delta(\rho)$ approximates the step function.
 Its role is to add just a small correction for positive $\rho$ and make the metric function $g_{\text{II}}(\rho)$
 approaching zero in the regime of large negative $\rho$. Depending on parameters $\rho_0$ and $b$ the function (\ref{metricII}) may develop some
 local minima. However, here   we are mostly interested in the case when $g_{\text{II}}(\rho)$  monotonically grows from $0$ at $\rho=-\infty$ to
 $1$ at $\rho=+\infty$  since in the non-monotonic case it would have some properties that are similar to that of metric I. Indeed,  in a non-monotonic case there 
 would appear a second non-symmetric peak in the potentials as for the metric I. Although the potentials would differ at infinity, the peak structure in the effective radial potential
 in both cases would be similar, leading to similar effects.  That is why we consider the monotonic case only.
 In Fig.~\ref{PlotMetric2}  it is shown a region in the parameter space $(b, \rho_0)$ for which the metric function (\ref{metricII}) is monotonic.
For small parameters, the boundary of the region can be  approximated by equation
\be
\rho_0 < 3\sqrt{3}b + 6\sqrt{3}b^3 - 15\sqrt{3}b^5+\dots\, .
\ee
 At $\rho=0$ the metric function (\ref{metricII})  takes value $g_{\text{II}}(\rho=0)=b^2/2$, so that for any $\rho<0$ in the inner region one has that  $g_{\text{II}}(\rho)<b^2/2$.
 Thus, in the inner region the spacetime represents an infinite tube in which the metric function  $g_{\text{II}}(\rho)$ is small. For large negative values
 one finds that $g_{\text{II}}(\rho)=\frac{(1+b^2)\rho_0^2}{4\rho^2}$ so that one has $\kappa_1=-2$ in this case. 
 Since the metric function  approaches $g_{\text{II}}\rightarrow 1+b^2$ at the infinity in the outer region, the proper physical time of an observer at infinity is given by $t_{\infty} = \sqrt{1+b^2}t$.

 \subsection{Metric III:  wormhole with semi-permeable  wall}  
 \label{sec:metric3disc}
 
 In the other two  test metrics the metric function grows to infinity when approaching the asymptotic infinity in the inner region. 
 First we consider the case when the metric function grows as $\sim(\rho)^2$ so that $\kappa_1=2$ and $\kappa_1=2\kappa_2$.
 The test metric then takes the form,
  \be
 g_{\text{III}}(\rho)=\left(g_{\text{sch}}(\rho)+b^2\right)\Delta^{-1}(\rho)\,  \  \  
 \lb{metric3}
 \ee
 One has $g_{\text{III}}(\rho=0)=2b^2$.  For large negative $\rho$ it behaves as $g_{\text{III}}(\rho)\sim (-\rho)^2$ as we wanted. Since the metric function $g_{\text{III}}\rightarrow 1+b^2$ at infinity in the inner region, the proper physical time of an observer at infinity is given by $t_{\infty} = \sqrt{1+b^2}\, t$.

\begin{figure}[h!]
    \centering
    \resizebox{0.6\linewidth}{!}{ 
            \begin{minipage}{0.45\linewidth}
                \includegraphics[width=\linewidth]{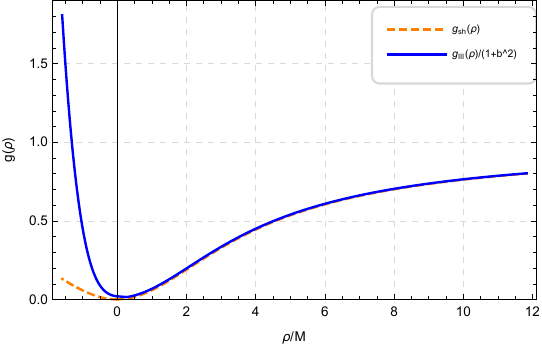}
            \end{minipage}
    }
    \caption{\footnotesize  The shape of metric function $g_{\text{III}}(\rho)$  in test metric III  for parameters $b=0.1$, $\rho_0=1$ and the comparison with the Schwarzschild metric function $g_{\text{sch}}(\rho)$.}
    \label{PlotMetric3}
\end{figure}

The shape of the metric function is shown in Fig.~\ref{PlotMetric3}.  It has a  minimum that is  located in the physical region and is determined by the equation 
\be
\dfrac{g'_{\text{sch}}(\rho)}{g_{\text{sch}}(\rho)+b^2} = \dfrac{\sqrt{\rho^2+\rho_0^2}-\rho}{\rho^2+\rho_0^2}\, .
\label{metric3minimum}
\ee
An analysis of this equation and an estimation of the travel time  \eqref{travel time} is given in appendix \ref{sec:apptraveltime3and4}.

\bigskip

 \bigskip
 
 \subsection{Metric IV:  wormhole with impenetrable wall}
 In the fourth test metric the metric function grows faster than in the case of the metric III when the asymptotic infinity in the inner region is approached.
 The metric function has the form,
  \be
 g_{\text{IV}}(\rho)=\left(g_{\text{sch}}(\rho)+b^2\right)\Delta^{-n}(\rho)\, , \  \  n>1
 \lb{metric4}
 \ee
 One has that $g_{\text{IV}}(\rho=0)=2^n b^2$. Asymptotically, at $\rho\rightarrow-\infty$, one finds that  $g_{\text{IV}}(\rho)\sim (-\rho)^{2n}$. So that in this case $\kappa_1=2n$ and hence $\kappa_1-2\kappa_2>0$ for $n>1$. At infinity in the inner region one has that $g_{\text{IV}}\rightarrow 1+b^2$. Hence, the proper physical time of a distant observer is given by $t_{\infty} = \sqrt{1+b^2}\, t$.
 
 Let us note an important feature of this metric: the coordinate travel time for a radial geodesic to infinity in the inner region  is finite, unlike the previous metrics that were considered. Indeed, since $g_{\text{IV}}(\rho)\sim (-\rho)^{2n}$ and $n>1$ one finds that asymptotically in the inner region, 
 \be
\Delta t_{\infty} \sim \int_{\rho_1}^{-\infty}\dfrac{d\rho}{(-\rho)^n} < \infty
\label{traveltimetoinfinity}
 \ee
This is similar to the behavior of geodesics in AdS space. Since the coordinate time $t$ is unbounded this implies that the radial massless geodesics should reflect back from the infinity. In fact, there are no massive geodesics or non-radial massless geodesics that could reach asymptotic infinity in the inner region, as the corresponding radial potentials have an impenetrable infinite wall. Metric function in this case has a shape similar to that shown in Fig.~\ref{PlotMetric3}, only it grows faster when the inner infinity is approached. 

The position of the minimum of the function $g_{\text{IV}}(\rho)$ is governed by an equation similar to \eqref{metric3minimum},
\be
\dfrac{g'_{\text{sch}}(\rho)}{g_{\text{sch}}(\rho)+b^2} = n\dfrac{\sqrt{\rho^2+\rho_0^2}-\rho}{\rho^2+\rho_0^2}\, .
\label{metric4minimum}
\ee
The analysis of this equation and an estimation of  the travel time is given in appendix \ref{sec:apptraveltime3and4}.

 \bigskip

\section{Shadows and constraints on parameters}
\label{sec:Shadow and constraints to the parameters}

In this section, we will discuss the optical properties of the  test metrics.  We will  also use the data available from
the EHT observations of Sgr$A^{*}$ and $M87^{*}$ to constrain the deformation parameters.
We will compute  the light shadow for each of the test metric and compare it with the observations.

The light shadow originates from a maximum in the effective radial potential for a light geodesic. Such a maximum corresponds to a photonic sphere formed by unstable closed 
light geodesics.
 The details of the computation, that is rather elementary, are given for instance in  \cite{Perlick:2021aok}, \cite{Vagnozzi:2022moj}. 
 For the Schwarzschild black hole there is only one such photon sphere which is located at $r = 3M$. For the $Z_2$ symmetric mimickers (see  \cite{Damour:2007ap}) there are two identical maxima  of the radial potential, one in the inner region and the other is in the outer region.

In the non-symmetric case, the radial potential typically exhibits two maxima of different heights: one in the physical region (located close to 
$3M$) and another in the inner region. When the maximum in the inner region is higher, an interesting situation arises in which a shadow, associated with the maximum of the potential in the physical region, appears first. At a later time, depending on the travel time through the throat, a distant observer will see the shadow gradually shrink. The shadow decreases to a size determined by the photon sphere in the inner region and may even vanish entirely in the case of an impenetrable wall.

Below, we first compute the radius of the light shadow in a general static, spherically symmetric spacetime. We then discuss the shadow and the corresponding observational constraints for each test metric using the EHT data.

 \bigskip

\subsection{Shadow radius in a generic spherically symmetric space-time}
\label{sec:shadowgeneral}
Let us consider a general static, spherical symmetric and asymptotically flat (in physical region) space-time, i.e. one admit a global, non-vanishing, time-like Killing vector and Killing vector associated with rotation. 
\be
ds^2 = -g(\rho)dt^2 + d\rho^2 + r^2(\rho)d\Omega^2
\ee
where $d\Omega$ is differential unit of solid angle. Since spacetime is spherical symmetric we consider motion in equatorial $\theta = \pi/2$ plane. The equation of motion for a massless geodesic is
\be
 \left(\dfrac{d\rho}{d\lambda}\right)^2 = \dfrac{L^2}{g(\rho)}\left(\dfrac{1}{b^2}\right. - \underbrace{\dfrac{g(\rho)}{r^2(\rho)}}_{V_{\text{eff}}(\rho)}\left.\vphantom{\dfrac{g(\rho)}{r^2(\rho)}}\right)\, ,
\ee
where $b = L/E$ is the impact parameter, $L$ is angular momentum and $E$ is energy. We are interested in the photon sphere of this metric. It is defined by the properties of circularity and instability \cite{Perlick:2021aok}, \cite{Vagnozzi:2022moj},
\be
\left.\dfrac{d\rho}{d\lambda}\right|_{\rho_{\text{ph}}} = 0,\quad
\left.\dfrac{d^2\rho}{d\lambda^2}\right|_{\rho_{\text{ph}}} = 0\, .
\ee
From these equations one finds the critical parameter $b_{\text{ph}}$ and the position $\rho_{\text{ph}}$ of the photon sphere,
\be
\dfrac{1}{b^2_{\text{ph}}} = \dfrac{g(\rho_{\text{ph}})}{r^2(\rho_{\text{ph}})}, \quad \left.\dfrac{d}{d\rho}V_{\text{eff}}(\rho)\right|_{\rho_{\text{ph}}} = 0
\ee
Let an observer be located at a sufficiently large radial coordinate 
$\rho_0>0$.  A light ray reaching the observer forms an angle $\alpha$
with the radial direction,
\be
\cot ^2 \alpha = \dfrac{1}{r^2(\rho_o)}\left.\left(\dfrac{d\rho}{d\varphi}\right)^2\right|_{\rho_o},\quad \sin ^2 \alpha = b^2\dfrac{g(\rho_o)}{r^2(\rho_o)}\, .
\ee
The shadow is formed by the rays coming from  the photon sphere.  The angular size of the shadow $\alpha_{sh}$ is found from the equation (see \cite{Perlick:2021aok})
\be
\sin ^2 \alpha_{sh} = b^2_{\text{ph}}\dfrac{g(\rho_o)}{r^2(\rho_o)} = \dfrac{r^2(\rho_{\text{ph}})}{g(\rho_{\text{ph}})}\dfrac{g(\rho_{o})}{r^2(\rho_{o})}\, .
\ee
One finds for the geometric size $R_{\text{sh}}$ of the shadow, 
\be
\frac{R_{\text{sh}}}{M} = \frac{r(\rho_o)}{M}\sin \alpha_{sh} = \frac{r(\rho_{\text{ph}})}{M}\sqrt{\dfrac{g(\rho_o)}{g(\rho_{\text{ph}})}}\, .
\label{shadowradius}
\ee
For a distant  observer  $g(\rho_o)$ is  approximated by the asymptotic value of the metric function. For the Schwarzschild black hole, as an example,  one has that $g_{\text{sch}}(r) = 1 -2M/r$ and  $r_{\text{ph}} = 3M$  so that  $R_{\text{sh}} = 3\sqrt{3}M$.

\bigskip

\subsection{Event horizon telescope data}

We will constrain the deformation parameters of the classical black hole metric using data from the EHT experiment. The EHT collaboration was the first to obtain images of compact objects located at the centers of the M87 galaxy \cite{EventHorizonTelescope:2019ggy} and the Milky Way galaxy \cite{EventHorizonTelescope:2022xqj}. In these images, a bright emission ring surrounds a dark region corresponding to the black hole shadow. From the perspective of a distant observer, the edge of this shadow delineates the photon region - the boundary of spacetime where closed spherical photon orbits exist, separating captured trajectories from those that scatter. Under certain conditions, the radius of the bright ring can serve as an approximate indicator of the actual shadow radius. In what follows, we employ the $1\sigma$
 constraints on 
$R_{\text{sh}}/M$  derived from the EHT data for M87$^\ast$ \cite{EventHorizonTelescope:2019ggy} and for  SgrA$^\ast$ \cite{EventHorizonTelescope:2022xqj},
\be
\text{M}87^{*}: \quad 4.31\leq R_{\text{sh}}/M \leq 6.08\nonumber\\
\text{SgrA}^{*}: \quad 4.55\leq R_{\text{sh}}/M \leq 5.22\nonumber
\label{constraintshadow}
\ee
In what follows, these experimental data will be used to constrain the deformation parameters of the four test metrics.

According to the previous discussions \cite{EventHorizonTelescope:2022xqj}, \cite{Vagnozzi:2022moj}, \cite{Johannsen:2010ru}, \cite{EventHorizonTelescope:2020qrl}, the shadow size and the deviation from circularity depend only weakly on the spin and the inclination with respect to the observer. Additionally, the estimated rotation of the observed
astrophysical objects is rather small \cite{Fragione:2020khu}. This allows us, following the
previous discussion, to use the data given above to place constraints on
the parameters of a static spherically symmetric metric.
% This allows us to use the data given above to place constraints on static and spherically symmetric metrics.}

% \textbf{We will use the data given above to place constraints on spherically symmetric metrics. This is justified for several reasons. We can neglect neglect the effect of spin and inclination with respect to the observer, because the size and shape of the shadow depend only weakly on these parameters \cite{EventHorizonTelescope:2022xqj}, \cite{Vagnozzi:2022moj}, \cite{Johannsen:2010ru}. For example, for the Kerr metric the deviation from circularity is within about $7\%$ \cite{EventHorizonTelescope:2020qrl}. At the same time, recent model-independent estimates of the spin parameter of $\text{Sgr A}^{\ast}$ \cite{Fragione:2020khu}, \cite{Fragione:2022oau} provide the following constraint on the spin parameter $J/M < 0.1$.}

\subsection{Shadow in metric I}
\label{sec:massless_potential1}
\begin{figure}[h!]
    \begin{minipage}[h!]{0.45\linewidth}
    \center{\includegraphics[width=8cm,height=5cm]{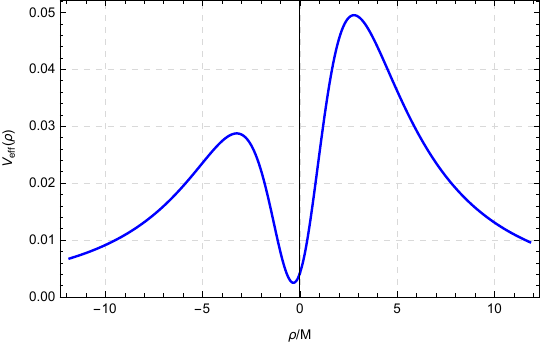}}\\ 
    \end{minipage}
    \hfill
    \begin{minipage}[h!]{0.45\linewidth}
    \center{\includegraphics[width=8cm,height=5cm]{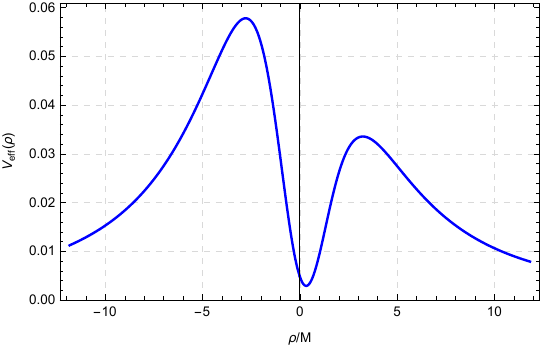}}\\
    \end{minipage}
    
    \caption{\footnotesize The effective radial potential for null geodesics for two sets of the parameters: $b = 0.1\, , \  a=0.08$ (\textbf{left}) and   $b=0.1\, , \ a=-0.08$ (\textbf{right}).}
    \label{PlotVeff1}
\end{figure}

In  the test metric I  the metric function is given by (\ref{13}).
Figure~\ref{PlotVeff1} shows the effective radial potential for null geodesics. We find that the sign of the parameter $a$
strongly affects the behavior of lightlike geodesics. We consider each case of the  signs in detail.

\bigskip

\bigskip
\noindent{\bf     The case  $a > 0 $. }
     In this case the maximum of the potential that  is located in the physical region $\rho>0$ is higher  than the maximum in the inner region. Thus, a light-like geodesic that was emitted at $\rho = \infty$ and whose impact parameter allows it to go above the maximum in the physical region will necessarily continue till $\rho=-\infty$. This case is essentially similar to  the classical black hole and is described in the same way as in Section~\ref{sec:shadowgeneral}. The only difficulty  is to find the position  of maximum of the effective potential $V_{\text{eff}}(\rho) = g(\rho)/r^2(\rho)$. It is more convenient to do this using the compact coordinate x, (\ref{compactcoord}), since the effective potential becomes polynomial if expressed in terms of $x$ and, thus, it is easier to analyze,
     \be
     V_{\text{eff}}(x) = \dfrac{(1-x^2)^2}{4}((x+a)^2+b^2)\, .
     \ee
One then finds,
     \be
     \dfrac{d}{dx}V_{\text{eff}}(x) = \frac{1}{2} \left(1-x^2\right) \left(-2 x \left((a+x)^2+b^2\right)+\left(1-x^2\right)(a+x)\right) = 0\, .
     \label{eqnshadow1}
     \ee
    
\begin{figure}[h!]
    \begin{minipage}[h!]{0.49\linewidth}
    \center{\includegraphics[width=8cm,height=6cm]{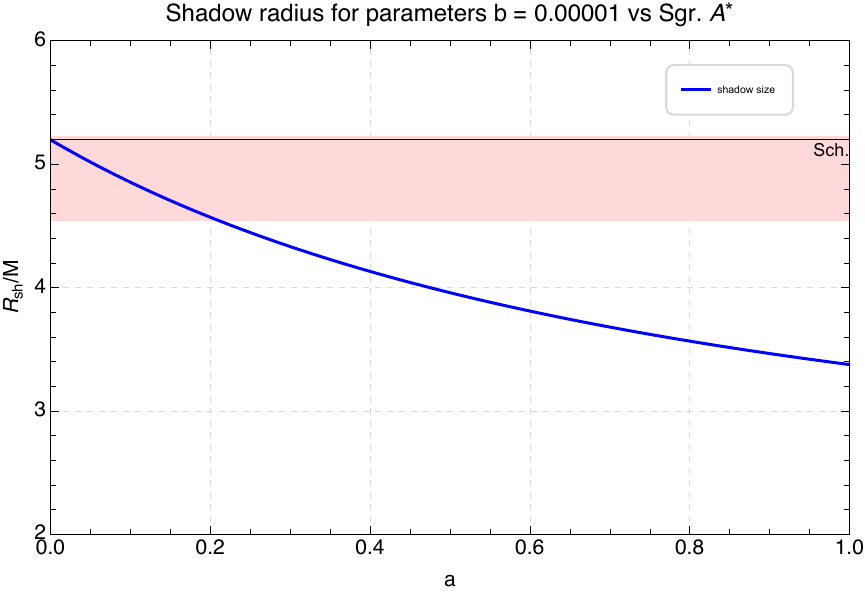}}\\ 
    \end{minipage}
    % \hfill
    \begin{minipage}[h!]{0.49\linewidth}
    \center{\includegraphics[width=8cm,height=6cm]{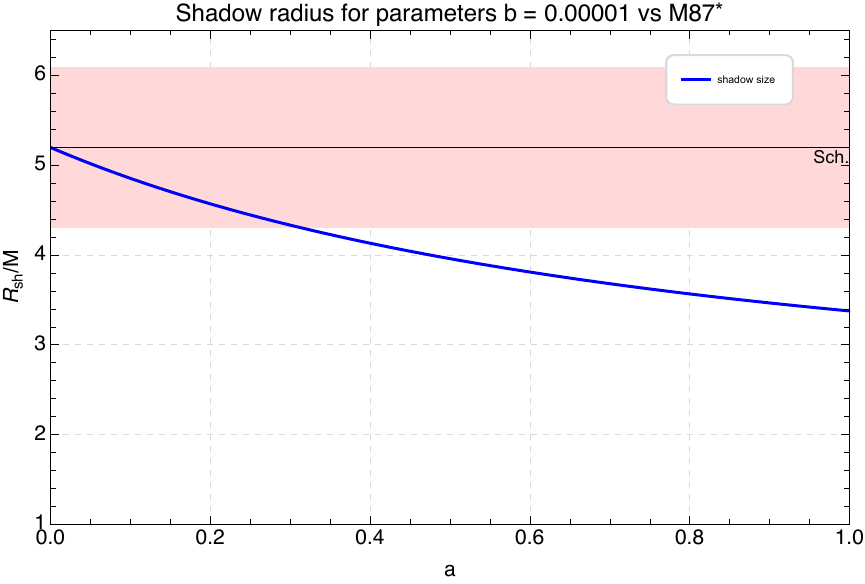}}\\
    \end{minipage}
    \begin{minipage}[h!]{0.49\linewidth}
    \center{\includegraphics[width=8cm,height=6cm]{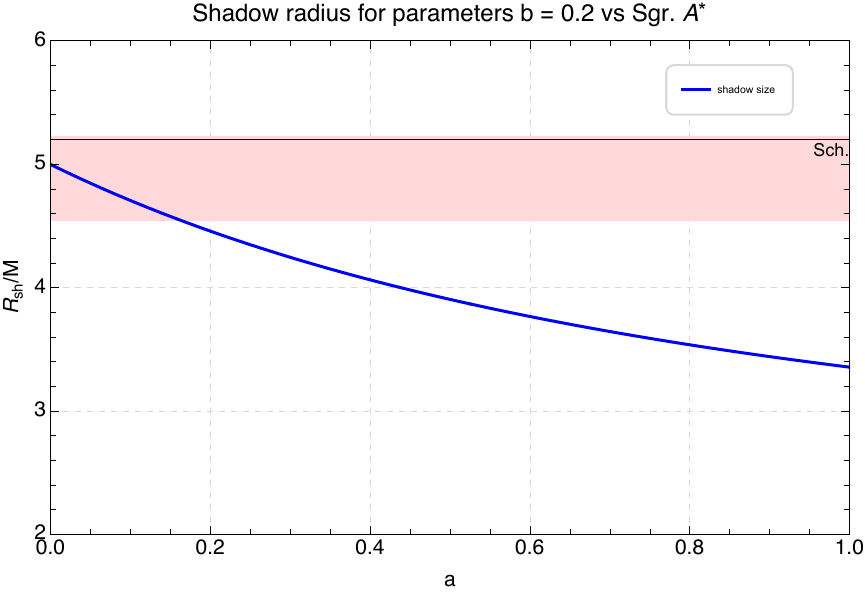}}\\
    \end{minipage}
    \hfill
    \begin{minipage}[h!]{0.49\linewidth}
    \center{\includegraphics[width=8cm,height=6cm]{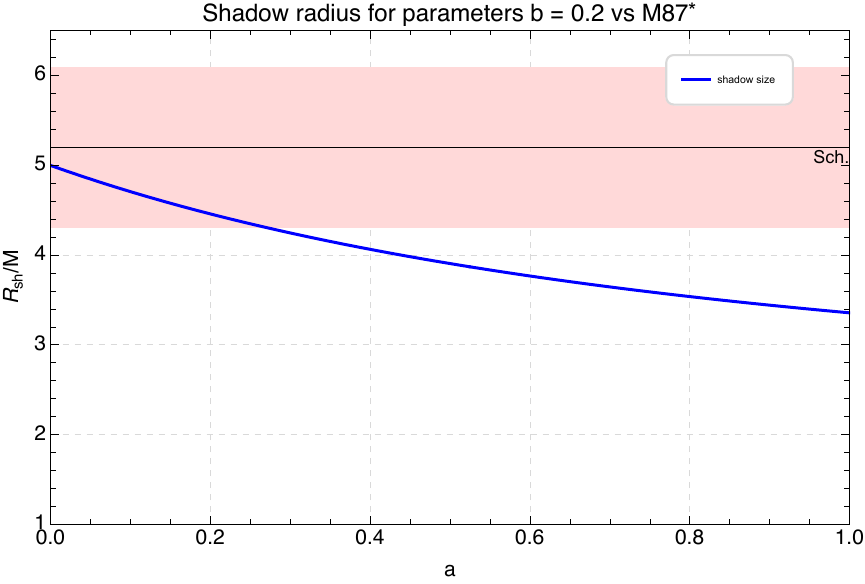}}\\
    \end{minipage}
    \begin{minipage}[h!]{0.49\linewidth}
    \center{\includegraphics[width=8cm,height=6cm]{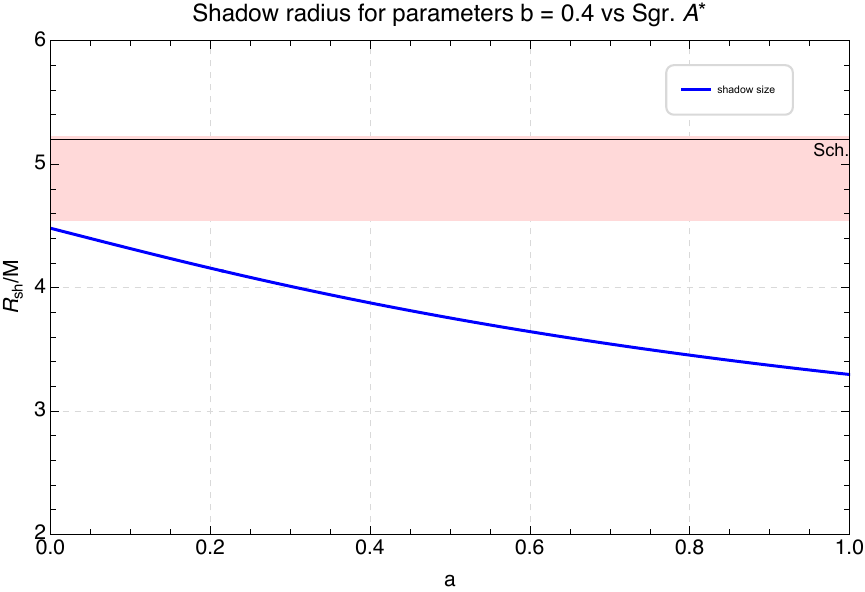}}\\
    \end{minipage}
    \hfill
    \begin{minipage}[h!]{0.49\linewidth}
    \center{\includegraphics[width=8cm,height=6cm]{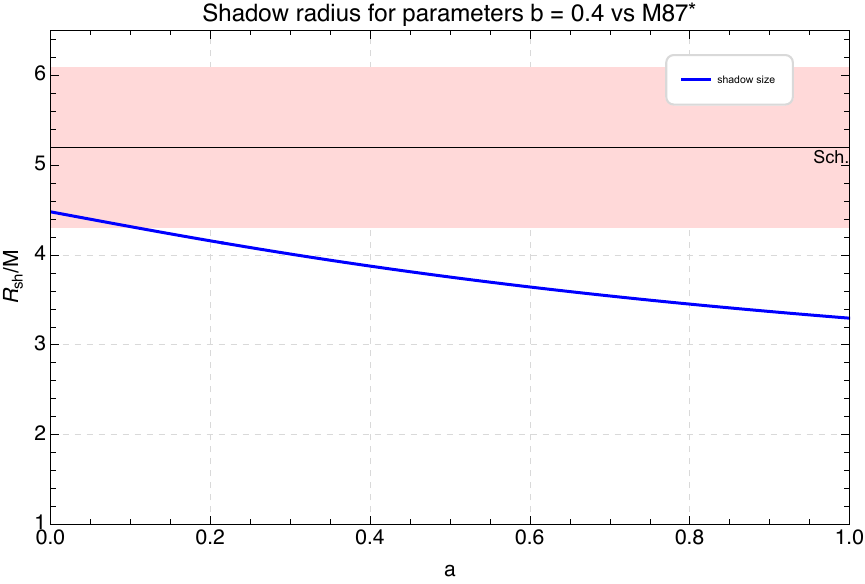}}\\
    \end{minipage}
    \caption{\footnotesize The dependence of the radius of the shadow (blue line) as a function of the parameter $a$ for various values of the parameter $b$ for the metric I. 
    The pink area is $1\sigma$-band for M87$^\ast$ (\textbf{right column}) and Sgr A$^\ast$ (\textbf{left column}). Solid gray line is the shadow size for the Schwarzschild black hole.}
    \label{Shadows1}
\end{figure}

Ignoring the factor $(1-x^2)$ that vanishes at asymptotic infinity, the right hand side of this equation is a cubic equation. It   can be solved exactly.
The solution can be presented as expansion for small values of parameter $a$ that controls the deviation from the $Z_2$ symmetry, 
     \be
     x_{\text{ph},\pm} = \pm\sqrt{\dfrac{1-2b^2}{3}} - \dfrac{1-5b^2}{3-6b^2}~a\pm\dfrac{\sqrt{3}}{18}~\dfrac{1-16b^2+b^4}{\left(1-2b^2\right)^{5/2}}~a^2 + \dots
     \ee
   Using this  and  equation \eqref{shadowradius}, one can find, perturbatively  in $a$, the size of the shadow formed by each maxima,
   
    \begin{gather}
    R_{\text{sh},\pm}/M = \sqrt{\dfrac{4\left((1+a)^2+b^2\right)}{\left(1-x_{\text{ph},\pm}^2\right)^2\left((x_{\text{ph},\pm}+a)^2+b^2\right)}} = \dfrac{3\sqrt{3}}{1+b^2}+\dfrac{3\left(\sqrt{3}\mp 3\sqrt{1-2b^2}\right)}{\left(1+b^2\right)^2}a+\nonumber\\
     +\dfrac{3\sqrt{3}\left(5+15b^4\mp 2\sqrt{3-6b^2}-b^2\left(22\mp 4\sqrt{3-6b^2}\right)\right)}{2\left(1+b^2\right)^3\left(1-2b^2\right)}a^2+\dots
    \label{shadowradius1}
    \end{gather}   
  For $a>0$ the maximum of the potential that lies in the physical region is the highest and the size of the shadow is $R_{\text{sh},+}$. For $a<0$ the maximum of the potential that lies in the inner region is the highest and the size of the shadow is $R_{\text{sh},-}$.

Fig.~\ref{Shadows1} shows  the shadow radius as function of parameter $a$ for various values of parameter $b$ as well as  the $1\sigma$-constraint from the Sgr A$^\ast$ \cite{EventHorizonTelescope:2022xqj} and M87$^\ast$ \cite{EventHorizonTelescope:2019ggy}. As the parameter $b$ increases, the curve describing the dependence of the shadow radius on the parameter $a$ goes down. For  a certain critical value of $b \lesssim0.376$, it completely goes outside the constraint region obtained from Sgr A$^\ast$ experiment. This critical value agrees with the constraint found earlier  in  \cite{Vagnozzi:2022moj}. 

\begin{figure}[h!]
    \centering
            \begin{minipage}{0.45\linewidth}
                \includegraphics[width=\linewidth]{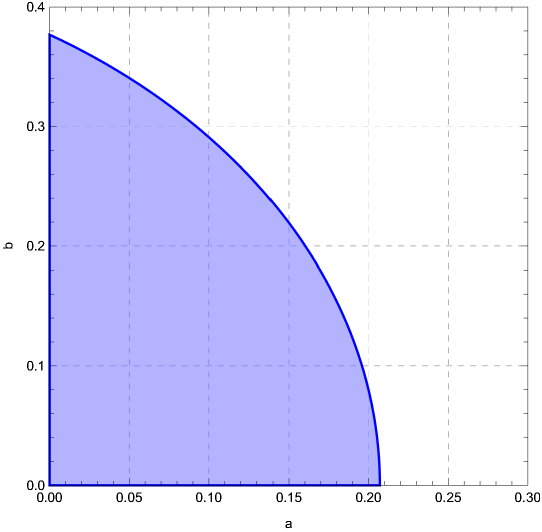}
            \end{minipage}
            \begin{minipage}{0.4\linewidth}
                \includegraphics[width=\linewidth]{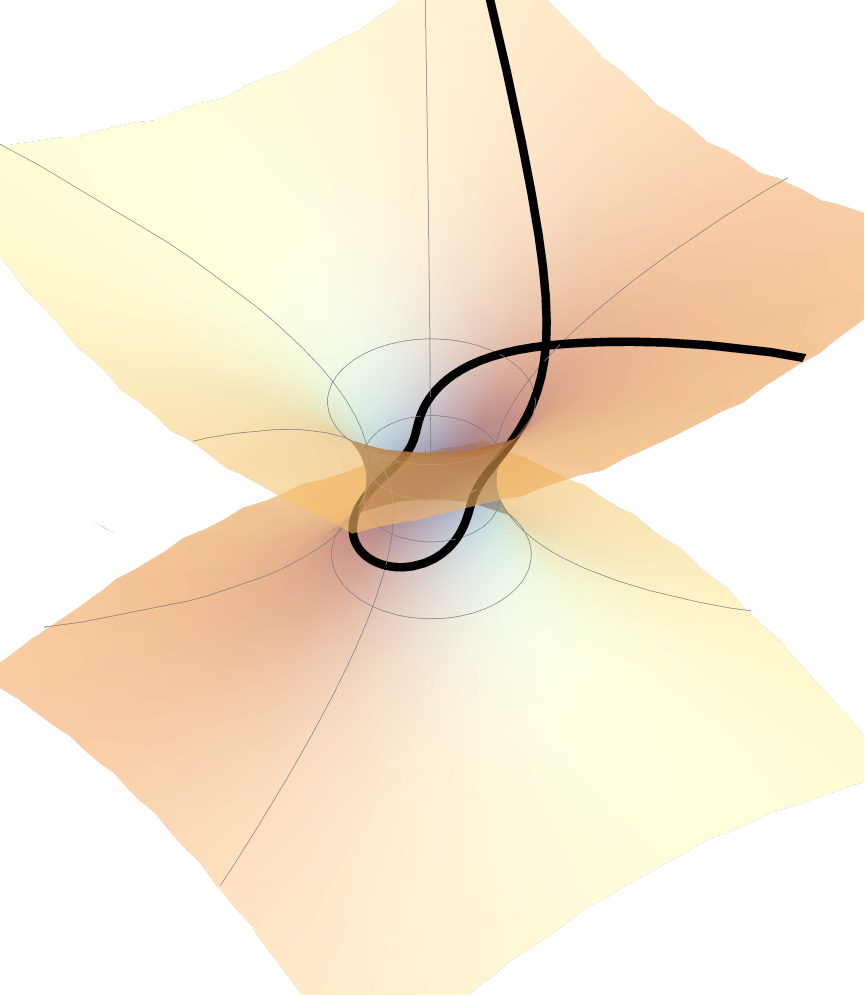}
            \end{minipage}
    \caption{\footnotesize \textbf{Left:}  the region of parameters $a>0$ and $b$ for which the shadow radius is in $1\sigma$-bands for Sgr A$^\ast$ experiment. \textbf{Right:}  the trajectory of a photon with $L=5$ in the background metric I with the parameters $a = -0.08$ and $b = 0.1$. The trajectory passes through the throat, enters the inner region, and, reflecting off the potential, returns back in the outer region.}
    \label{ShadowConstr1}
\end{figure}

\bigskip

\noindent{\bf  The case $a<0$.}  In this case the maximum of the radial potential in the inner region is the highest. So that  there exist light-like geodesics whose impact parameter allow them to go above  the maximum of the potential in physical region ($\rho>0$), but  below the maximum that lies  in the inner region ($\rho<0$). 
Such a trajectory is shown in Fig.~\ref{ShadowConstr1}. It can be seen that the geodesic starts in the outer region, passes through the throat into the inner region, but then returns back to the outer region. If we neglect the travel time \eqref{travel time} due to penetrating  through the throat, the shadow should be formed by the maximum of the potential in the inner region.

However, since the travel time can be considerably large, a distant observer will initially see the shadow due to the maximum of the potential in the physical region.
Only later, when the light-like geodesics whose impact parameter allows them to penetrate into the inner region  will reach the maximum in the inner region
and come back, the observer will see that  the shadow size will decrease to the one determined by the maximum of the potential in the inner region. The travel time in this case is estimated as
$\Delta t\sim 8M\ln{1/b}$, it can be considerably large depending on how small is parameter $b$.

\begin{figure}[h!]
    \begin{minipage}[h!]{0.49\linewidth}
    \center{\includegraphics[width=8cm,height=6cm]{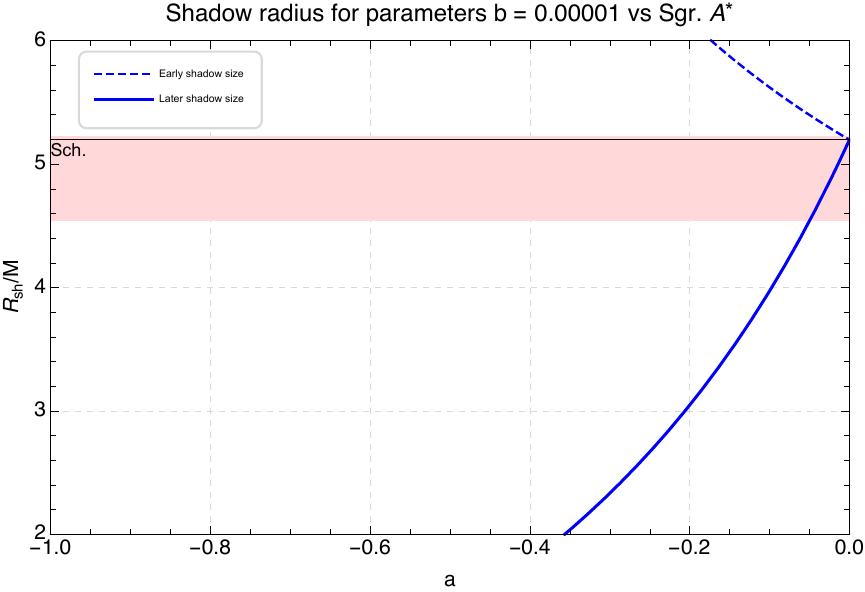}}\\ 
    \end{minipage}
    % \hfill
    \begin{minipage}[h!]{0.49\linewidth}
    \center{\includegraphics[width=8cm,height=6cm]{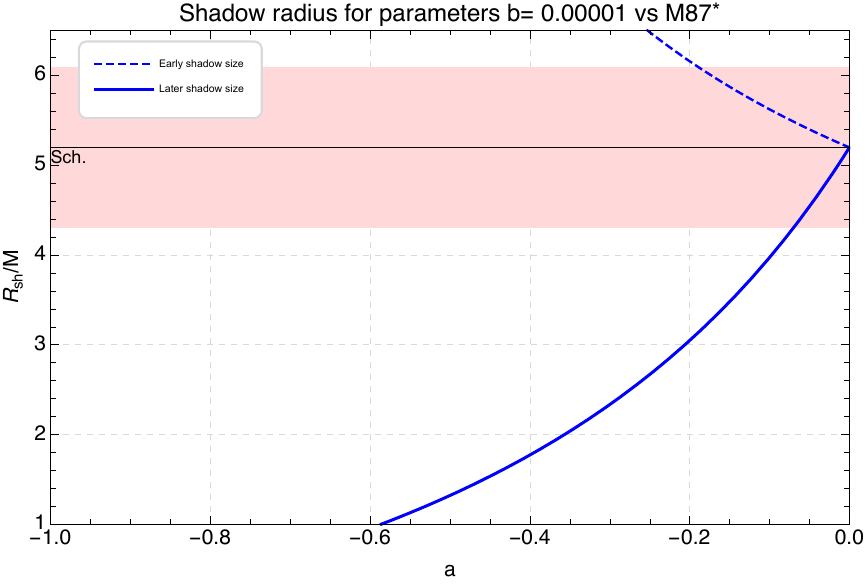}}\\
    \end{minipage}
    \begin{minipage}[h!]{0.49\linewidth}
    \center{\includegraphics[width=8cm,height=6cm]{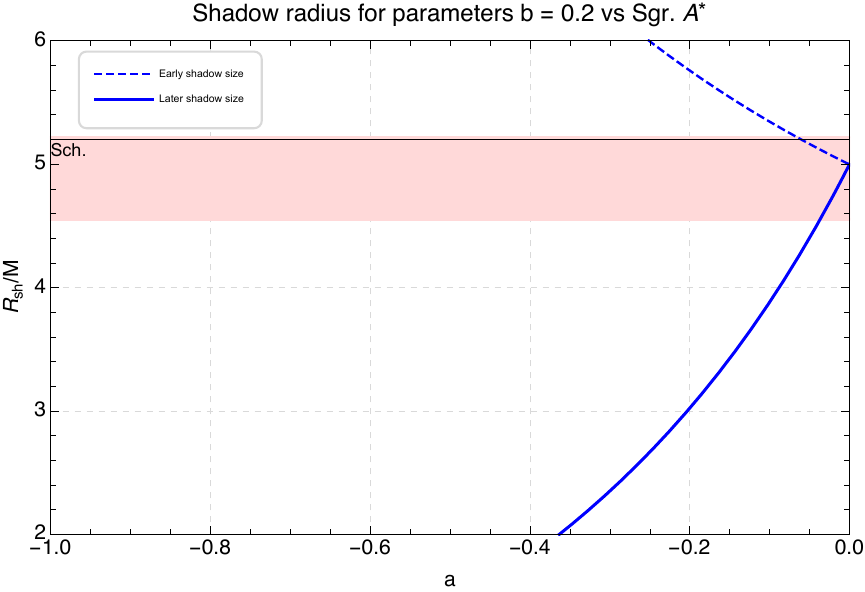}}\\
    \end{minipage}
    \hfill
    \begin{minipage}[h!]{0.49\linewidth}
    \center{\includegraphics[width=8cm,height=6cm]{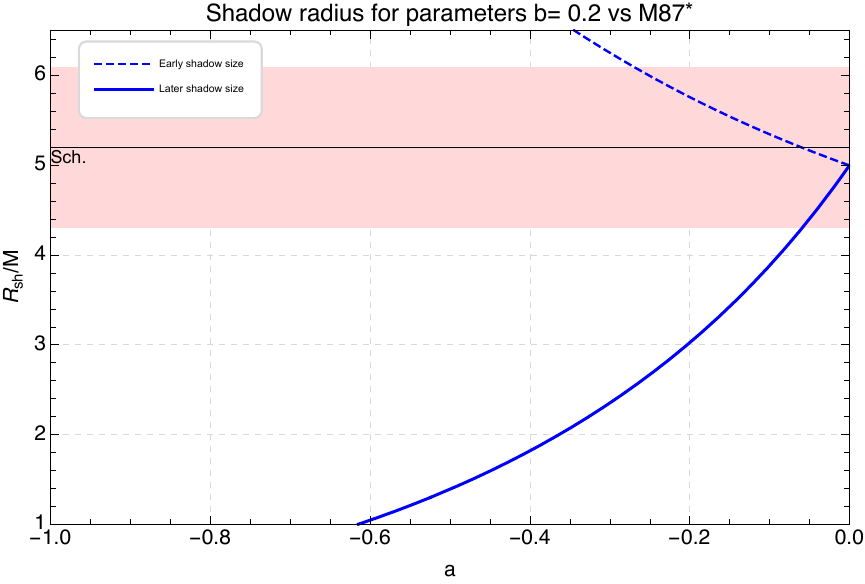}}\\
    \end{minipage}
    \begin{minipage}[h!]{0.49\linewidth}
    \center{\includegraphics[width=8cm,height=6cm]{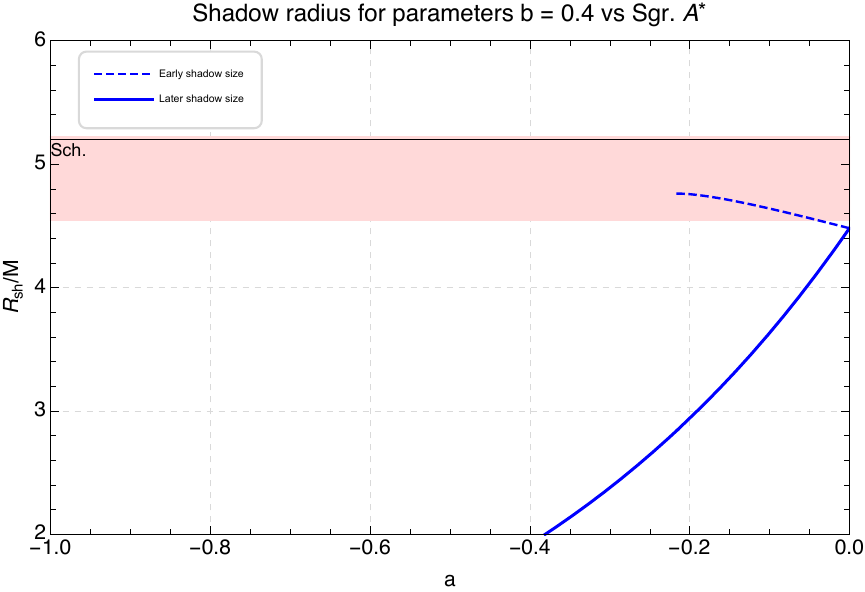}}\\
    \end{minipage}
    \begin{minipage}[h!]{0.49\linewidth}
    \center{\includegraphics[width=8cm,height=6cm]{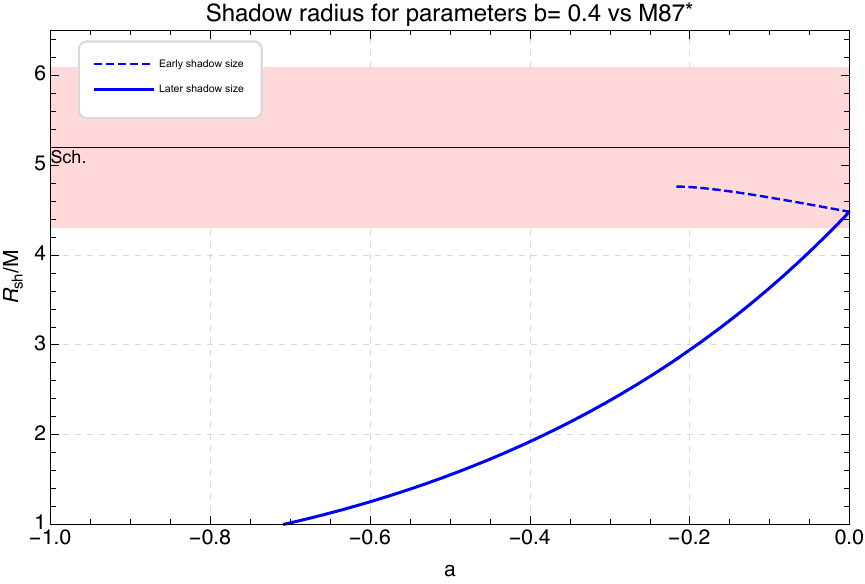}}\\
    \end{minipage}
    \caption{\footnotesize The dependence of the radius  of the late shadow (blue solid lines) and the early shadow (blue dashed lines) on the parameter $a$ for various values of parameter $b$ for metric I. The pink area is $1\sigma$-bands for M87$^\ast$ (\textbf{right column}) and Sgr A$^\ast$ (\textbf{left column}). Solid gray line is the shadow size for the Schwarzschild black hole.}
    \label{Shadows12}
\end{figure}

\begin{figure}[h!]
    \centering
            \begin{minipage}{0.4\linewidth}
                \includegraphics[width=\linewidth]{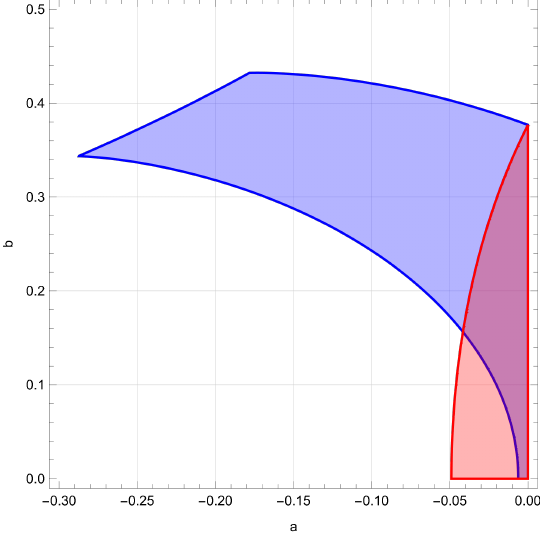}
            \end{minipage}
    \caption{\footnotesize The region of parameters $a$ and $b$ for which the shadow radius is in $1\sigma-$bands for Sgr A$^\ast$ experiment. The blue region is for the early-time shadow and the red region is for the late-time shadow.}
    \label{Const12}
\end{figure}

It should be noted that the shadow size does not decrease directly from the larger value to the smaller one. Instead, the transition begins at a radius located between the two shadow radii. At this stage, the shadow starts to brighten in both directions. This behavior can be explained by the fact that both shadows are associated with the maxima of the effective potential, which correspond to unstable stationary points. Consequently, the travel time from these points to any other location is infinite, whereas the travel time from the turning point of the potential lying between the two maxima is finite. Hence, there exists a particular impact parameter, approximately equal to the average of the impact parameters corresponding to the two maxima, for which the travel time is minimal. For this value of the impact parameter, the first bright ring appears inside the initial shadow, corresponding to photons reflected by the potential in the inner region. As this ring expands, the apparent shadow size decreases, and the late-time shadow fully emerges.

In Fig.~\ref{Shadows12}, we present the shadow radius together with the  $1\sigma$-constraints from Sgr A$^\ast$
 \cite{EventHorizonTelescope:2022xqj} and M87$^\ast$
 \cite{EventHorizonTelescope:2019ggy}. We refer to the shadow formed by the maximum in the physical region, when photons have not yet reached the maximum
  in the inner region, as the early-time shadow. In the figure, it is shown by a blue dashed line. This line terminates at specific values of the parameter 
$a$, since for certain parameter ranges one of the maxima disappears, leaving only the maximum located in the physical region. The shadow formed by the maximum in the inner region, which does not appear immediately in observations, is referred to as the late-time shadow. In Fig.~\ref{Shadows12}, it is indicated by a solid blue line.

The region of constraint on the parameters $a$ and $b$, obtained by using  the data from Sgr A$^\ast$ observations \cite{EventHorizonTelescope:2022xqj}, is presented in Fig.~\ref{Const12}. It is worth noting that, due to the fact that  with different parameters the shadow can be either only an early shadow or transitioning into a late shadow, two regions for negative parameter $a$ are marked in the figure. The blue region is for the early-time shadow, while the red region is for the late-time shadow. The intersection of these regions contains those parameters at which the transition from the early-time shadow to the late-time shadow can be observed, while both of these shadows will remain within the $1\sigma$-bands for Sgr A$^\ast$ experiment.

The parameter space constrained by the Sgr A$^\ast$ observations \cite{EventHorizonTelescope:2022xqj} is shown in Fig.~\ref{Const12}. It is worth noting that, depending on the parameter values, the shadow can correspond either to an early-time shadow or to a transition into a late-time shadow. As a result, two regions for negative values of the parameter 
$a$. The blue region corresponds to the early-time shadow, while the red region represents the late-time shadow. The intersection of these regions identifies the parameter values for which the transition from the early-time to the late-time shadow can occur, with both shadows remaining within the 
$1\sigma$-bands of the Sgr A$^\ast$ observations.

\subsection{Shadow in metric II}
In the test metric II the respective metric function is (\ref{metricII}).
We will consider only the monotonic version of metric II. This means that the parameters  $b$ and $\rho_0$
must be constrained to lie within the region shown in Fig.~\ref{PlotMetric2}.
 Since $g_{\text{II}}(\rho)$ is monotonic the radial potential $V_{\text{null}}(\rho)=\frac{g(\rho)}{r^2(\rho)}$ is monotonically growing in the inner region and hence there could be only one maximum lying in the physical region. This is similar to the Schwarzschild metric. In the limit $\rho\rightarrow -\infty$ the potential behaves as $V_{\text{null}}(\rho)\sim 1/\rho^4$.
It is convenient to use  the compact coordinate $x$ \eqref{compactcoord} that we introduced earlier. The effective radial potential then takes the form
\be
V_{\text{null}}(x) = \dfrac{1}{4}\left(1-x^2\right)^2\left(x^2+b^2\right)\Delta\left(\rho(x)\right)
\ee
and a position of the photon sphere is determined by the following equation 
\be
\dfrac{d}{dx}V_{\text{null}}(x) = \Delta(\rho(x))\left[\dfrac{1}{2}x\left(1-x^2\right)\left(1-2b^2-3x^2\right)+\left(x^2+b^2\right)\Delta^{-1}\left(\rho(x)\right)\dfrac{d}{d\rho}\Delta\left(\rho(x)\right)\right] = 0
\label{eqnshadow2}
\ee
For the last term in this equation we find,
\be
\Delta^{-1}\left(\rho(x)\right)\dfrac{d}{d\rho}\Delta\left(\rho(x)\right) = \dfrac{\sqrt{\rho^2(x)+\rho_0^2} - \rho(x)}{\rho^2(x)+\rho_0^2}\, .
\label{usefull3.2}
\ee
Equation \eqref{eqnshadow2} will be solved numerically to find the position of the photon sphere. This solution and \eqref{shadowradius} yield the shadow radius,
\be
R^2_{\text{sh}}/M^2 = \dfrac{4\left(1+b^2\right)}{\left(1-x_{\text{ph}}^2\right)^2\left(x_{\text{ph}}^2+b^2\right)\Delta\left(\rho(x_{\text{ph}}\right)}\, ,
\ee
where $x_{\text{ph}}$ is the solution of the \eqref{eqnshadow2}.

\begin{figure}[h!]
    \centering
            \begin{minipage}{0.5\linewidth}
                \includegraphics[width=\linewidth]{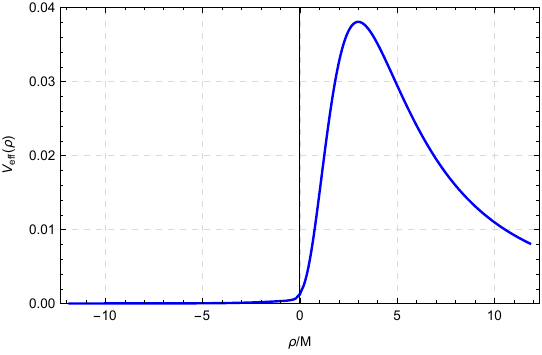}
            \end{minipage}
    \caption{\footnotesize The effective radial  potential for a light geodesic in metric II for  parameters $b = 0.1$ and $\rho_0 = 0.3$.}
    \label{Potential2}
\end{figure}

\begin{figure}[h!]
    \begin{minipage}[h!]{0.49\linewidth}
    \center{\includegraphics[width=8cm,height=6cm]{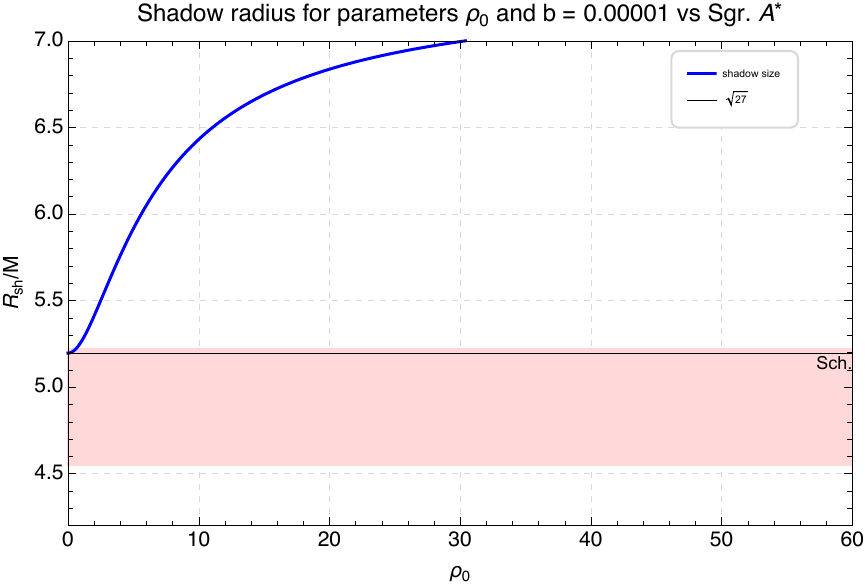}}\\ 
    \end{minipage}
    % \hfill
    \begin{minipage}[h!]{0.49\linewidth}
    \center{\includegraphics[width=8cm,height=6cm]{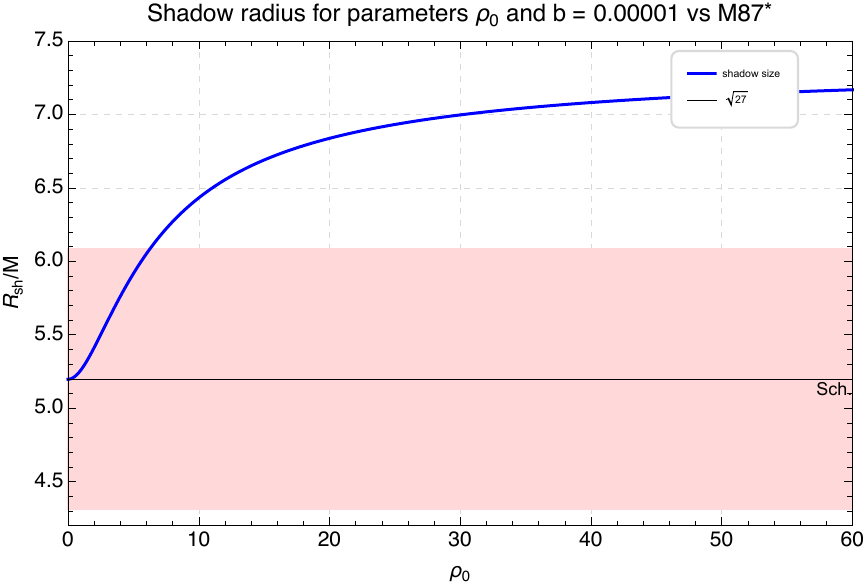}}\\
    \end{minipage}
    \begin{minipage}[h!]{0.49\linewidth}
    \center{\includegraphics[width=8cm,height=6cm]{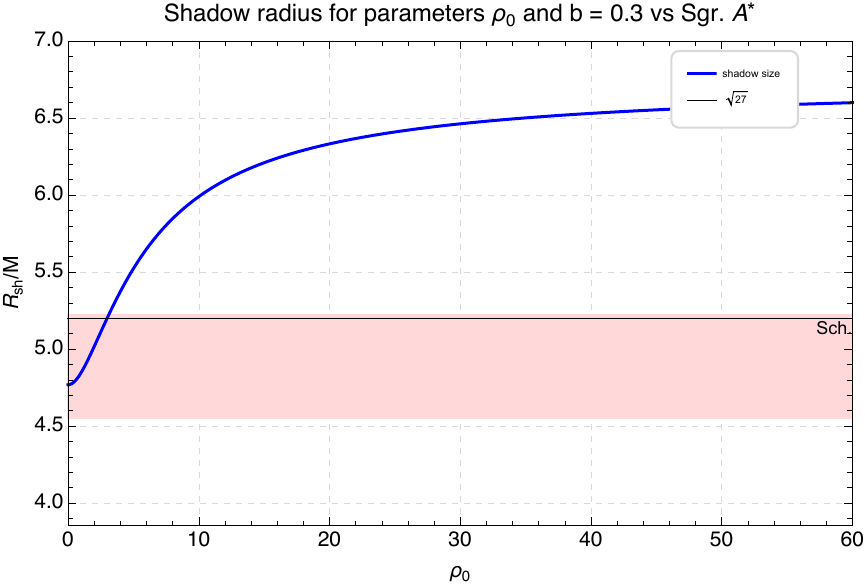}}\\
    \end{minipage}
    \hfill
    \begin{minipage}[h!]{0.49\linewidth}
    \center{\includegraphics[width=8cm,height=6cm]{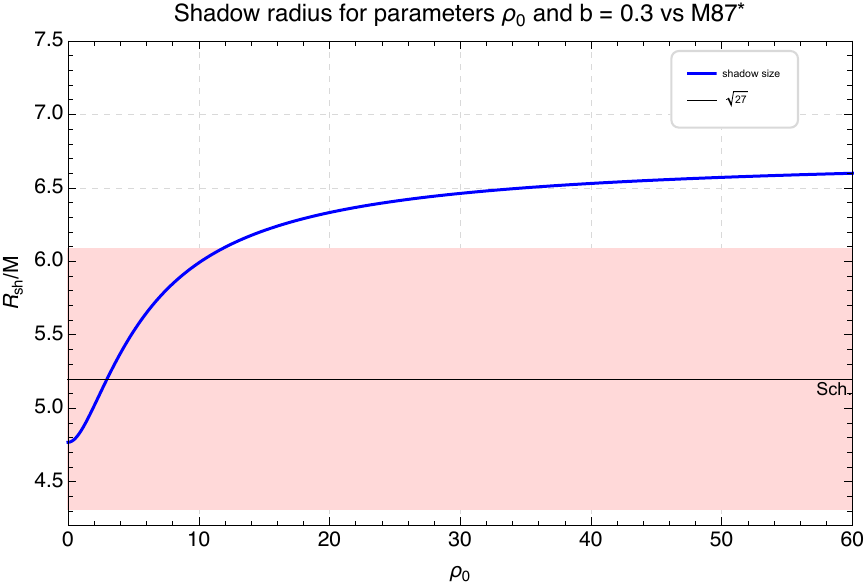}}\\
    \end{minipage}
    \begin{minipage}[h!]{0.49\linewidth}
    \center{\includegraphics[width=8cm,height=6cm]{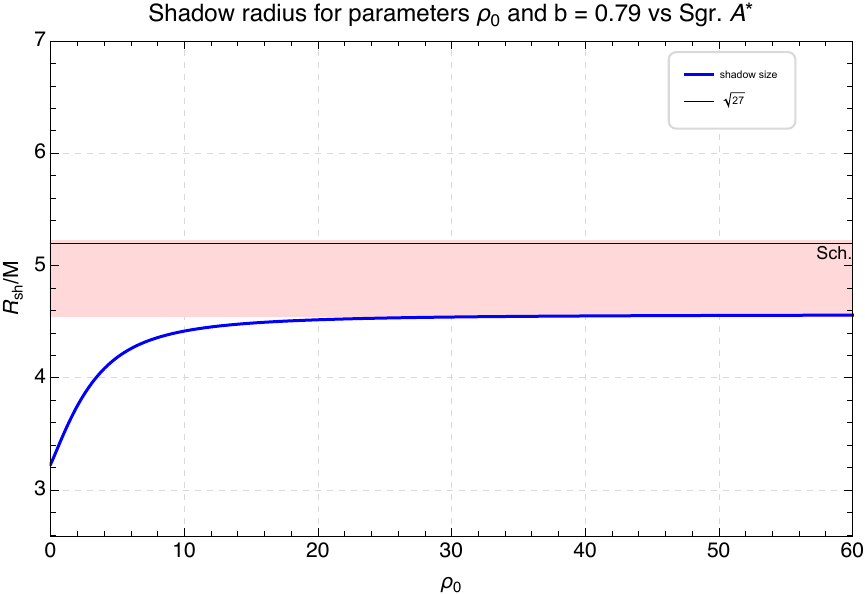}}\\
    \end{minipage}
    \begin{minipage}[h!]{0.49\linewidth}
    \center{\includegraphics[width=8cm,height=6cm]{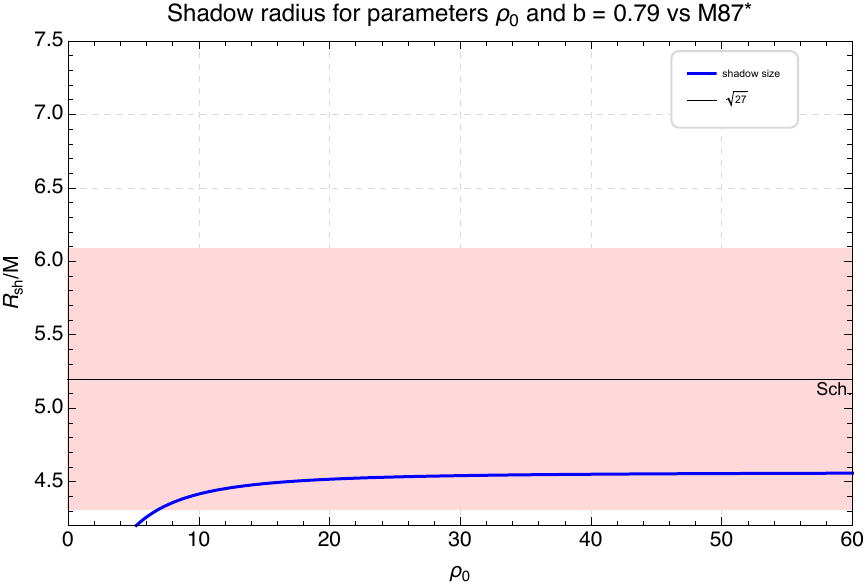}}\\
    \end{minipage}
    \caption{\footnotesize The dependence of the radius of the shadow (blue line) on the parameter $\rho_0$ for various values  of parameter $b$ for metric II. The pink area is $1\sigma$-bands  for M87$^\ast$ (\textbf{right column}) and Sgr A$^\ast$ (\textbf{left column}).  Solid gray line is the shadow size in the Schwarzschild case.}
    \label{Shadows2}
\end{figure}

\begin{figure}[h!]
    \begin{minipage}[h!]{0.49\linewidth}
    \center{\includegraphics[width=7cm,height=6cm]{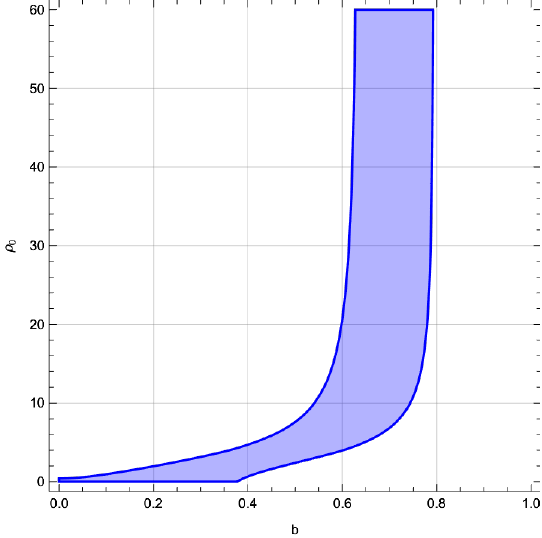}}\\ 
    \end{minipage}
    % \hfill
    \begin{minipage}[h!]{0.49\linewidth}
    \center{\includegraphics[width=7cm,height=6cm]{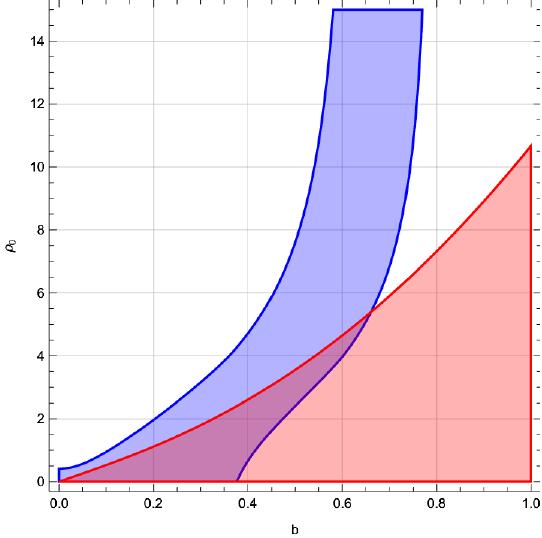}}\\
    \end{minipage}
   \caption{\footnotesize \textbf{Left:} Region  in space of parameters $\rho_0$ and $b$ for which the shadow size is within the $1\sigma$-bands for Sgr A$^\ast$.
   \textbf{Right:}  Intersection of this region and the region in which  the metric function is monotonic.}
    \label{Const2}
\end{figure}

In Fig.~\ref{Shadows2} we present the  shadow size as function of the parameters $\rho_0$ and $b$. It can be seen that the graph increases as a function of the parameter $\rho_0$ and the graph  as the whole goes down  when the parameter $b$ increases. Therefore, there is a sufficiently large range of parameters (the left panel in Fig.~\ref{Const2} for which the shadow size is within $1\sigma$-bands).

Since we are interested in the monotonic case, the emergent constraints on parameters $\rho_0$ and $b$  should be combined with those that come from the condition of the monotonicity of the metric function, shown in the right panel of Fig.~\ref{Const2}. It can be seen that the parameters of interest lie in the intersection of these two regions.


\subsection{Shadow in metric III}

\begin{figure}[h!]
    \begin{minipage}[h!]{0.49\linewidth}
    \center{\includegraphics[width=8cm,height=6cm]{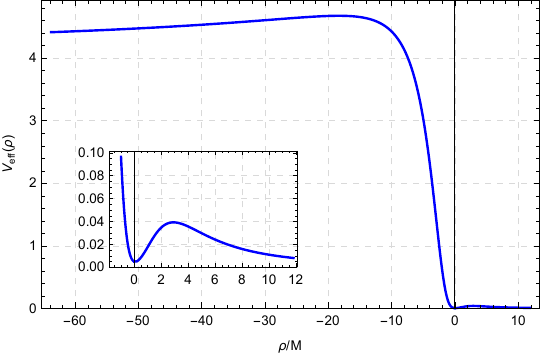}}\\ 
    \end{minipage}
    % \hfill
    \begin{minipage}[h!]{0.49\linewidth}
    \center{\includegraphics[width=8cm,height=6cm]{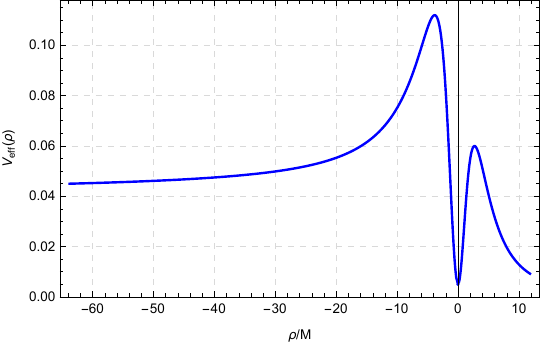}}\\
    \end{minipage}
    \caption{\footnotesize \textbf{Left:}  the effective radial potential for  a light geodesic in metric III for the parameters $b = 0.1$ and $\rho_0 = 1$. The insert shows the presence of a maximum in the physical region. \textbf{Right:}  the radial potential  for the parameters $b = 0.1$ and $\rho_0 = 10$. The potential has maximum both in the physical (outer) and  in the inner regions.  Both maxima are sharp in this case.
    }
    \label{Potential3}
\end{figure}

In the test  metric III the respective metric function is given by (\ref{metric3}).
This metric is similar to the Schwarzschild metric in the physical region ($\rho>0$). It grows in the inner region ($\rho<0$) as $g_{\text{III}}(\rho)\sim (-\rho)^{2}$. 
The respective radial potential  approaches a constant value,
\be
V_{\text{null}}(\rho) = \frac{g_{\text{III}}(\rho)}{r^2(\rho)} \sim \dfrac{4\left(1+b^2\right)}{\rho_0^2}\, ,
\ee
in the limit $\rho\rightarrow -\infty$.
This potential still has a local maximum lying in the physical region similarly to the Schwarzschild case and the other maximum in the inner region as shown in Fig.~\ref{Potential3}. 
Since the two maxima of the radial potential have different heights, the resulting shadow decreases over time, similar to what was described for metric I.

 A light ray, whose impact parameter is such that it goes above  the maximum of the potential in the outer region, when falling into the wormhole, will be reflected off the potential in the inner region and then come to a distant observer. The light rays, that go over the both maxima of the radial potential, will eventually escape to the asymptotic infinity in the
 inner region. The eventual late-time shadow is, thus, due to the unstable circular orbits at the position of the highest maximum of the potential that lies in the inner region.

The shadow formed by the maximum of the potential that lies in the physical region will be referred to as the early-time  shadow, similar to the case of metric I, while the shadow formed by the maximum in the inner region will be called the late-time shadow. 

\begin{figure}[h!]
    \begin{minipage}[h!]{0.49\linewidth}
    \center{\includegraphics[width=8cm,height=6cm]{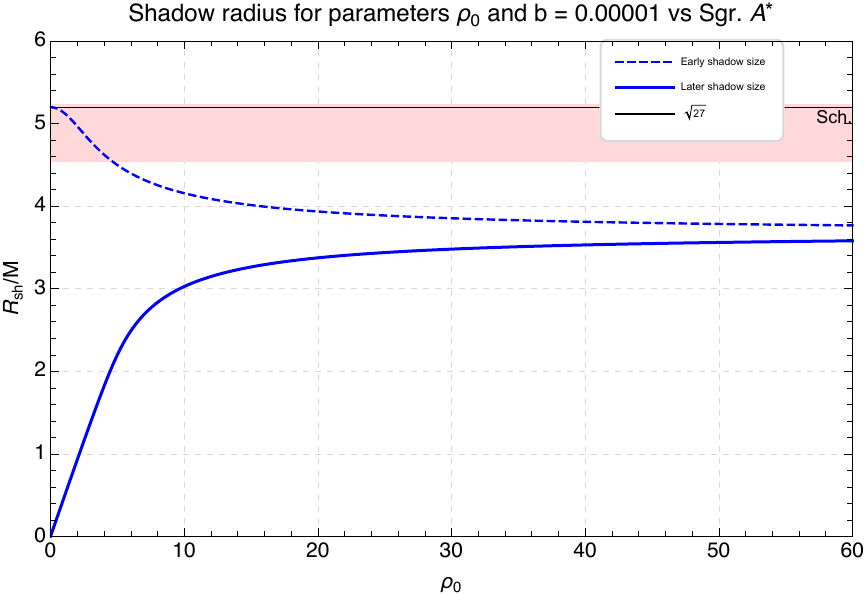}}\\ 
    \end{minipage}
    % \hfill
    \begin{minipage}[h!]{0.49\linewidth}
    \center{\includegraphics[width=8cm,height=6cm]{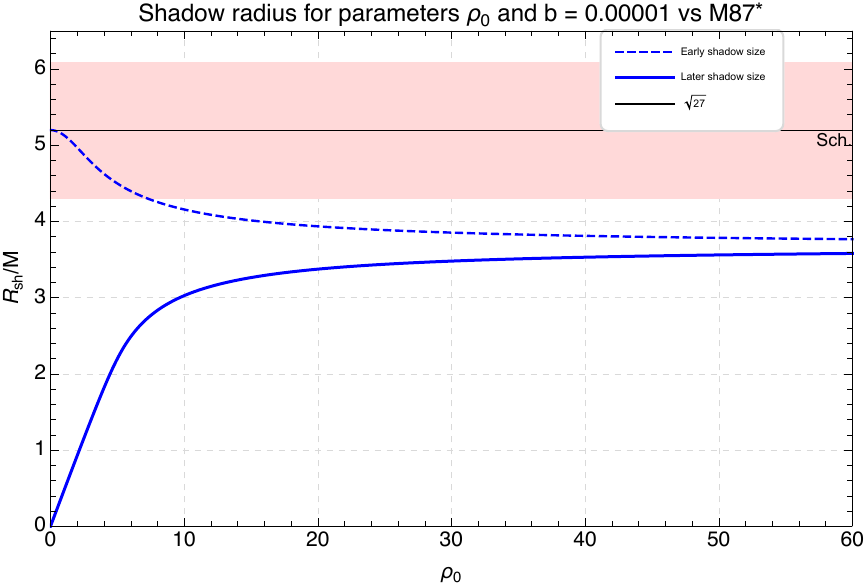}}\\
    \end{minipage}
\begin{minipage}[h!]{0.49\linewidth}
    \center{\includegraphics[width=8cm,height=6cm]{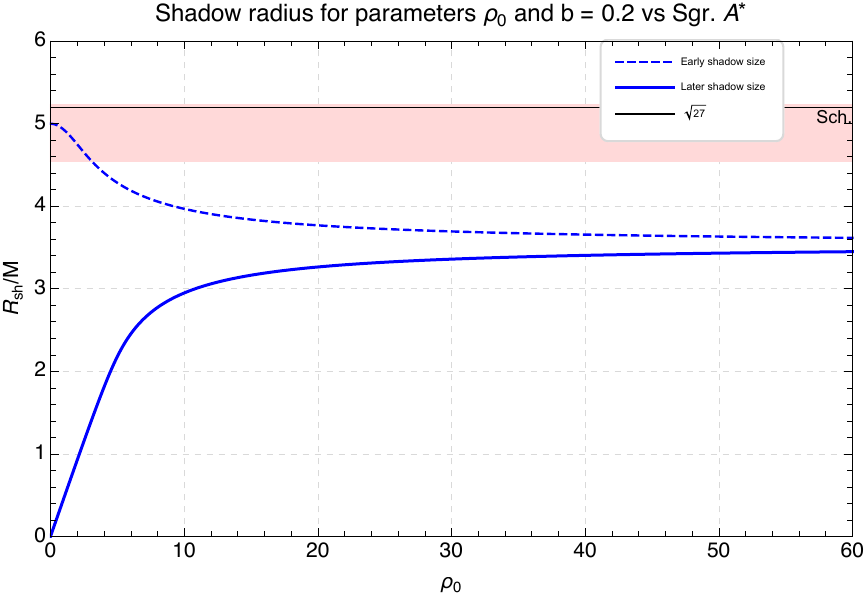}}\\
    \end{minipage}
    \hfill
    \begin{minipage}[h!]{0.49\linewidth}
    \center{\includegraphics[width=8cm,height=6cm]{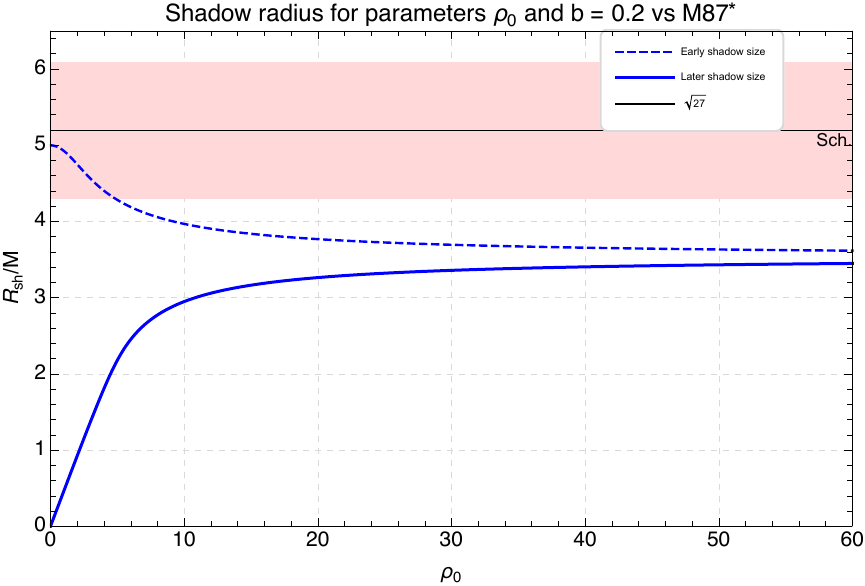}}\\
    \end{minipage}
    \begin{minipage}[h!]{0.49\linewidth}
    \center{\includegraphics[width=8cm,height=6cm]{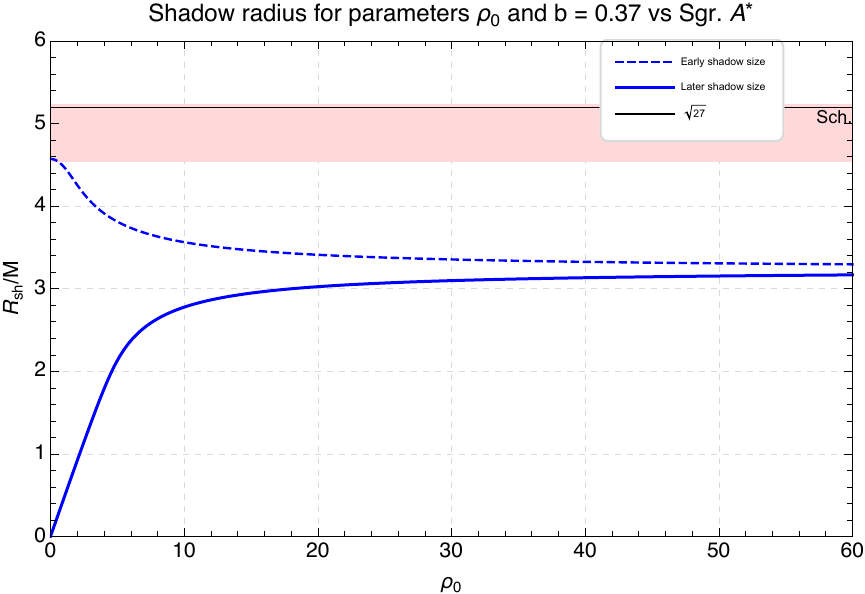}}\\
    \end{minipage}
    \begin{minipage}[h!]{0.49\linewidth}
    \center{\includegraphics[width=8cm,height=6cm]{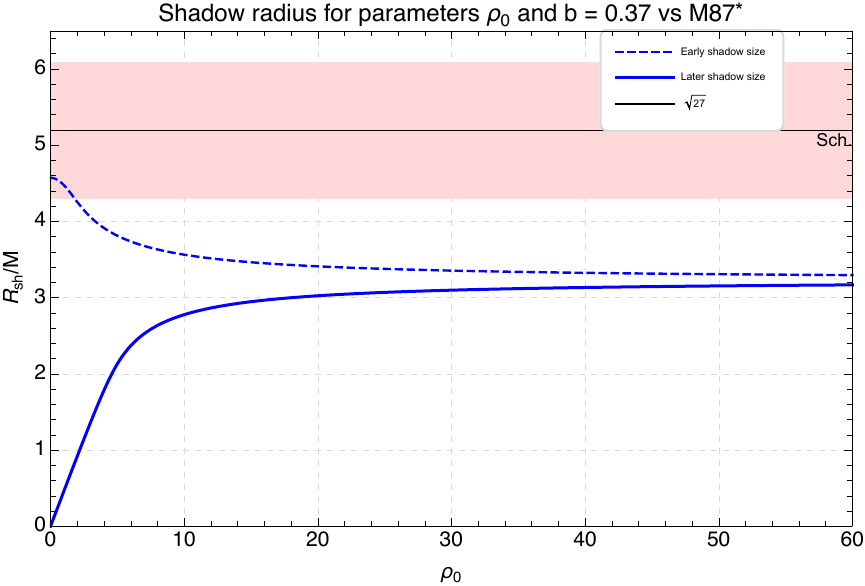}}\\
    \end{minipage}
    \caption{\footnotesize The  dependence of the radius of the late-time shadow (blue solid lines) and the early-time shadow (blue dashed lines) on the parameter $\rho_0$ 
    for various values of the  parameter $b$ for metric III. The pink area is  $1\sigma$-bands for M87$^\ast$ (\textbf{right column}) and Sgr A$^\ast$ (\textbf{left column}). Solid gray line is the shadow size in the Schwarzschild black hole case.}
    \label{Shadows3}
\end{figure}

in the coordinate $x$ \eqref{compactcoord}. The effective radial potential for null geodesics, when expressed in terms of compact radial coordinate $x$, see  \eqref{compactcoord}, takes the form,
\be
V_{\text{null}}(x) = \dfrac{1}{4}\left(1-x^2\right)^2\left(x^2+b^2\right)\Delta^{-1}\left(\rho(x)\right)\, .
\ee
The position of each photon sphere is determined by the following equation, 
\be
\dfrac{d}{dx}V_{\text{null}}(x) = \Delta^{-1}(\rho(x))\left[\dfrac{1}{2}~x\left(1-x^2\right)\left(1-2b^2-3x^2\right)-\left(x^2+b^2\right)\Delta^{-1}\left(\rho(x)\right)\dfrac{d}{d\rho}\Delta\left(\rho(x)\right)\right] = 0
\label{eqnshadow3}
\ee
Here the equation \eqref{usefull3.2} has to be used for the derivative of the function $\Delta (\rho(x))$.  The equation above is quite similar to the equation \eqref{eqnshadow2}.
One finds two positions for the local maximum of the radial potential. The value  of radius $R_{\text{sh}}$ for the early-time and late-time shadows is determined by equation,
\be
R^2_{\text{sh}}/M^2 = \dfrac{4\left(1+b^2\right)\Delta\left(\rho(x_{\text{ph}})\right)}{\left(1-x_{\text{ph}}^2\right)^2\left(x_{\text{ph}}^2+b^2\right)}\, ,
\ee
where $x_{\text{ph}}$ is a solution of the \eqref{eqnshadow3}, it can be found numerically.

In general the situation with the early-time shadow and the late-time shadow here is similar to that we described for metric I. We will not repeat it here.

Fig.~\ref{Shadows3} demonstrates that the late-time shadow does not fall within the $1\sigma$-bands for any parameters $b$ and $\rho_0$. 
This simply shows that what is observed today as a shadow can not be interpreted as the late-time shadow in metric III. 
However, there is still some possibility that it is an early-time shadow in metric III, as is seen from our Fig.~\ref{Shadows3}.

\begin{figure}[h!]
    \centering
            \begin{minipage}{0.4\linewidth}
                \includegraphics[width=\linewidth]{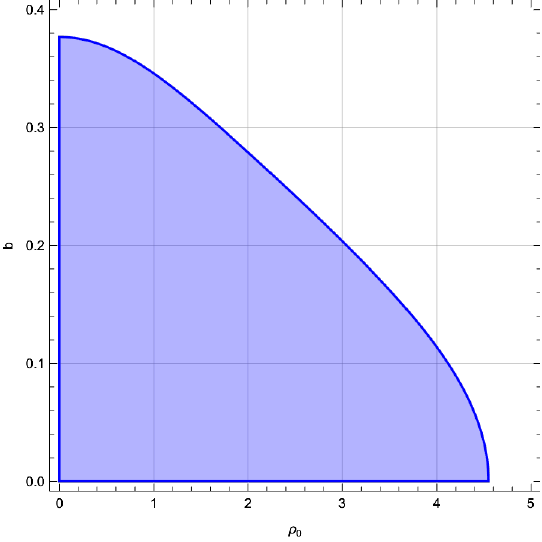}
            \end{minipage}
    \caption{\footnotesize The region of parameters $b$ and $\rho_0$ for which the early shadow radius is in $1\sigma$-bands for Sgr A$^\ast$ experiment. }
    \label{Const3}
\end{figure}

Thus, the main constraint on the parameters $b$ and $\rho_0$  arises from the requirement that the early-time shadow lies within the $1\sigma$-bands for Sgr A$^\ast$. 
This constraint is shown in Fig.~\ref{Const3}

\subsection{Shadow in metric IV}

\begin{figure}[h!]
\centering
    \begin{minipage}[h!]{0.49\linewidth}
    \center{\includegraphics[width=8cm,height=6cm]{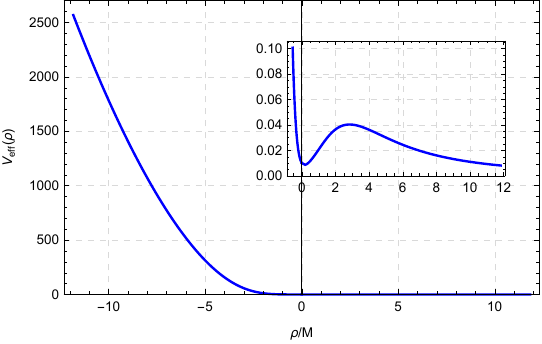}}\\ 
    \end{minipage}
    \caption{\footnotesize The effective radial  potential  for the null geodesics in metric IV for $n=2$ and the values of parameters $b=0.1$, $\rho_0 = 1$. 
    The inset presents the same potential for $\rho>0$, which makes it possible to see the peak present in the physical region.} 
    \label{Potential4}
\end{figure}

In metric IV the metric function is given by (\ref{metric4}).
This metric function has similar behavior to that of in the metric III. The main difference is that the effective radial potential for the light geodesics is unbounded in the inner region, meaning that the potential has an impenetrable wall at $\rho_0\rightarrow -\infty$. Indeed, (we remind that $n>1$)
\be
V_{\text{null}}(\rho) = \dfrac{g_{\text{III}}(\rho)}{r^2(\rho)}
\rightarrow \dfrac{4\left(1+b^2\right)}{\rho_0^2}\left(\dfrac{2\rho}{\rho_0}\right)^{2(n-1)}\, .
\ee
The radial potential has a single local maximum that lies in the physical region, as shown in Fig.~\ref{Potential4}. Similar to the case of metrics I and III, this metric possesses the property of a gradually shrinking  shadow. The main difference is that, 
due to the presence of an impenetrable wall in the asymptotic inner region, the shadow will shrink continuously until its size becomes zero. 
So that there is no a late-time shadow in this case.

The effective potential, if expressed in terms of the compact coordinate $x$, takes the form,
\be
V_{\text{eff}}(x) = \dfrac{1}{4}\left(1-x^2\right)^2\left(x^2+b^2\right)\Delta^{-n}\left(\rho(x)\right),\quad n>1
\ee
The position of the photon sphere is determined by the following equation, 
\be
\dfrac{d}{dx}V_{\text{eff}}(x) = \Delta^{-n}(\rho(x))\left[\dfrac{1}{2}~x\left(1-x^2\right)\left(1-2b^2-3x^2\right)-n\left(x^2+b^2\right)\Delta^{-1}\left(\rho(x)\right)\dfrac{d}{d\rho}\Delta\left(\rho(x)\right)\right] = 0
\label{eqnshadow4}
\ee
Here \eqref{usefull3.2} has to be used. Unlike the similar equation for metric III, the present  equation  has only one solution for the  photon sphere $x_{\text{ph}}$
that can be found numerically.
 Using this solution and the formula for the radius of shadow one finds,
 \be
 R^2_{\text{sh}}/M^2 = \dfrac{4\left(1+b^2\right)\Delta^n\left(\rho(x_{\text{ph}})\right)}{\left(1-x_{\text{ph}}^2\right)^2\left(x_{\text{ph}}^2+b^2\right)}\, .
 \ee

\begin{figure}[h!]
    \begin{minipage}[h!]{0.49\linewidth}
    \center{\includegraphics[width=8cm,height=6cm]{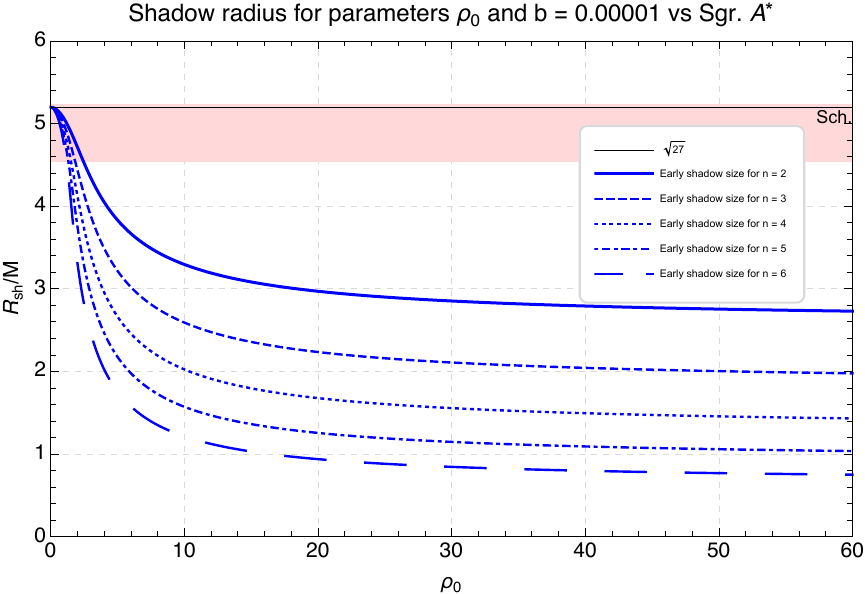}}\\ 
    \end{minipage}
    % \hfill
    \begin{minipage}[h!]{0.49\linewidth}
    \center{\includegraphics[width=8cm,height=6cm]{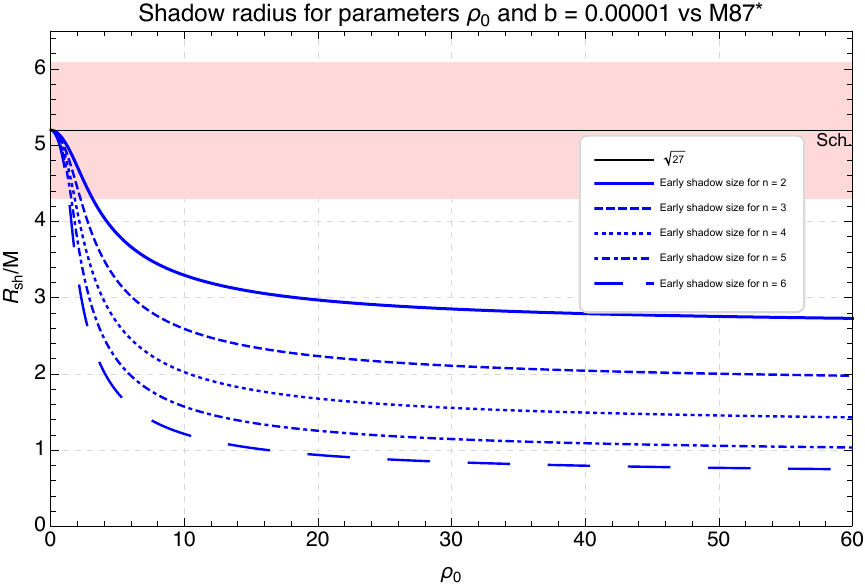}}\\
    \end{minipage}
    \begin{minipage}[h!]{0.49\linewidth}
    \center{\includegraphics[width=8cm,height=6cm]{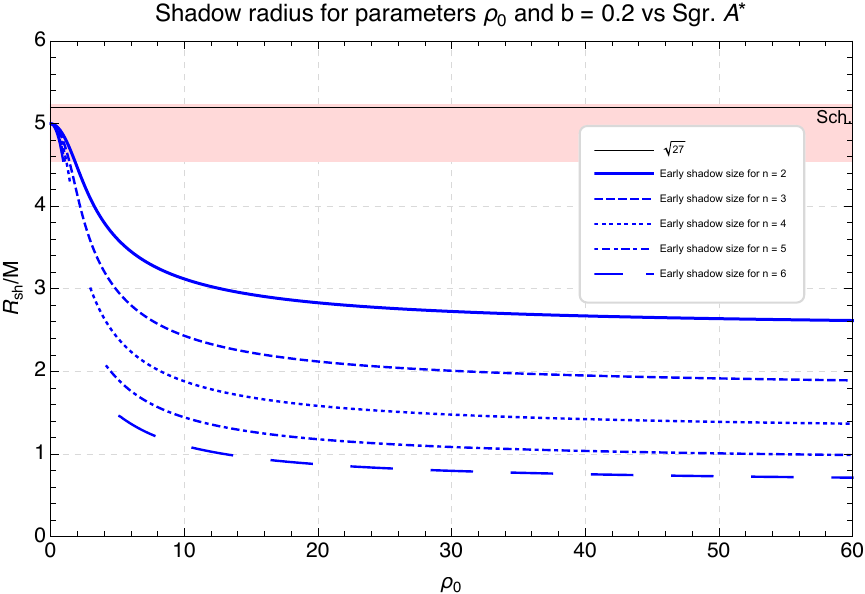}}\\
    \end{minipage}
    \hfill
    \begin{minipage}[h!]{0.49\linewidth}
    \center{\includegraphics[width=8cm,height=6cm]{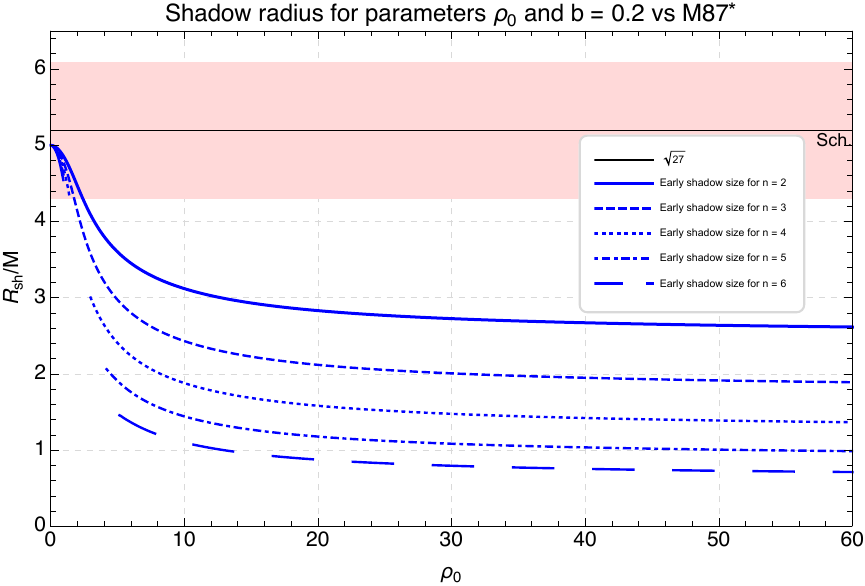}}\\
    \end{minipage}
    \begin{minipage}[h!]{0.49\linewidth}
    \center{\includegraphics[width=8cm,height=6cm]{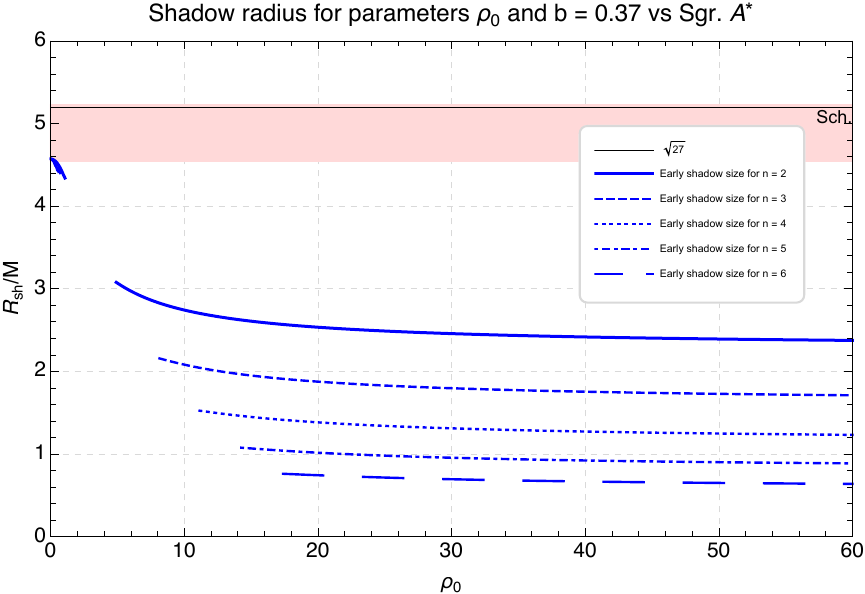}}\\
    \end{minipage}
    \begin{minipage}[h!]{0.49\linewidth}
    \center{\includegraphics[width=8cm,height=6cm]{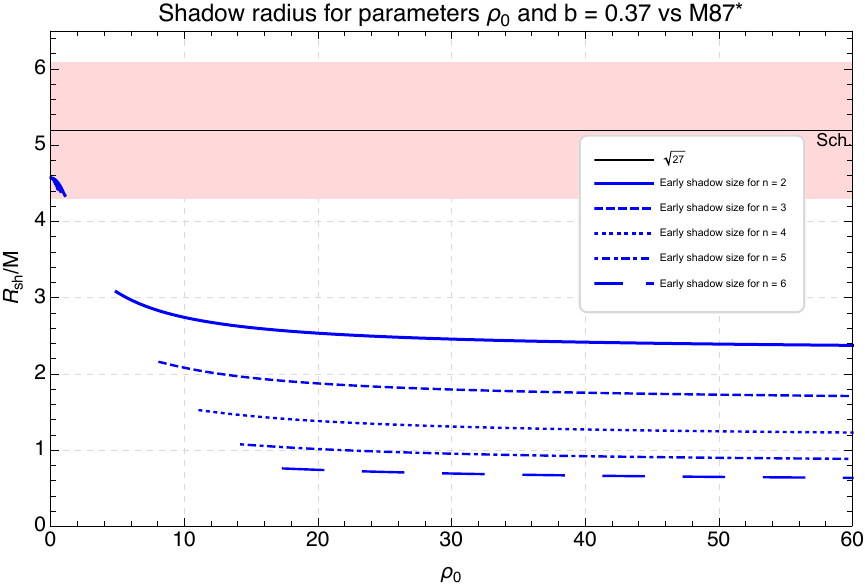}}\\
    \end{minipage}
    \caption{\footnotesize The  dependence of the radius of the early-time shadow on the parameter $\rho_0$  for various values of $n$ and $b$ for metric IV. The pink area is $1\sigma$-bands for M87$^\ast$ (\textbf{right column}) and Sgr A$^\ast$ (\textbf{left column}). Solid gray line is the shadow size in the Schwarzschild black hole case.}
    \label{Shadows4}
\end{figure}

\begin{figure}[h!]
    \centering
            \begin{minipage}{0.35\linewidth}
                \includegraphics[width=\linewidth]{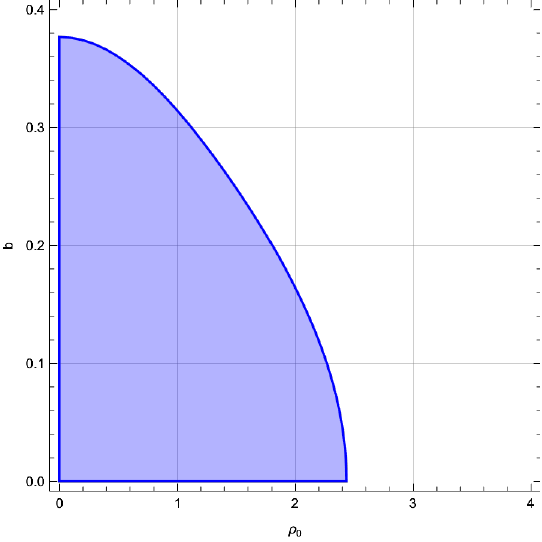}
            \end{minipage}
    \caption{\footnotesize The region of parameters $b$ and $\rho_0$ for which the early shadow radius is in $1\sigma - $bands for Sgr A$^\ast$ experiment. }
    \label{Const4}
\end{figure}

\noindent Fig.~\ref{Shadows4}  shows the observational constraints on the parameters in metric IV.
It should be noted that the gaps in the last two plots are due to the fact that for these values of parameters, the effective radial  potential does not have a maximum in the physical region. 
Thus, the main constraint on the parameters $b$ and $\rho_0$ will come from the limitations on the early-time shadow to be within the $1\sigma$-bands for Sgr A$^\ast$. This constraint is shown in Fig.~\ref{Const4}.

\section{Quasinormal modes and waveforms}
\label{sec:Quasinormal Modes}

Much information can be gained by studying the perturbations due to  various fields in a given spacetime background. Among these, the most interesting are the gravitational perturbations - perturbations of the metric itself in its own background - which allow one, for instance, to investigate the spectra of gravitational waves emitted by binary systems (see, e.g., \cite{Shibata:1999hn}, \cite{Shibata:2006bs}, \cite{Pretorius:2005gq}, \cite{Aylott:2009ya}). Interest in such studies has been greatly stimulated by recent advances in gravitational-wave astronomy, beginning with the first direct detection of gravitational waves by LIGO \cite{LIGOScientific:2016aoc}.

Nevertheless, the perturbations of lower-spin fields on various backgrounds are also of considerable interest (see \cite{Decanini:2015yba} for spin~$0$, \cite{Dolan:2006vj}, \cite{Jing:2005pk} for spin~$1/2$, and \cite{Folacci:2020ekl}, \cite{Ruffini:1972pw} for spin~$1$). In the present work, for simplicity, we restrict our attention to the study of scalar ($s = 0$) perturbations.

Unlike most macroscopic physical systems, perturbations in the background of gravitational objects are inherently dissipative, as these systems are not time-symmetric \cite{Berti:2009kk}. A key feature of such systems is the existence of quasinormal modes. Because the system is dissipative, the QNMs correspond to the eigenvalues of a non-Hermitian operator. In general, these eigenvalues are complex numbers, whose imaginary parts characterize the decay rate of the perturbations. Consequently, the sign of the imaginary part of a QNM determines the stability of the spacetime under the corresponding type of perturbation. The associated eigenfunctions are typically non-normalizable and do not form a complete set (see, e.g., \cite{Ching:1998mxl}, \cite{Nollert:1998ys} for more details).

Let us consider a general spherically symmetric static  metric,
\be
ds^2 = -g(\rho)dt^2+d\rho^2+r^2(\rho)d\Omega^2\, ,
\ee
where $d\Omega^2 = d\theta^2+\sin^2\theta ~d\varphi^2$ is metric on two-dimensional  sphere. In such a background, the study of linearized perturbations of arbitrary spin reduces to the analysis of the following differential equation \cite{Berti:2009kk} in the frequency domain
\be
\dfrac{d^2\Psi_s}{dz_{\ast}^2} + \left(\omega^2-V_s(z_\ast)\right)\Psi_s = 0\, ,
\label{QNMeqn}
\ee
where $V_s(z_\ast)$ is some effective radial potential the form  of which depends on  spin $s$ of the field.
We defined a tortoise coordinate $z_{\ast}$ as follows,
\be
z_\ast=\int_0^\rho \frac{d\rho}{\sqrt{g(\rho)}}+ {\rm constant}\, .
\lb{z1}
\ee
For metrics I, II and III the tortoise coordinate changes in the limits, $-\infty <z_\ast<+\infty$. In the case of metric IV the integral in (\ref{z1}) converges when $\rho\rightarrow -\infty$
and we can choose the
integration constant in such a way that $z_\ast$ changes in the limits: $0 \leq z_\ast<+\infty$. 

The quasi-normal modes are solutions to equation (\ref{QNMeqn}) subject to certain boundary conditions. These conditions are in nature the dissipative conditions: the respective modes 
are escaping through all possible boundaries. In the case of  metrics I, II and III  that we consider in this paper the tortoise coordinate changes in the limits, $-\infty<z_{\ast}<+\infty$. 
The standard conditions to be imposed are  to make the perturbations  out-going for $z_{\ast}\rightarrow +\infty$ and in-going for $z_\ast\rightarrow -\infty$.
In the case of metric IV the situation is quite different: the tortoise coordinate changes in the limits   $0 \leq z_\ast<+\infty$. So that we impose the Dirichlet boundary condition at $z_\ast=0$.  The QNM conditions then are as follows,
\begin{equation}
    \begin{cases}
        \Psi_s(z_{\ast}) \sim e^{i{\omega} z_{\ast}},\quad\quad z_{\ast} \rightarrow\infty\quad \text{(Metrics I - IV)}\, , \nonumber \\
        \nonumber\\
        \Psi_s(z_{\ast}) \sim e^{-i\bar{\omega} z_{\ast}},~~\quad z_{\ast} \rightarrow -\infty\quad \text{(Metrics I - III)}\, ,\\
        \\
        \Psi_s(z_{\ast}) \sim 0 ,~~\quad\quad\quad z_{\ast} \rightarrow 0 \quad \text{(Metric IV)}\, ,
    \end{cases}
    \label{boundary}
\end{equation}
where we have introduced $\bar{\omega} = \sqrt{\omega^2 - V_{-\infty}}$ since the effective potential for perturbation can tend to a non - zero constant at $z_{\ast}\rightarrow -\infty$. The choice of the Dirichlet condition for  metric IV  at $z_{\ast}\rightarrow -\infty$ is due to the behavior of the   potential  $V_{(s)}(z_\ast)$ at $z_{\ast}\rightarrow -\infty$. We will discuss this in more detail later.

Let us note another important feature of QNMs: they correspond to the poles of the Green's function associated with Eq.~\eqref{QNMeqn} \cite{Nollert:1992ifk, Andersson:1996cm, Andersson:1995zk, Berti:2009kk}. At late times, the so-called ringdown stage, the behavior of perturbations is completely governed by these poles. Consequently, during this stage, the dynamics of the perturbations are fully determined by the QNMs. Therefore, by studying the spectrum and properties of QNMs, one can gain a deeper understanding of the late-time behavior of perturbations in curved spacetime.

\subsection{Computational methods}

As mentioned above, the computation of QNMs reduces to solving an eigenvalue problem \eqref{QNMeqn} with boundary conditions \eqref{boundary}. There are various methods for dealing with this problem: the WKB approximate method \cite{Schultz}, \cite{Konoplya:2004ip}, Leaver's continued fractional method \cite{Leaver:1985ax}, the matrix method \cite{Lin:2019mmf}, the monodromy method \cite{Motl:2003cd}, the hyperboloidal approach \cite{Jaramillo:2020tuu}. In this work, we will use the hyperboloidal approach for metrics I, II, and IV, and the matrix method for metric III.  In the case of metric III,  the effective radial potential tends to a constant when $z_\ast\rightarrow -\infty$. This results in an effective mass for the  scalar perturbations. In this situation, the hyperboloidal approach turns out to be more suitable for finding QNMs, as it reduces the problem to an eigenvalue problem for a certain matrix. This is in contrast to the matrix method, where one generally has to solve a non-polynomial equation that is  the determinant of a certain matrix.

In the following sections, we will briefly discuss the main features of both methods and consider how they work  in each particular metric. See appendix \ref{Convergence tests} for details on the convergence tests.

\subsubsection{Hyperboloidal approach}
\label{sec:hyp}

Let us begin the discussion of the methods that we use with the hyperboloidal approach (more information can be found in \cite{Zenginoglu:2007jw}, \cite{Zenginoglu:2011jz}, \cite{Jaramillo:2020tuu}, \cite{PanossoMacedo:2023qzp}). The main idea of this methods is to employ, alongside with  compactification, a specially chosen time coordinate $\tau$ that automatically imposes the desired radiation boundary conditions. The necessity of introducing a new time coordinate $\tau$ can be explained as follows: when an infinite region is compactified into a finite domain, the characteristics of the corresponding differential equation behave poorly near the boundaries of the compactified region. Introducing a new time coordinate makes the characteristics regular at the boundaries, restoring their outgoing behavior at the respective boundaries. 

The new time coordinate can be viewed as something intermediate between the usual coordinate time $t$ and the retarded time $u = t -z_{\ast}$ \cite{Jaramillo:2020tuu}. Indeed, a surface of $t=\text{const}$ is spacelike and remains so as $z_{\ast}\rightarrow\infty$. On the other hand, a slice of retarded time $u=\text{const}$ forms a null hypersurface, which also remains null as $z_{\ast}\rightarrow\infty$. The hyperboloidal time coordinate $\tau$ retains the property that its slices $\tau=\text{const}$ are spacelike hypersurfaces, but as $z_{\ast}\rightarrow\infty$, these hypersurface asymptotically approach null $u=\text{const}$.

We obtain a time coordinate whose $\tau=\text{const}$ slice penetrate both future null infinities (in the case of a wormhole) or the future null infinity and the event horizon (in the case of a black hole) \cite{Zenginoglu:2011jz}. This allows to implement the appropriate out-going boundary conditions at both boundaries geometrically. As a result, the physical degrees of freedom do not need to satisfy any special boundary conditions (these will be automatically satisfied) as long as they are regular at the boundaries. 

Let there be some compactified coordinate $x\in[a,~b]$, and we wish to introduce a hyperboloidal time coordinate in the following way:
\be
t=\tau-h(x)\, ,
\label{hypertime}
\ee
where $h(x)$ is the height function. There are several ways to introduce the height function; we will consider the minimal gauge approach \cite{PanossoMacedo:2023qzp}. Let us examine the tortoise coordinate in the compact coordinates $z_{\ast}(x)$. Since we are dealing with the wormhole type spacetime, we assume that the tortoise coordinate may have singularities only at the infinities in the physical and inner regions (as in metric I and II), or only in the physical region (as in metric IV). We will explicitly isolate the singularities in the tortoise coordinate:
\be
z_{\ast}(x) = \begin{cases}
    z_{\ast}^{(+,~\text{sing})}(x) + z_{\ast}^{(-,~\text{sing})}(x) +z_{\ast}^{(\text{reg})}(x),\quad \text{Metric I and II}\nonumber\\
    \\
    z_{\ast}^{(+,~\text{sing})}(x)  + z_{\ast}^{(\text{reg})}(x),\quad\quad\quad\quad\quad\quad~~~~ \text{Metric IV}\nonumber
\end{cases}
\ee
where $z_{\ast}^{(\pm,~\text{sing})}(x)$ are singular parts as $\rho\rightarrow\pm\infty$ and $z_{\ast}^{(\text{reg})}(x)$ is regular part of the tortoise coordinate. 

\medskip

\noindent We can use two strategies to introduce the height functions. 

\medskip

\noindent $\bullet$   The in-out strategy: let us consider the equation for an outgoing null geodesic in advanced null coordinate $v=t+z_{\ast}$ and then integrate it near infinity in the physical region.
    \be
    \dfrac{dv}{dx} = \left.2z_{\ast,x}(x)\right|_{x\rightarrow b}\sim 2z_{\ast,x}^{(+,~\text{sing})}(x)\Rightarrow \left.v\right|_{x\rightarrow b}\sim \tau+2z_{\ast}^{(+,~\text{sing})}(x)\, ,
    \ee
    where $z_{,x}$ denotes differentiation with respect to $x$, and we have chosen the coordinate $\tau$ as the constant of integration. Returning from the coordinate $v$ to the usual time coordinate $t=v-z_{\ast}(x)$, we obtain
    \be
    t=\tau - \left(-2z_{\ast}^{(+,~\text{sing})}(x) + z_{\ast}(x)\right)\Rightarrow h^{\text{in-out}}(x) = z_{\ast}^{(-,~\text{sing})}(x) - z_{\ast}^{(+,~\text{sing})}(x) +z_{\ast}^{(\text{reg})}(x)\, .
    \label{inout}
    \ee

\medskip

\noindent $\bullet$  The out-in strategy is analogous to the in-out strategy, but now we will instead consider the equation for an ingoing geodesic in terms of the retarded coordinates $u=t-z_{\ast}(x)$ and then integrate it near infinity in the inner region,
    \be
    \dfrac{du}{dx} = \left.-2z_{\ast,x}(x)\right|_{x\rightarrow a}\sim -2z_{\ast,x}^{(-,~\text{sing})}(x)\Rightarrow \left.v\right|_{x\rightarrow a}\sim \tau-2z_{\ast}^{(-,~\text{sing})}(x)
    \ee
    Returning from the coordinate $u$ to the time coordinate $t=u+z_{\ast}(x)$, we obtain
    \be
    t=\tau - \left(2z_{\ast}^{(-,~\text{sing})}(x) - z_{\ast}(x)\right)\Rightarrow h^{\text{out-in}}(x) = z_{\ast}^{(-,~\text{sing})}(x) - z_{\ast}^{(+,~\text{sing})}(x) - z_{\ast}^{(\text{reg})}(x)
    \ee

\medskip

\noindent It is evident that both approaches differ only in the sign of the regular term. In the case where this term is zero, both strategies yield the same result for the height function. 

Now we can rewrite the equation describing the scalar perturbations,
\be
-\Psi_{,tt}+\Psi_{,z_{\ast}z_{\ast}}-V(z_{\ast})\Psi=0\, .
\ee
in terms of the coordinates $\tau$ and $x$. Since,
\be
\partial_{t}=\partial_{\tau},\quad \partial_{z_{\ast}} = \dfrac{1}{z_{\ast,x}(x)}\partial_{x} + \dfrac{h_{,x}(x)}{z_{\ast,x}(x)}\partial_{\tau},
\ee
 we obtain the main equation (after multiplying by $z_{\ast,x}$), which we will use in the following analysis,
\be
-p_{\tau\tau}(x)\partial_{\tau}^2\Psi+2p_{\tau x}(x)\partial_{\tau}\partial_x\Psi+p_{xx}(x)\partial_x^2\Psi + p_{\tau}(x)\partial_{\tau}\Psi+p_x(x)\partial_x\Psi-\hat{V}(x)\Psi=0\, ,
\label{hyperdiffeq}
\ee
where we have used the following notations,
\be
 &&       p_{\tau \tau}(x) = z_{\ast,x}(x)-\dfrac{h_{,x}(x)^2}{z_{\ast,x}(x)},\quad p_{\tau x}(x) = \dfrac{h_{,x}(x)}{z_{\ast,x}(x)},\quad p_{\tau}(x) = \partial_x\left(\dfrac{h_{,x}(x)}{z_{\ast,x}(x)}\right)\nonumber\\
        \\
 &&       p_{xx}(x) = \dfrac{1}{z_{\ast,x}(x)},\quad p_x(x) = \partial_x\left(\dfrac{1}{z_{\ast,x}(x)}\right),\quad \hat{V}(x) = z_{\ast,x}(x)~V(x) \nonumber
        \label{pfunction}
\ee
The equation can be reduced to a system of equations that are first order in time derivative,
\begin{gather}
    \partial_{\tau}\underbrace{\begin{pmatrix}
            \Psi(t,~x)\\[4pt]
            \Phi(t,~x)
        \end{pmatrix}}_{U} = i\underbrace{\dfrac{1}i{}\begin{pmatrix}
            0 & 1\\[4pt]
         \mathbf{L}_1 & \mathbf{L}_2
        \end{pmatrix}}_{\mathbf{L}}\begin{pmatrix}
            \Psi(t,~x)\\[4pt]
            \Phi(t,~x)
        \end{pmatrix}\, , \nonumber 
        %[6pt]
        \end{gather}
        where  $\Phi(\tau,~x)=\partial_{\tau}\Psi(\tau,~x)$ and we introduced the operators, 
        \begin{gather}
     \mathbf{L}_1 = \dfrac{1}{p_{\tau\tau}(x)}\left[\partial_x\left(p_{xx}(x)\partial_x\right) - \hat{V}(x)\right],\quad \mathbf{L}_2 = \dfrac{1}{p_{\tau\tau}(x)}\left[2p_{\tau x}(x)\partial_x + \partial_xp_{\tau x}(x)\right] \, .\nonumber
        \label{L1L2hyper}
\end{gather}
 Since $p_{\tau\tau}(x)$ appears in the denominator, it must remain positive $p_{\tau\tau}(x)>0$.  
 This consideration serves as a criterion for selecting an appropriate height function $h(z_\ast)$.

One can move to the frequency domain $U(\tau,~x) = e^{i\omega \tau}\hat{U}(x)$ and obtain an eigenvalue problem for the operator $\mathbf{L}$,
\be
\mathbf{L}\hat{U}=\omega\hat{U}
\label{mainhyper}
\ee
whose eigenvalues correspond to QNMs. It is worth noting that the operator $\mathbf{L}$ is non-self-adjoint in the energy norm (see \cite{Jaramillo:2020tuu} for more details). In what follows, the search for quasinormal modes will reduce to finding the eigenvalues of the operator $\mathbf{L}$. 

We can introduce a suitable discretization of the interval $[a,~b]$ into $N$ interpolation point $x_i$ with $i=1,\dots,N$; in this case, the differential system \eqref{mainhyper} is reduced to a matrix equation, and the problem of finding QNMs becomes a task of determining the eigenvalues of the corresponding discrete version of the matrix $\textbf{L}$. For more details on the numerical method, see the appendix \ref{sec:appNumApp}.

\subsubsection{Matrix method}
\label{sec:matrixmethod}

As mentioned above, for metric III we will use the matrix method (see more details in \cite{Lin:2019mmf}, \cite{Lin:2023rkd}, \cite{Lin:2016sch}) to compute the QNMs. Since the potential approaches a constant  $V_{-\infty}$ at $\rho\rightarrow -\infty$ in the inner region, one can say that the scalar field acquires an effective mass in this region,
\be
-\Psi_{,tt}+\Psi_{,z_{\ast}z_{\ast}} + V_{-\infty}\Psi = 0\, .
\ee
As a result, the boundary conditions take the following form ($\Psi(t,z_{\ast}) = e^{-i\omega t}\psi(z_{\ast})$)
\be
\begin{cases}
    \psi\sim e^{i\omega z_{\ast}},\quad\quad\quad\quad z_{\ast}\rightarrow\infty  \\
    \psi\sim e^{-i\sqrt{\omega^2-V_{-\infty}}z_{\ast}}, \quad z_{\ast}\rightarrow -\infty
\end{cases}
\label{boundarymatrixmathod}
\ee
Let us introduce a suitable compact coordinate $x\in [a,b]$ ($z_{\ast}(x\rightarrow a)\rightarrow -\infty$ and $z_{\ast}(x\rightarrow b)\rightarrow\infty$) and rewrite the scalar perturbation equation in terms of this coordinate,
\be
p_0(x)\partial_x^2\psi(x) + \lambda_0(x)\partial_x\psi(x) + s_0(x)\psi(x)=0\, .
\ee
To account for the boundary conditions \eqref{boundarymatrixmathod}, we need to analyze the behavior of the solution near the both boundaries and choose the  forms that satisfy the required conditions. We can factor out from the function $\psi(x) = A(x)R(x)$ a term $A(x)$ that encodes the boundary behavior at $x=a$ and $x=b$, and rewrite the equation in terms of the
 function $R(x)$,
\be
\bar{p}_0(\omega, x)\partial_x^2R(x) + \bar{\lambda}_0(\omega, x)\partial_xR(x) + \bar{s}_0(\omega, x)R(x)=0\, .
\ee
Since the singular behavior is contained in the factor $A(x)$, we expect the function $R(x)$ to be regular at the boundaries,
\be
R(a)=C_0\quad\text{and}\quad R(b)=C_1 \, .
\ee

For convenience, we perform one more transformation \cite{Lin:2019mmf}:
\be
F(x)=R(x)f(x),\quad f(a)=f(b)=0\, ,
\ee
where the function $f(x)$ is chosen to vanish at the boundaries, for instance, one can choose $f(x) = (x-a)(x-b)$. As a result, we obtain a differential equation with the simplest possible boundary conditions,
\be
\bar{\bar{p}}_0(\omega, x)\partial_x^2F(x) + \bar{\bar{\lambda}}_0(\omega, x)\partial_xF(x) + \bar{\bar{s}}_0(\omega, x)F(x)=0,\quad F(a)=F(b)=0\, .
\label{matrixmethodequation}
\ee
It is worth noting that this redefinition is not a strictly necessary step, but it helps to eliminate the arbitrariness of the constants $C_0$ and $C_1$ in the boundary conditions. 

We have obtained an eigenvalue problem for a differential equation with boundary conditions \eqref{matrixmethodequation}. In order to find  numerically  the corresponding eigenvalues $\omega_{\text{QNM}}$, the interval $[a,~b]$ must be divided into $N$ interpolation points $x_i$ with $i=1,\dots,N$. Along with this, the function $F(x)$ is also discretized on the grid as $f_i=F(x_i)$. The differential equation \ref{boundarymatrixmathod} is then projected onto the grid, resulting in a matrix version of the equation
\be
\mathcal{M}(\omega)\mathcal{F}=0\, ,
\ee
where $\mathcal{M}(\omega)$ is the discretized version of the differential operator $\bar{\bar{p}}_0(\omega, x)\partial_x^2 + \bar{\bar{\lambda}}_0(\omega, x)\partial_x + \bar{\bar{s}}_0(\omega, x)$, and $\mathcal{F}$ is the column vector composed of the values $f_i$. The boundary condition $F(a)=F(b)=0$ implies $f_1=f_N=0$. This allows us to replace the first and last rows of the matrix $\mathcal{M}(\omega)$ with $1$, thereby reducing the matrix equation to a new form:
\be
\bar{\mathcal{M}}(\omega)\mathcal{F}=0,\quad\bar{\mathcal{M}}_{i,j}=\begin{cases}
    \delta_{i,j},\quad i=1~\text{or}~N\nonumber\\
    \\
    M_{i,j},\quad i=2,\dots,N-1
\end{cases}
\ee
The resulting matrix equation states that the vector $\mathcal{F}$ is an eigenvector of the matrix $\bar{\mathcal{M}}(\omega)$. Therefore, the following condition must be satisfied:
\be
\det{\bar{\mathcal{M}}(\omega)}=0\, .
\label{mainmatrixmethod}
\ee
This is our main equation, which we will solve numerically. 
For more details on the numerical method, see appendix \ref{sec:appNumApp}.

\subsection{Remarks on  the symmetric case.}
\label{sec:symworm}

\begin{figure}[h!]
    \begin{minipage}[h!]{0.49\linewidth}
    \center{\includegraphics[width=8cm,height=6cm]{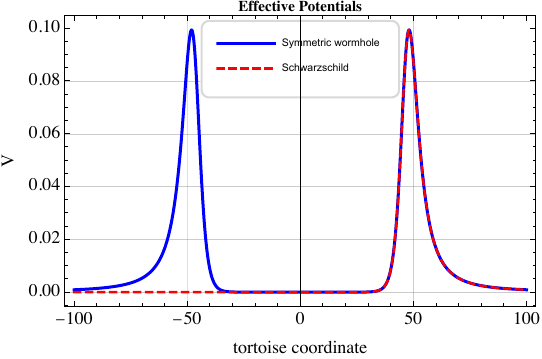}}\\
    \end{minipage}
    \hfill
    \begin{minipage}[h!]{0.49\linewidth}
    \center{\includegraphics[width=8cm,height=6cm]{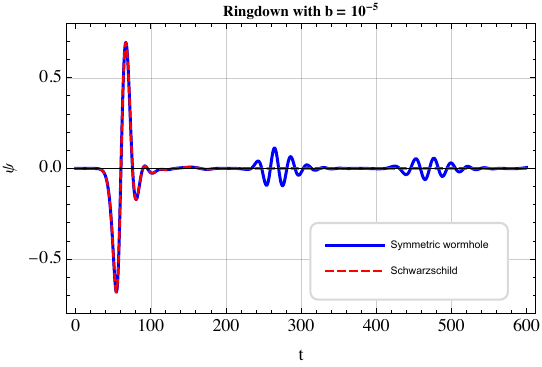}}\\
    \end{minipage}
     \caption{\footnotesize \textbf{Left:} a comparison of the effective radial potential for the Schwarzschild metric (red dashed line) and for the symmetric wormhole metric with $b=10^{-5}$ (blue solid line). \textbf{Right:}  a comparison of the ringdown signals for the Schwarzschild metric (red dashed line) and the symmetric wormhole metric with $b=10^{-5}$ (blue solid line).}
    \label{sympotandecho}
\end{figure}

Before presenting the result of our analysis  for the four test  metrics, we briefly discuss the $Z_2$ symmetric potential that corresponds to the DS metric \cite{Damour:2007ap}, equivalent to metric I with $a=0$, since it shows the typical behavior of QNMs and ringdown signals common to all cases. 
The echo effect arises from the presence of a trapping cavity in the effective potential, which temporarily confines the initial perturbation. Due to dissipation and leakage through the potential barriers, the trapped signal is gradually released, producing a sequence of secondary bursts, or echoes. The left panel of Fig.~\ref{sympotandecho} compares the potentials of the Schwarzschild metric (red dashed line) and a symmetric wormhole with value $b=10^{-5}$ (blue solid line) of the deformation parameter. While the peaks in the physical region for both metrics coincide, the wormhole potential has the second separated peak thus forming a cavity. The right panel of Fig.~\ref{sympotandecho} shows ringdown signals for the wormhole (blue solid line) and for the Schwarzschild metric (red dashed line). In both cases, the initial Gaussian signal was located near the maxima. Because the peaks have the same shape \cite{Cardoso:2008bp}, \cite{Konoplya:2017wot}, the first signal coincide and are governed by the Schwarzschild QNMs \cite{Cardoso:2016rao}, \cite{Barausse:2014tra}, but later the wormhole signal develops echoes absent in the Schwarzschild case. Thus, if only the first signal is detected, wormholes and black holes cannot be distinguished (one must wait for the echoes). 

Previous studies \cite{Cardoso:2016rao}, \cite{Bueno:2017hyj}, \cite{Volkel:2018hwb} have shown that due to the presence of the cavity, the QNMs of the ECOs include long-lived modes. These QNMs describe the behavior of the signals in the ringdown stage at late time (starting from the first echo). As already noted, the primary signal is described by the QNMs of a black hole, since it is formed by scattering on a single potential peak and therefore does not probe the full global structure of the potential. However, these black hole modes are not present in the wormhole QNM spectrum \cite{Cardoso:2016rao}, \cite{Cardoso:2017njb}, \cite{Cardoso:2017cqb}, \cite{Mark:2017dnq}. The typical behavior of the QNMs for a double-peaked potential is that the real part of the modes scales as $1/L$, while the imaginary part scales as $1/L^{2l+3}$, where $L$ is the distance between the potential peaks and $l$ is the angular momentum of the perturbations \cite{Cardoso:2016rao} (explicit calculation for the double-delta potential can be found in the appendix \ref{app:doubledelta}). This behavior can be explained by the fact that a cavity without dissipation would have a discrete set of normal frequencies with similar scaling $1/L$ in terms of its size $L$. The weak leakage leads to small imaginary parts of these modes. 

\begin{figure}[h!]
  \centering
    \begin{minipage}[h!]{0.49\linewidth}
    \center{\includegraphics[width=8cm,height=6cm]{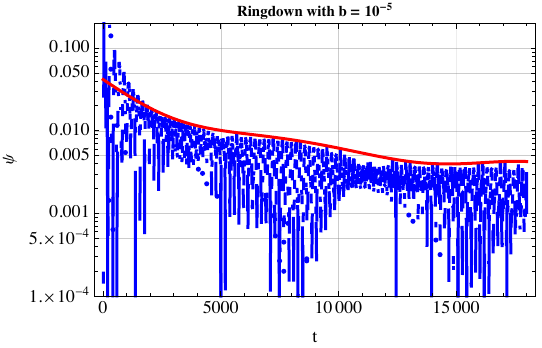}}\\
    \end{minipage}
     \caption{\footnotesize The behavior of the ringdown stage in logarithmic scale for times up to $18000$ (blue solid line). The red line represents the sum over the first six longest-lived QNMs.}
    \label{logringdownsym}
\end{figure}

During the ringdown phase, far from the source the signal can be expressed as a sum over QNMs (see $\omega_n$ \cite{Berti:2009kk}, \cite{Nollert:1992ifk}, \cite{Andersson:1996cm}, \cite{Andersson:1995zk}, \cite{Berti:2006wq}), 
\begin{equation}
    \psi(t, z_{\ast}\rightarrow\infty) = \sum_{n=0}^{\infty}C_n e^{-i\omega_n(t-z_{\ast})}\, .
    \label{sumqnm}
\end{equation}
with coefficients $C_n$ depending on the initial signal. The imaginary part of each QNM determines its decay time $\tau_n=-1/\text{Im}\omega_n$, so with increasing time the contributions with larger imaginary parts disappear. For the symmetric wormhole, the spectrum includes modes with very small imaginary parts (see \cite{Bueno:2017hyj} or Table~\ref{QNM1} for $a=0$) $\tau_n\sim\{5\cdot 10^7,2\cdot 10^6, 3\cdot 10^5, 62000, 16000, 5000, 1700, 660, 300, 162,\dots\}$, the fundamental mode ($n=1$) has the longest decay time.
 At late times, only the lowest-lying modes remain, and the decay is governed by them. This behavior can be seen in Fig.~\ref{logringdownsym}, where the ringdown signal (blue solid line) is shown for long time $t<18000$, together with the amplitude decay (red solid line). To reconstruct the curve, the first six modes with the smallest imaginary parts were used. At earlier times this sum does not reproduce the signal well, since many short-lived modes are still present. As they decay, only six modes under consideration remain, and they describe the signal damping.
The signal, thus, decays much longer than the one produced by perturbations of  the Schwarzschild black hole. This is a distinguishing feature of the wormhole mimickers.

\begin{figure}[h!]
    \begin{minipage}[h!]{0.49\linewidth}
    \center{\includegraphics[width=8cm,height=6cm]{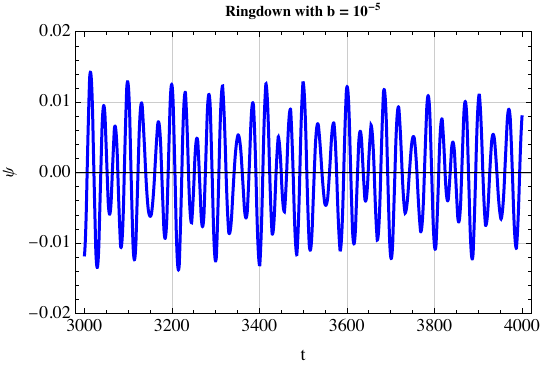}}\\
    \end{minipage}
    \hfill
    \begin{minipage}[h!]{0.49\linewidth}
    \center{\includegraphics[width=8cm,height=6cm]{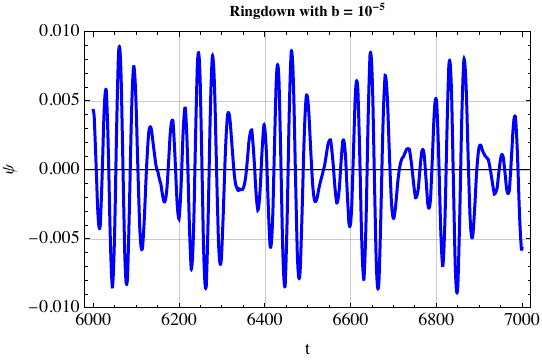}}\\
    \end{minipage}
    \begin{minipage}[h!]{0.49\linewidth}
    \center{\includegraphics[width=8cm,height=6cm]{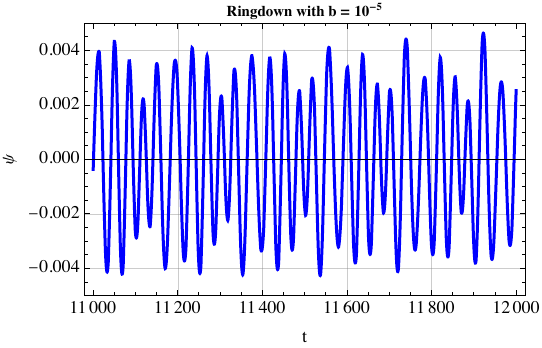}}\\
    \end{minipage}
    \begin{minipage}[h!]{0.49\linewidth}
    \center{\includegraphics[width=8cm,height=6cm]{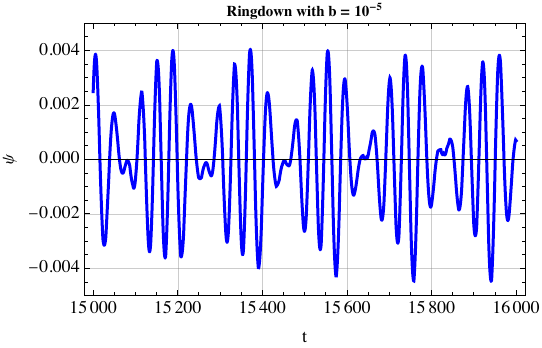}}\\
    \end{minipage}
    \caption{\footnotesize The panels show parts of the same ringdown signal at different time intervals. This demonstrates the long stages of clear beats (\textbf{right}) changing to long stages of signals overlapping (\textbf{left}).}
    \label{ringdownsymreappearence}
\end{figure}

Another interesting phenomenon appears when one examines the echo signal over a sufficiently long period of time. For a certain interval, the echo bursts gradually blur and overlap, making it difficult to distinguish individual echoes within the ringdown signal. However, after some time, the echoes reorganize and a sequence of isolated beats re-emerges. The stage of well-defined beats is then followed again by a blurred stage, and these two regimes alternate periodically. A similar effect was observed for hairy black holes in \cite{Guo:2022umh}.

This behavior is illustrated in Fig.~\ref{ringdownsymreappearence}. The four panels display different time intervals of the same ringdown signal, which initially exhibited distinct echo bursts (as in Fig.~\ref{sympotandecho}). In the interval $3000 < t < 4000$, the individual echoes merge into a single, blurred signal. In the interval $6000 < t < 7000$, the separate echo bursts reappear, forming a clear pattern of beats. Later, in the interval $11000 < t < 12000$, the echoes again coalesce into a structureless waveform, while in $15000 < t < 16000$, the beat pattern re-emerges once more.

This effect can be understood from the temporal evolution of the quasinormal mode sum~\eqref{sumqnm}. As time progresses, the more rapidly decaying modes drop out of the superposition, causing alternating stages of blurring and reappearance of the beats. The characteristic duration of each such stage can be roughly estimated from the decay times of the neighboring modes,
$\Delta t_{\text{reappear}} \sim \tau_{n+1}-\tau_{n}$.

This leads to an interesting situation. On the one hand, the observation of the primary ringdown signal provides information about the QNMs of the black hole, which are effectively mimicked by the corresponding wormhole geometry. On the other hand, by analyzing the signal at later times and detecting a sequence of echoes, one can extract information about the QNMs of the wormhole itself. Detecting multiple echoes grants access to the modes with shorter decay times, whereas obtaining information about the long-lived modes requires observing the ringdown signal over a sufficiently long time interval, comparable to their characteristic decay times.

\subsection{Results for the test metrics}

Here we present our results of applying the above-described methods to the study of the test metrics under consideration. 

\subsubsection{Metric I}
\label{sec:m1}

Above, we have introduced a convenient compactified coordinate \eqref{compactcoord}. In terms of this  coordinate, metric I takes the following form:
\be
ds^2=-\left(\left(x+a\right)^2+b^2\right)dt^2+\dfrac{16dx^2}{\left(1-x^2\right)^4}+\dfrac{4}{\left(1-x^2\right)^2}d\Omega^2\, ,
\ee
where we set $M=1$.
In terms of this coordinate, the equation for the tortoise coordinate reads,
\be
z_{\ast,x}=\dfrac{4}{\left(1-x^2\right)^2\sqrt{\left(x+a\right)^2+b^2}}\, .
\ee
Respectively,  the effective potential is
\be
V=\dfrac{1}{4}\left(1-x^2\right)^2\left[l(l+1)\left(\left(x+a\right)^2+b^2\right) + \dfrac{1}{2}\left(1-x^2\right)\left((x+a)^2+(x+a)x+b^2\right)\right]\, . \nonumber
\label{potm1}
\ee
The hyperboloidal method relies on a combination of these two functions \eqref{pfunction}, so that the resulting potential contains a square root in the denominator
\be
\hat{V}(x)\sim \frac{1}{2\sqrt{\left(x+a\right)^2+b^2}}\, .
\ee 
As we are interested in exploring regimes with small values of parameters, this leads to the function becoming large near zero, which in turn causes significant numerical errors. Therefore, in order to apply this method for finding QNMs, we  use a yet different compactified coordinate that eliminates the square root in the denominator,
\be
x(\sigma)=b\sinh\sigma-a,\quad\sigma\in \left[\arsinh\dfrac{a-1}{b},~\arsinh\dfrac{a+1}{b}\right]\, .
\ee
In terms of this coordinate, the metric takes the following form,
\be
ds^2=-b^2\cosh^2\sigma dt^2 + \dfrac{16b^2\cosh^2\sigma d\sigma^2}{\left(1-\left(b\sinh\sigma - a\right)^2\right)^4}+\dfrac{4}{\left(1-\left(b\sinh\sigma - a\right)^2\right)^2}\, d\Omega^2\, .
\ee
The tortoise coordinate in terms of this  new compact coordinate takes the following form,
\be
z_{\ast,\sigma} = \dfrac{4}{\left(1-\left(b\sinh\sigma - a\right)^2\right)^2}\, .
\ee
Respectively,  the potential \eqref{pfunction} takes the form,
\be
\hat{V}(\sigma)=\dfrac{1}{4}\left[l(l+1)b^2\cosh^2\sigma+\dfrac{1}{2}\left(1-\left(b\sinh\sigma - a\right)^2\right)\left(2b^2\sinh^2\sigma-ab\sinh\sigma +b^2\right)\right]\, .
\ee

Next, we need to construct the height function, using the minimal gauge method described above. To do this, we split the tortoise coordinate into singular and regular parts:
\be
z_{\ast,\sigma} = \underbrace{\dfrac{2-\left(b\sinh\sigma - a\right)}{\left(1-(b\sinh\sigma - a)\right)^2}}_{z_{\ast,\sigma}^{(+,~\text{sing})}(\sigma)}+\underbrace{\dfrac{2+\left(b\sinh\sigma - a\right)}{\left(1+(b\sinh\sigma - a)\right)^2}}_{z_{\ast,\sigma}^{(-,~\text{sing})}(\sigma)}\, .
\ee
In our case, the regular part turns out to be zero, and therefore both strategies (in-out and out-in) yield the same result for the height function. Since only the derivative of the height function with respect to the compact coordinate are needed to construct the functions \eqref{pfunction} and operators $\textbf{L}_1$ and $\textbf{L}_2$ \eqref{L1L2hyper}, we will write down only the derivative:
\be
h_{,\sigma} = \dfrac{2+\left(b\sinh\sigma - a\right)}{\left(1+(b\sinh\sigma - a)\right)^2}-\dfrac{2-\left(b\sinh\sigma - a\right)}{\left(1-(b\sinh\sigma - a)\right)^2}\, .
\ee

Explicit expressions for the functions \eqref{pfunction} are listed in appendix \ref{pfunctionmetric1}. They are used to construct the operator $\textbf{L}$. By using this operator and substituting it into the evolution equation \eqref{L1L2hyper}, one can compute the waveform, and by substituting it into the eigenvalue problem \eqref{mainhyper} and solving it, one can obtain the QNMs.

\bigskip

\begin{figure}[h!]
    \begin{minipage}[h!]{0.49\linewidth}
    \center{\includegraphics[width=8cm,height=6cm]{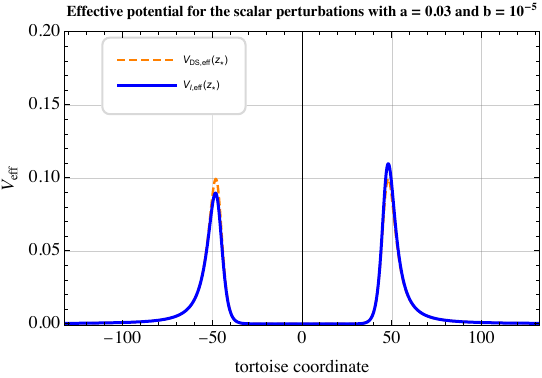}}\\
    \end{minipage}
    \hfill
    \begin{minipage}[h!]{0.49\linewidth}
    \center{\includegraphics[width=8cm,height=6cm]{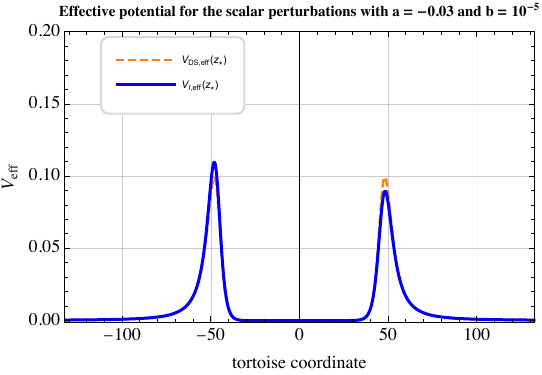}}\\
    \end{minipage}
    \begin{minipage}[h!]{0.49\linewidth}
    \center{\includegraphics[width=8cm,height=6cm]{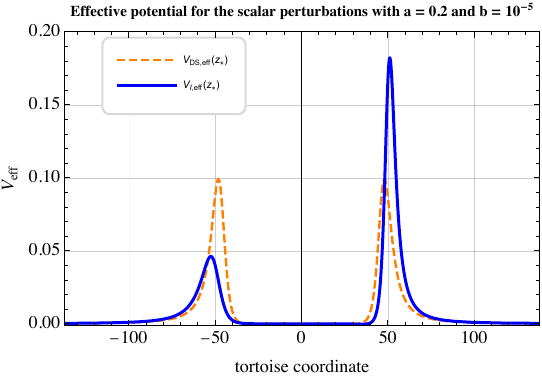}}\\
    \end{minipage}
    \begin{minipage}[h!]{0.49\linewidth}
    \center{\includegraphics[width=8cm,height=6cm]{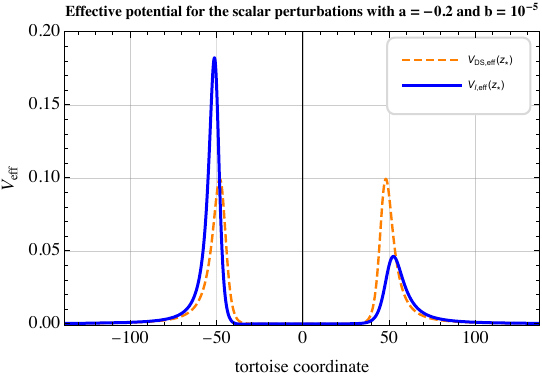}}\\
    \end{minipage}
    \caption{\footnotesize The plots show the effective potential for metric I and its comparison with the symmetric case. For all panels, the following parameters are used: $l=1$ and $b=10^{-5}$.}
    \label{Potm1}
\end{figure}

%\begin{itemize}
    %\item 
\noindent $\bullet$     \textbf{Waveform}. We begin by presenting the results obtained for the waveform through numerical integration of the equation \eqref{L1L2hyper}. Here, we show the results for parameters $l=1$ and $b=10^{-5}$ in order to compare them with those already available in the literature \cite{Bueno:2017hyj} for the symmetric DS wormhole. In Fig.~\ref{Potm1}, one can see the non-symmetric potential for scalar perturbations plotted in terms of the tortoise coordinate, along with a comparison to the symmetric case. It is evident that the sign of the parameter $a$ determines which of the maxima is higher (similarly to the effective potential of the massless geodesic discussed in Section \ref{sec:massless_potential1}). As we will see later, this will affect the resulting signal. The absolute value of the parameter $a$ determines the difference in the heights of the two peaks, as well as the distance between them, which is consistent with the formula for the travel time \eqref{traveltime1}.
   
\begin{figure}[h!]
    \begin{minipage}[h!]{0.49\linewidth}
    \center{\includegraphics[width=8cm,height=6cm]{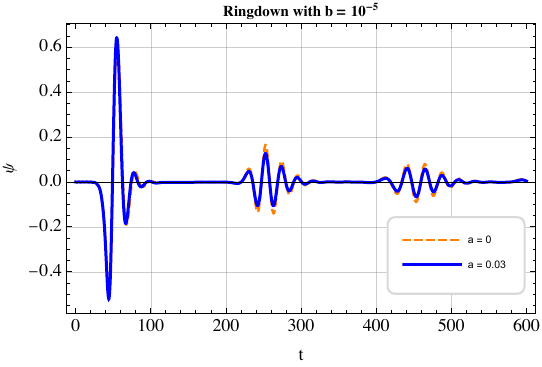}}\\
    \end{minipage}
    \hfill
    \begin{minipage}[h!]{0.49\linewidth}
    \center{\includegraphics[width=8cm,height=6cm]{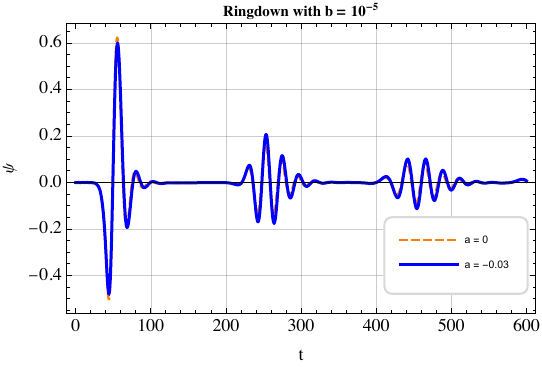}}\\
    \end{minipage}
    \begin{minipage}[h!]{0.49\linewidth}
    \center{\includegraphics[width=8cm,height=6cm]{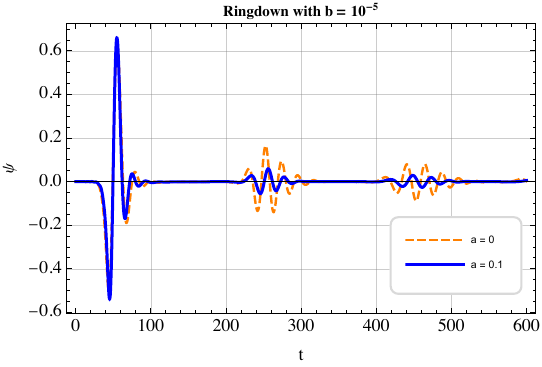}}\\
    \end{minipage}
    \begin{minipage}[h!]{0.49\linewidth}
    \center{\includegraphics[width=8cm,height=6cm]{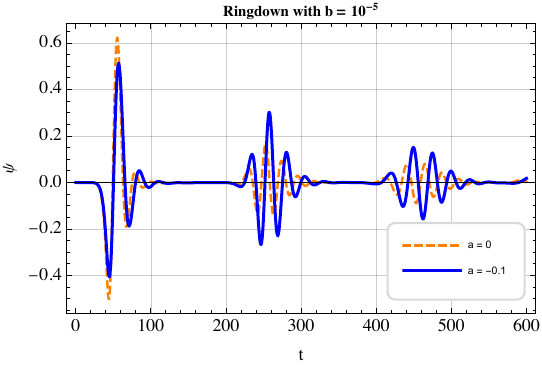}}\\
    \end{minipage}
    \begin{minipage}[h!]{0.49\linewidth}
    \center{\includegraphics[width=8cm,height=6cm]{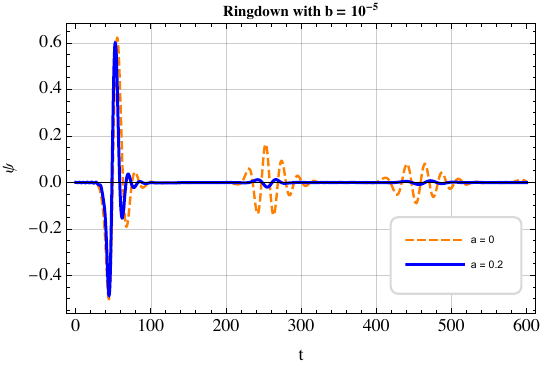}}\\
    \end{minipage}
    \begin{minipage}[h!]{0.49\linewidth}
    \center{\includegraphics[width=8cm,height=6cm]{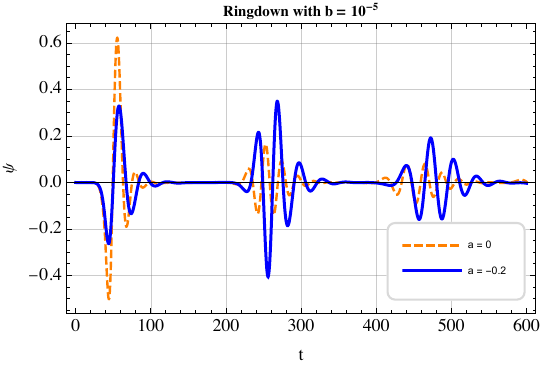}}\\
    \end{minipage}
    \caption{\footnotesize The plots show the ringdown stages for various values of the parameter $a$ as a function of time at infinity, compared with the symmetric case. In all plots, the parameters $b=10^{-5}$ and $l=1$ are used. The symmetric case is represented by the dashed orange line, while the non-symmetric case is shown as the solid blue line.}
    \label{Ringdown1}
    \end{figure}

Fig.~\ref{Ringdown1} shows the ringdown stage for metric I in comparison with the  $Z_2$ symmetric case. By analogy with the symmetric case \cite{Bueno:2017hyj}, \cite{Cardoso:2016rao} (see Sec.~\ref{sec:symworm}), for metric I the ringdown signal exhibits an echo effect - repetitions of the slightly smeared - out signals of decreasing amplitude. This occurs due to the presence of a second  peak in the potential, which reflects part of the signal that has already been scattered by the first peak. As a result, part of the perturbation becomes trapped, between the two peaks and gradually leaks out over time, manifesting as additional bursts in the signal. It is evident that the time interval between successive bursts is approximately equal to twice the travel time between the two peaks. The ringdown signals that we present here are obtained for an initial Gaussian perturbation located near the maximum of the potential in the physical region. The observer is located at infinity in the physical region.

As is seen from the resulting signals, for positive values of the parameters $a$, increasing $a$ leads to a significant decrease in the amplitude of the echo signals compared to the symmetric case. At the same time, the amplitude of the primary signal changes slightly.   This is because the peak of the potential located in the inner region - responsible for generating the echoes - becomes lower than in the symmetric case as the positive parameter $a$ increases.

   In contrast, for negative values of the parameter$a$, increasing its absolute value leads to a growth in the amplitude of the subsequent echo signals, while the amplitude of the primary signal decreases relative to the symmetric case. For example, when $a = -0.2$, the amplitudes of the primary and first echo signals become nearly equal. The explanation is analogous: as the absolute value of $a$ increases, the inner potential peak becomes higher compared to the symmetric configuration, thereby enhancing the reflection within the cavity.
   
\bigskip

    %\item
 \noindent $\bullet$       \textbf{QNMs}. Similar to the symmetric case discussed earlier (see Sec.~\ref{sec:symworm}), the present case also contains the long-lived QNMs with small imaginary parts. As noted before, their appearance is related to the structure of the potential, which has two peaks that a trapping region (see Fig.~\ref{Potm1}).
     A typical behavior of the real part is proportional to $1/L$ and of the imaginary part is proportional to $1/L^{2l+3}$, where $L$ is the distance between the two peaks of the potential and $l$ is the angular momentum \cite{Cardoso:2016rao}. 
    \begin{figure}[h!]
    \begin{minipage}[h!]{0.49\linewidth}
    \center{\includegraphics[width=8cm,height=6cm]{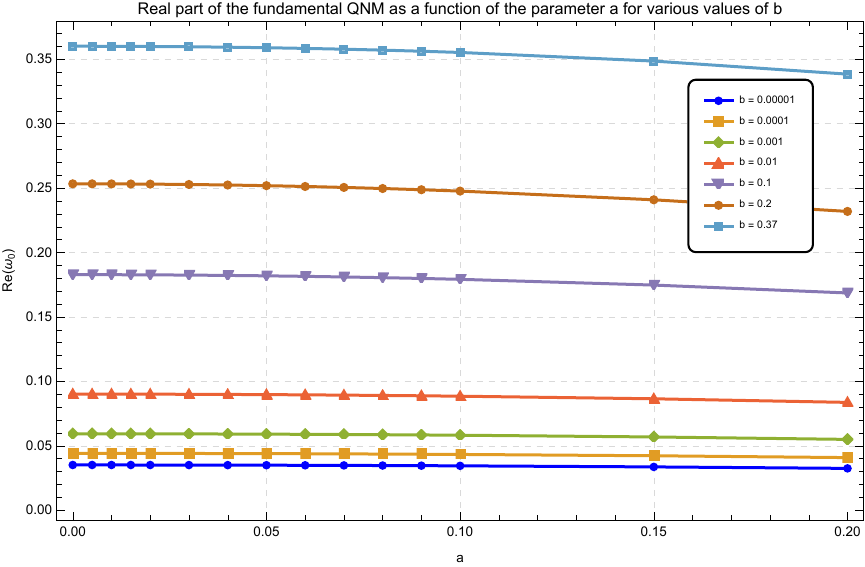}}\\
    \end{minipage}
    \hfill
    \begin{minipage}[h!]{0.49\linewidth}
    \center{\includegraphics[width=8cm,height=6cm]{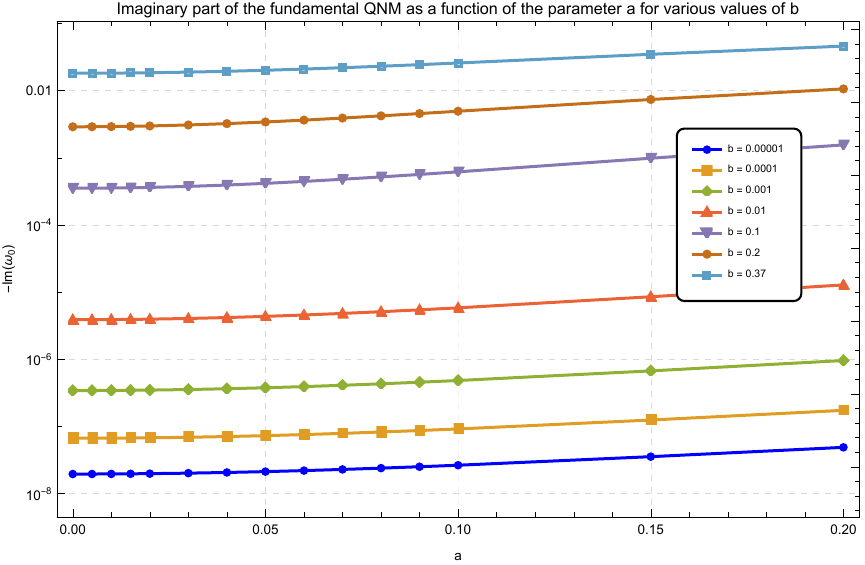}}\\
    \end{minipage}
    \caption{\footnotesize \textbf{The left plot} shows the dependence of the real part of the fundamental QNM on the parameter $a$, with different lines corresponding to different values of the parameter $b$. \textbf{The right plot} shows the dependence of the imaginary part of the fundamental QNM as a function of the parameter $a$. Different lines represent different values of $b$. The right plot is presented in a logarithmic scale along the vertical axis.}
    \label{QNMfunca}
    \end{figure}

    Fig.~\ref{QNMfunca} shows plots illustrating the dependence of the real and imaginary parts of the fundamental mode as functions of the parameter $a$, for various values of the parameter $b$. It is important to note that the following results are presented for angular number $l=1$. The parameter ranges are chosen in such  a way that their values lie within the allowed region, as shown in Fig.~\ref{Const2}. We have chosen a logarithmic scale for the imaginary part to provide a more convenient visual representation.

\begin{figure}[h!]
    \begin{minipage}[h!]{0.49\linewidth}
    \center{\includegraphics[width=8cm,height=6cm]{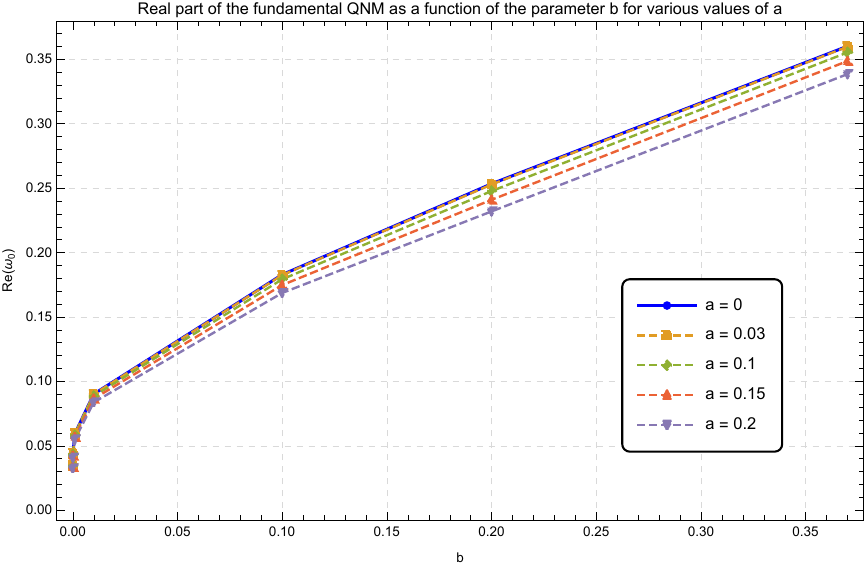}}\\
    \end{minipage}
    \hfill
    \begin{minipage}[h!]{0.49\linewidth}
    \center{\includegraphics[width=8cm,height=6cm]{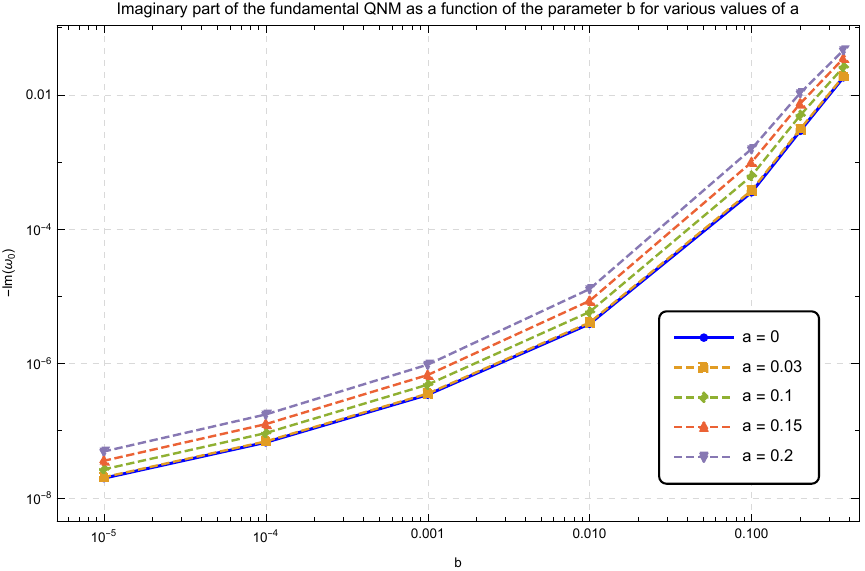}}\\
    \end{minipage}
    \caption{\footnotesize \textbf{The left plot} shows the dependence of the real part of the fundamental QNM on the parameter $b$, with different lines corresponding to different values of the parameter $a$. \textbf{The right plot} show the dependence of the imaginary part of the fundamental QNM as a function of the parameter $b$. Different lines represent different values of $a$. The right plot is presented in a log-log scale.}
    \label{QNMfuncb}
    \end{figure}

    Fig.~ \ref{QNMfuncb} shows plots illustrating the dependence of the real and imaginary parts of the fundamental modes as functions of the parameter $b$, for various values of the parameter $a$ are presented. It can be seen that as $b$ increases, the real and imaginary parts increase too. This behavior is explained by the fact that increasing $b$ reduce the distance between the peaks of the potential, making the cavity narrower. As a result, the lifetime of the QNM trapped in the potential well decreases, that is manifested  in the growth of the imaginary part. At the same time, the normal mode frequency of the well increases, which, as discussed earlier, leads to an increase in the real part of the QNM.

\begin{figure}
    \begin{minipage}[h!]{0.49\linewidth}
    \center{\includegraphics[width=8cm,height=6cm]{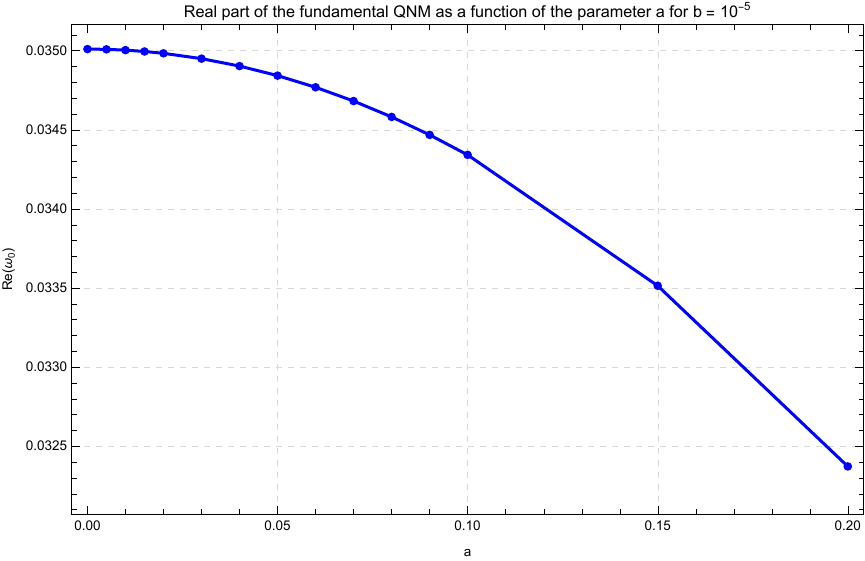}}\\
    \end{minipage}
    \begin{minipage}[h!]{0.49\linewidth}
    \center{\includegraphics[width=8cm,height=6cm]{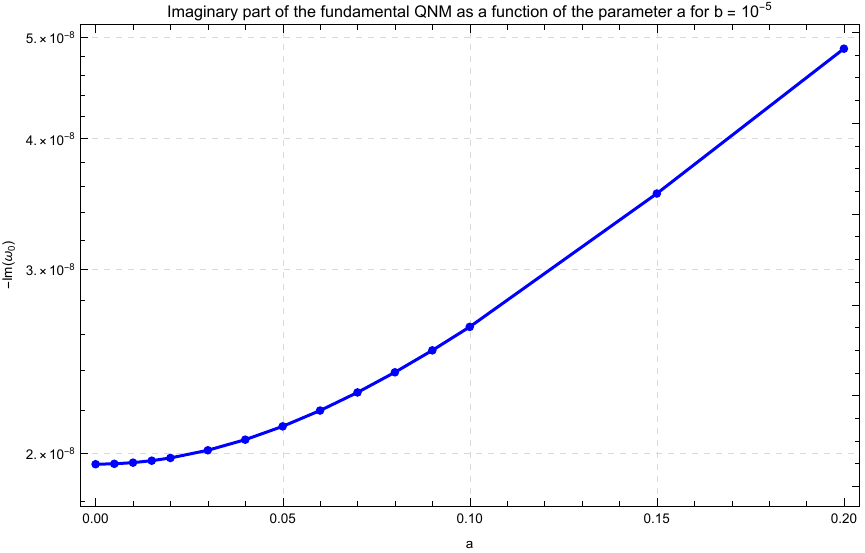}}\\
    \end{minipage}
    \begin{minipage}[h!]{0.49\linewidth}
    \center{\includegraphics[width=8cm,height=6cm]{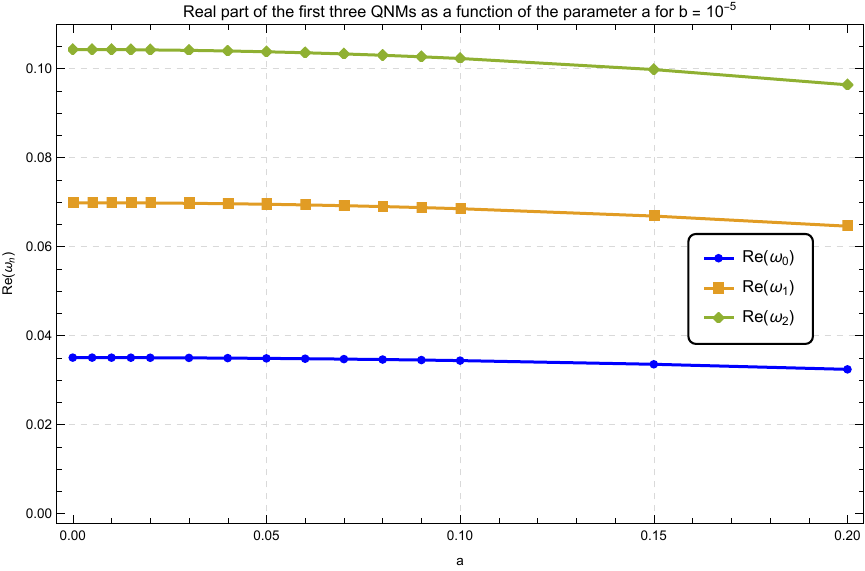}}\\
    \end{minipage}
    \begin{minipage}[h!]{0.49\linewidth}
    \center{\includegraphics[width=8cm,height=6cm]{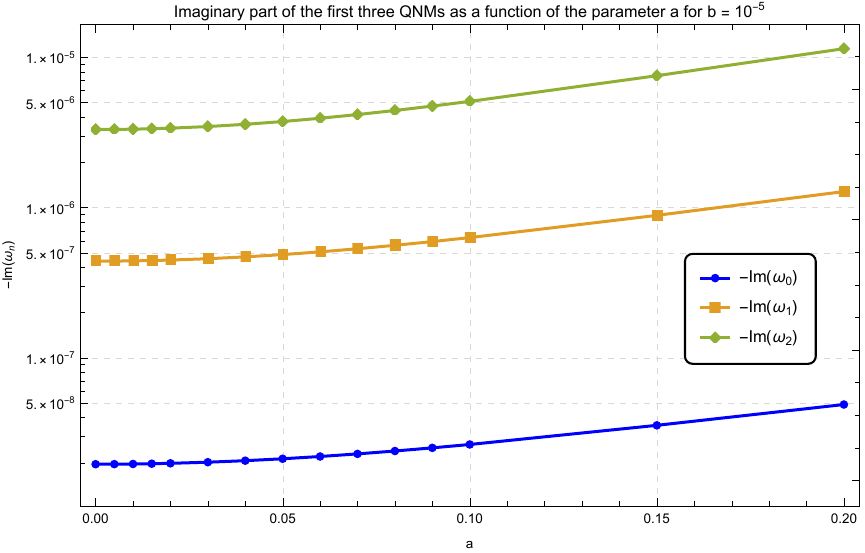}}\\
    \end{minipage}
    \caption{\footnotesize \textbf{The top plots} show the dependence of the real and imaginary parts of the fundamental mode on the parameter $a$, for fixed value $b=10^{-5}$. \textbf{The bottom plots} present analogous dependencies for the first three modes. The graphs for the imaginary parts are plotted in logarithmic scale for better visualization.}
    \label{QNM11}
    \end{figure}

    Fig.~\ref{QNM11} presents plots showing a similar dependence of the fundamental mode (top plots) and the first three modes (bottom plots) on the parameter $a$ for a fixed value of the other deformation parameter, $b=10^{-5}$. It is observed that as the parameter $a$ increases, the real part of the modes decreases. This is due to the fact that increasing $a$ leads to an increase in the distance between the potential peaks $L$ \eqref{traveltime1}, which in turn reduces the normal mode frequency of the potential well, $\omega_n\sim \pi n/L$, and thus decreases the real part of the QNM. The imaginary part, however, does not decrease - as one might expect from the increased distance $L$. It turns out that the imaginary part also depends strongly on the heights or the potential maxima, which are themselves sensitive to changes in the parameter $a$. This behavior can be clearly seen in a simple, exactly solvable example with a double delta potential 
    (see the appendix \ref{app:doubledelta} for more details),
    \be
    \text{Re}(\omega_n) \sim \dfrac{\pi n}{L}-\dfrac{\pi n }{L^2}\left(\dfrac{1}{V_1}+\dfrac{1}{V_2}\right)+\dfrac{\pi n }{L^3}\left(\dfrac{1}{V_1}+\dfrac{1}{V_2}\right)^2,\quad \text{Im}(\omega_n)\sim -\dfrac{\pi^2 n^2 }{L^3}\left(\dfrac{1}{V_1^2}+\dfrac{1}{V_2^2}\right)\, ,
    \ee
    where $V_1$ and $V_2$ are the heights of the potential peaks. We can observe similar behavior for all considered values of the parameters $b$ in Fig.~\ref{QNMfunca}.

    This distinguishes the situation from the previously described behavior in Fig.~\ref{QNMfuncb}. Since the potential heights depend weakly on the parameter $b$, the entire dependence on $b$ is essentially encoded in the width of the trapping region. As a result, the naive analysis based on the lifetime in the potential cavity holds in the case.
  %  \newpage

    Next, we present a table~\ref{QNM1} with the first three QNMs calculated for $l=1$, for selected values of the parameters $a$ and $b$. The QNMs for $a=0$ are in exact agreement with the previously computed modes in \cite{Bueno:2017hyj}, \cite{Volkel:2018hwb}.

%\end{itemize}
    \begin{table}[h!]
    \begin{center}
        \begin{tabular}{|c|c|p{2cm}|p{2cm}|p{2cm}|p{2cm}|p{2cm}|p{2cm}|}
             \hline
             \multirow{2}{2em}{$b$} & \multirow{2}{2em}{$n$} & \multicolumn{6}{|c|}{$M\omega_n$}\\
             \cline{3-8}
             & & $a = 0$ & $a = 0.03$ & $a = 0.05$ & $a = 0.07$ & $a = 0.1$ & $a = 0.2$\\
             \hline\hline
             \multirow{3}{2em}{$10^{-5}$} & $0$ & $0.03501 - 1.953\cdot 10^{-8}i$ & $0.03495 - 2.014\cdot 10^{-8}i$ & $0.0348 - 2.123\cdot 10^{-8}i$ & $0.03468 - 2.287\cdot 10^{-8}i$ & $0.03434 - 2.643\cdot 10^{-8}i$ & $0.03237 - 4.876\cdot 10^{-8}i$\\
             \cline{2-8}
             & $1$ & $0.0698 - 4.391\cdot 10^{-7}i$ & $0.0697 - 4.558\cdot 10^{-7}i$ & $0.0695  - 4.857\cdot 10^{-7}i$ & $0.0692 - 5.311\cdot 10^{-7}i$ & $0.0685 - 6.300\cdot 10^{-7}i$ & $0.06455 - 1.276\cdot 10^{-6}i$\\
             \cline{2-8}
             & $2$ & $0.1043 - 3.305\cdot 10^{-6}i$ & $0.1041 - 3.458\cdot 10^{-6}i$ & $0.1037 - 3.734\cdot 10^{-6}i$ & $0.1033 - 4.156\cdot 10^{-6}i$ & $0.1022 - 5.083\cdot 10^{-6}i$ & $0.0963 - 1.141\cdot 10^{-5}i$\\
             \hline
    %     \end{tabular}
    % \end{center}
    % % \end{table}
    % % \begin{table}
    % \begin{center}
    %     \begin{tabular}{|c|p{2em}|p{2cm}|p{2cm}|p{2cm}|p{2cm}|p{2cm}|p{2cm}|}
             \hline
             \multirow{3}{2em}{$10^{-4}$} & $0$ & $0.0440-6.674\cdot 10^{-8}i$ & $0.0439 - 6.893\cdot 10^{-8}i$ & $0.0438 - 7.285\cdot 10^{-8}i$ & $0.0436 - 7.879\cdot 10^{-8}i$ & $0.0432 - 9.165\cdot 10^{-8}i$ & $0.0407 - 1.735\cdot 10^{-7}i$\\
             \cline{2-8}
             & $1$ & $0.0876 - 1.672\cdot 10^{-6}i$ & $0.0874 - 1.743\cdot 10^{-6}i$ & $0.0871 - 1.869\cdot 10^{-6}i$ & $0.0867 - 2.063\cdot 10^{-6}i$ & $0.0859 - 2.487\cdot 10^{-6}i$ & $0.0810 - 5.327\cdot 10^{-6}i$\\
             \cline{2-8}
             & $2$ & $0.1302 - 1.421\cdot 10^{-5}i$ & $0.1300 - 1.497\cdot 10^{-5}i$ & $0.1296 - 1.634\cdot 10^{-5}i$ & $0.1290 - 1.845\cdot 10^{-5}i$ & $0.1277 - 2.312\cdot 10^{-5}i$ & $0.1203 - 5.591\cdot 10^{-5}i$\\
             \hline\hline
             \multirow{3}{2em}{$10^{-3}$} & $0$ & $0.0592 - 3.419\cdot 10^{-7}i$ & $0.0591 - 3.542\cdot 10^{-7}i$ & $0.0589 - 3.762\cdot 10^{-7}i$ & $0.0587 - 4.097\cdot 10^{-7}i$ & $0.0581 - 4.826 \cdot 10^{-7}i$ & $0.0549 - 9.578\cdot 10^{-7}i$\\
             \cline{2-8}
             & $1$ & $0.1169 - 1.027\cdot 10^{-5}i$ & $0.1167 - 1.078\cdot 10^{-5}i$ & $0.1164 - 1.170\cdot 10^{-5}i$ & $0.1158 - 1.312\cdot 10^{-5}i$ & $0.1147 - 1.625\cdot 10^{-5}i$ & $0.1081 - 3.806\cdot 10^{-5}i$\\
             \cline{2-8}
             & $2$ & $0.1721 - 1.045\cdot 10^{-4}i$ & $0.1718 - 1.112\cdot 10^{-4}i$ & $0.1712 - 1.235\cdot 10^{-4}i$ & $0.1704 - 1.424\cdot 10^{-4}i$ & $0.1686 - 1.843\cdot 10^{-4}i$ & $0.1584 - 4.797\cdot 10^{-4}i$\\
             \hline\hline
             \multirow{3}{2em}{$10^{-2}$} & $0$ & $0.0900-3.880\cdot 10^{-6}i$ & $0.0899 - 4.048\cdot 10^{-6}i$ & $0.0896 - 4.351\cdot 10^{-6}i$ & $0.0892 - 4.813\cdot 10^{-6}i$ & $0.0884-5.830\cdot 10^{-6}i$ & $0.0836 - 1.279\cdot 10^{-5}i$\\
             \cline{2-8}
             & $1$ & $0.1729 - 1.568\cdot 10^{-4}i$ & $0.1725 - 1.669\cdot 10^{-4}i$ & $0.1720-1.852\cdot 10^{-4}i$ & $0.1712 - 2.134\cdot 10^{-4}i$ & $0.1694 - 2.760\cdot 10^{-4}i$ & $0.1590 - 7.117\cdot 10^{-4}i$\\
             \cline{2-8}
             & $2$ & $0.2459 - 0.00181i$ & $0.2454 - 0.00194i$ & $0.2445 - 0.00216i$ & $0.2432-0.0025i$ & $0.2403-0.0032i$ & $0.2247-0.00702i$\\
             \hline\hline
             \multirow{3}{2em}{$0.1$} & $0$ & $0.1827 - 3.522\cdot 10^{-4}i$ & $0.1824 - 3.750\cdot 10^{-4}i$ & $0.1818-4.160\cdot 10^{-4}i$ & $0.1810-4.789\cdot 10^{-4}i$ & $0.1791-6.177\cdot 10^{-4}i$ & $0.1685-0.001556i$\\
             \cline{2-8}
             & $1$ & $0.2973-0.01017i$ & $0.2967 - 0.01068i$ & $0.2958-0.01158i$ & $0.2943-0.01289i$ & $0.2912-0.01555i$ & $0.2733-0.0284i$\\
             \hline\hline
             \multirow{3}{2em}{$0.37$} & $0$ & $0.3601-0.01817i$ & $0.3597-0.01887i$ & $0.3589-0.02011i$ & $0.3578-0.02195i$ & $0.3553-0.0258i$ & $0.3384-0.04606i$\\
             \cline{2-8}
             & $1$ & $0.454-0.103i$ & $0.4542-0.1032i$ & $0.4546-0.1035i$ & $0.4552-0.104i$ & $0.4566-0.1052i$ & $0.4683-0.1151i$\\
             \hline
        \end{tabular}
    \end{center}
    \caption{\footnotesize Scalar QNMs of  metric I with $l=1$.}
    \label{QNM1}
    \end{table}
    
\subsubsection{Metric II}
\label{sec:m2}

In terms of the compact coordinate \eqref{compactcoord},  metric II is presented as follows,
\be
ds^2 = -\left(x^2+b^2\right)\Delta\left(\rho(x)\right)dt^2 + \dfrac{16dx^2}{\left(1-x^2\right)^4} + \dfrac{4}{\left(1-x^2\right)^2}d\Omega^2,\quad \Delta(\rho)=\dfrac{1}{2}\left(1+\dfrac{\rho}{\sqrt{\rho^2+\rho_0^2}}\right)\, 
\ee
where $\rho(x)$ is the  function given  in  \eqref{rhoincompactcoord}. The derivative with respect to $x$ of the tortoise coordinate takes the form,
\be
z_{\ast,x}(x) = \dfrac{4}{\left(1-x^2\right)^2\sqrt{\left(x^2+b^2\right)\Delta\left(\rho(x)\right)}}\, .
\ee
One finds the asymptotic behavior, 
\begin{gather}
\Delta\left(\rho(x)\right)\sim\begin{cases}
    1-\dfrac{\rho_0^2}{4}\left(1-x\right)^2,\quad x\rightarrow1\\[1.5ex]
    \dfrac{\rho_0^2}{4}\left(1+x\right)^2,\quad x\rightarrow -1\\
\end{cases}\Rightarrow
z_{\ast,x}(x)\sim\begin{cases} 
     \dfrac{4}{\left(1-x^2\right)\sqrt{1+b^2}},\quad x\rightarrow 1\\[1.5ex]
     \dfrac{2}{\sqrt{1+b^2}\rho_0(1+x)^3},\quad x\rightarrow -1
     \end{cases}
\label{asymptotDelta}
\end{gather}
the following relation holds:
\be
z_{\ast,x}(x)\sim\begin{cases} 
    \dfrac{4}{\left(1-x^2\right)^2\sqrt{1+b^2}},\quad x\rightarrow 1\\
    \\
    \dfrac{2}{\sqrt{1+b^2}\rho_0(1+x)^3},\quad x\rightarrow -1
\end{cases}
\ee
For similar reasons as described in the Sec.~\ref{sec:m1}, the compact coordinate \eqref{compactcoord} is not fully suitable for the numerical application of the hyperbolic method (due to the square root in the denominator). Therefore, the further analysis will be carried out using a new radial coordinate,
\be
x(\sigma) = b\sinh \sigma,\quad \sigma\in\left[-\arsinh\dfrac{1}{b},~\arsinh\dfrac{1}{b}\right]\, .
\label{comphyper2}
\ee
In this case, the metric takes the following form,
\be
ds^2 = -b^2\cosh^2\sigma\Delta\left(\rho(b\sinh\sigma)\right)dt^2+\dfrac{16b^2\cosh^2\sigma d\sigma^2}{\left(1-b^2\sinh^2\sigma\right)^4}+\dfrac{4}{\left(1-b^2\sinh^2\sigma\right)^2}d\Omega^2\, .
\ee
The derivative of the tortoise coordinate with respect to variable $\sigma$ takes the following form,
\be
z_{\ast,\sigma}(\sigma) = \dfrac{4}{\left(1-b^2\sinh^2\sigma\right)^2\sqrt{\Delta\left(\rho(b\sinh\sigma)\right)}}\, .
\label{tortm2}
\ee
The effective potential for the scalar perturbations \eqref{pfunction} takes the following form
\begin{gather}
 \hat{V}(\sigma) = \sqrt{\Delta\left(\rho(b\sinh\sigma)\right)}\left[l(l+1)b^2\cosh^2\sigma+\dfrac{1}{2}\left(1-b^2\sinh^2\sigma\right)\left(2b^2\sinh^2\sigma + b^2\right)+\right.\nonumber\\
 \\
 \left.+\dfrac{b^3\cosh^2\sigma\sinh\sigma}{1-b^2\sinh^2\sigma}\Delta^{-1}\left(\rho(b\sinh\sigma)\right)\dfrac{d}{d\rho}\Delta\left(\rho(b\sinh\sigma)\right)\right]\, , \nonumber
\end{gather}
here we use \eqref{usefull3.2}.

Next, we need to construct the height function for metric II. We use the minimal gauge method described in the Sec.~\ref{sec:hyp}. 
Unlike the case of metric I, the regular part of the tortoise coordinate derivative is not zero, and therefore the in-out and out-in strategies will yield different height functions. In this work, we chose the in-out strategy to obtain the height function \eqref{inout}. This requires knowledge of the singular part of the tortoise coordinate in a vicinity of infinity in the physical region ($\sigma=\arsinh1/b$),
\begin{gather}
z_{\ast,\sigma}^{+,~\text{sing}}(\sigma) = \dfrac{2-b\sinh\sigma}{\left(1-b\sinh\sigma\right)^2}\, .
\end{gather}
Thus, the height function takes the following form,
\be
&&h_{,\sigma} = z_{\ast,\sigma}^{-,~\text{sing}}(\sigma) - z_{\ast,\sigma}^{+,~\text{sing}}(\sigma) + z_{\ast,\sigma}^{\text{reg}}(\sigma) = z_{\ast,\sigma}(\sigma)-2z_{\ast,\sigma}^{+,~\text{sing}}(\sigma) = \nonumber\\
\\
&&=\dfrac{4}{\left(1-b^2\sinh^2\sigma\right)^2\sqrt{\Delta\left(\rho(b\sinh\sigma)\right)}} - \dfrac{2-b\sinh\sigma}{\left(1-b\sinh\sigma\right)^2}\, .\nonumber
\ee
%\end{gather}

Explicit expressions for the functions \eqref{pfunction} used in the construction of the differential operator $\textbf{L}$ can be found in the appendix \ref{pfunctionmetric2}.
We can obtain the waveform by solving the evolution equation \eqref{L1L2hyper}, and the QNMs by solving the eigenvalue problem \eqref{mainhyper}. It should be noted that this metric is considered under the condition (Fig.~\ref{Const2}) that the resulting function $g_{\text{II}}(\rho)$ is monotonic. This is because, in the case of non-monotonic function, we would obtain results similar to those for the first metric, as there would also be two maxima separated by a distance $L \sim 4M\ln1/b^2$. As mentioned earlier, all observable effects are sensitive precisely to the heights of these maxima and the distance between them.

%\begin{itemize}
    %\item 
    
  \bigskip
    
\begin{figure}[h!]
    \begin{minipage}[h!]{0.49\linewidth}
    \center{\includegraphics[width=8cm,height=6cm]{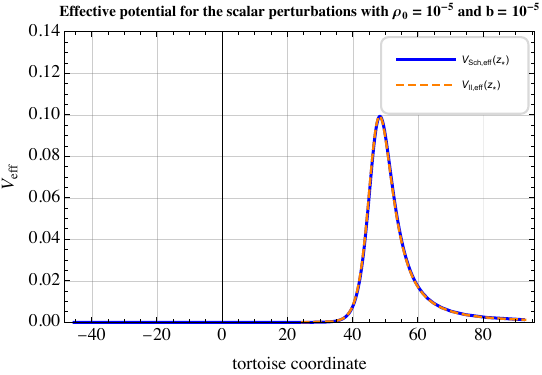}}\\
    \end{minipage}
    \begin{minipage}[h!]{0.49\linewidth}
    \center{\includegraphics[width=8cm,height=6cm]{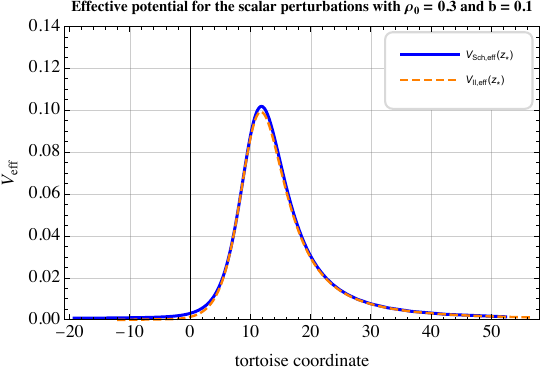}}\\
    \end{minipage}
    \caption{\footnotesize Both plots show a comparison between the effective radial potential for metric II (blue solid line) for certain values of $b$ and $\rho_0$ and that of the Schwarzschild case (orange dashed line). \textbf{The left plot} is generated for the parameters $b = 10^{-5}$ ans $\rho_0 = 10^{-5}$. \textbf{The right plot} is generated for the parameters $b = 0.1$ and $\rho_0=0.3$}
    \label{Potm2}
    \end{figure}

 \noindent $\bullet$   \textbf{Waveform}. We begin by presenting the results for the waveform. We restrict ourselves to the values of the parameters for which the metric function is monotonic (these values are shown in Fig.~\ref{Const2} as the intersection of the red and blue regions).  It occurs that the respective  effective potential for the  scalar perturbations is also monotonic in the inner region. This implies that for small values of $b$ and $\rho_0$, the potential closely approximates the effective potential of the Schwarzschild black hole, as shown in the left panel of Fig.\ref{Potm2}. As the parameters increase, the effective potential starts to deviate strongly from that of the Schwarzschild spacetime, as can be seen in the right panel of Fig.~\ref{Potm2}. Here, we consider only the case $l=1$.
    
\begin{figure}[h!]
    \begin{minipage}[h!]{0.49\linewidth}
    \center{\includegraphics[width=8cm,height=6cm]{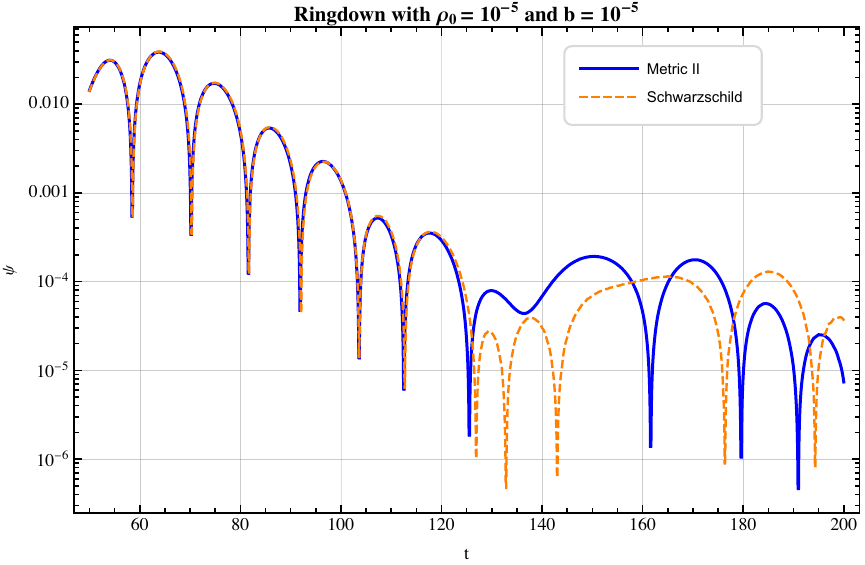}}\\
    \end{minipage}
    \begin{minipage}[h!]{0.49\linewidth}
    \center{\includegraphics[width=8cm,height=6cm]{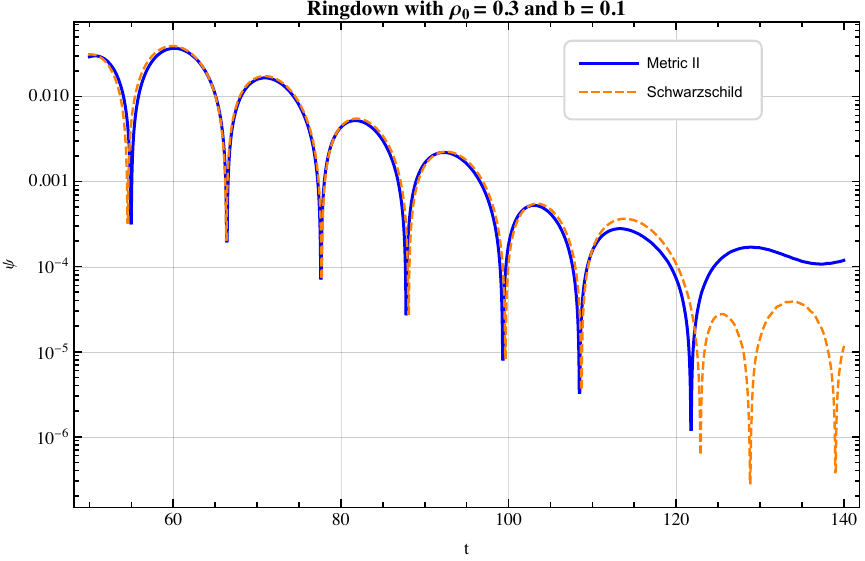}}\\
    \end{minipage}
    \caption{\footnotesize Both plots show ringdown signals for metric II (blue solid line) in comparison with the ringdown signal of the Schwarzschild metric (orange dashed line). \textbf{The left plot} presents the signal for parameters $b = 10^{-5}$ and $\rho_0=10^{-5}$. \textbf{The right plot} shows the signal for parameters $b=0.1$ and $\rho_0=0.3$.}
    \label{Ringdownm2}
    \end{figure}

    Fig.~\ref{Ringdownm2} presents the waveforms corresponding to the two potentials shown above in Fig.~\ref{Potm2} for the parameters $b = 10^{-5}$, $\rho_0=10^{-5}$, and $b=0.1$, $\rho_0=0.3$. It can be seen that there is a segment in the ringdown signals for both metric II and the Schwarzschild black hole that overlap, indicating that metric II can mimic the Schwarzschild metric, and their fundamental QNMs coincide. As is seen from our results, the ringdown signal for metric II is sustained by the fundamental QNM much longer than in the Schwarzschild black hole case
    %(as shown in Fig.~\ref{Ringdownm21}), 
    meaning that for the Schwarzschild metric, overtones start to contribute to the signal rather quickly, whereas in the case of metric II, they do not. It is also worth noting the difference between these waveforms and that of metric I. Since the effective potential in that case had a two-peak structure, we observed an echo effect. In the present case, this effect is absent due to the lack of a second peak in the effective potential.
 
\bigskip

    %\item
\noindent $\bullet$     \textbf{QNMs}.  Since the effective radial potential does not have  the two-peak structure with a long valley between the peaks,  the QNMs are not expected to have small imaginary part, contrary to what was observed for metric I. Furthermore, as discussed in the analysis of the waveform, the numerical results closely resemble those obtained in the Schwarzschild case. 
This observation is further supported by the shape of the effective potential, which closely approximates that of the Schwarzschild spacetime. A WKB analysis confirms that, under such conditions, the QNM spectrum (at least for the fundamental mode) should indeed coincide with the Schwarzschild case to a good approximation.

    The table~\ref{tab:QMNm2} presents several fundamental QNMs for selected values of the parameters.
%\end{itemize}
    
\begin{table}[h!]
    \begin{center}
        \begin{tabular}{|c|p{2cm}|c|p{2cm}|c|p{2cm}|}
             \hline
             \multicolumn{6}{|c|}{$M\omega_0$}\\
             \hline\hline
             \multicolumn{2}{|c|}{$b=10^{-5}$} & \multicolumn{2}{|c|}{$b=10^{-4}$} & \multicolumn{2}{|c|}{$b=0.1$}\\
             \hline\hline
             $\rho_0=10^{-5}$ & $0.292936 - 0.0976592i$  & $\rho_0=5\cdot 10^{-5}$ & $0.292937-0.0976594i$ & $\rho_0=0.15$ & $0.293848 - 0.098612i$\\
             \hline
             $\rho_0 = 1.5\cdot 10^{-5}$ & $0.292934 - 0.097662i$ & $\rho_0=5\cdot 10^{-5}$ & $0.292937-0.0976594i$ & $\rho_0=0.2$ & $0.294863-0.0983469i$\\
             \hline
             $\rho_0=2\cdot 10^{-5}$ & $0.292927 - 0.0976683i$ & $\rho_0=10^{-4}$ & $0.292937 - 0.0976596i$ & $\rho_0=0.3$ & $0.296442-0.0982661i$\\
             \hline
             $\rho_0=4\cdot 10^{-5}$ & $0.292915-0.097674i$ & $\rho_0=2\cdot 10^{-4}$ & $0.292936-0.0976597i$ & $\rho_0=0.4$ & $0.296663-0.0981466i$\\
             \hline
        \end{tabular}
    \end{center}
    \caption{\footnotesize The table presents the values of the fundamental QNMs for metric II for the  parameters $b=10^{-5}$, $b=10^{-4}$ and $b=0.1$, and the corresponding values of $\rho_0$ for which the metric is monotonic and the effective potential has only a single maximum.}
    \label{tab:QMNm2}
    \end{table}
    
\subsubsection{Metric III}
\label{sec:metricIII}

\begin{figure}[h!]
    \begin{minipage}[h!]{0.49\linewidth}
    \center{\includegraphics[width=8cm,height=6cm]{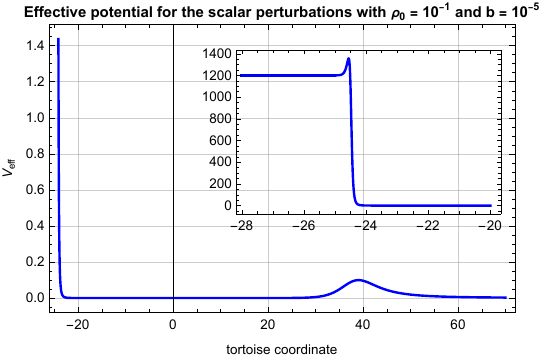}}\\
    \end{minipage}
    \begin{minipage}[h!]{0.49\linewidth}
    \center{\includegraphics[width=8cm,height=6cm]{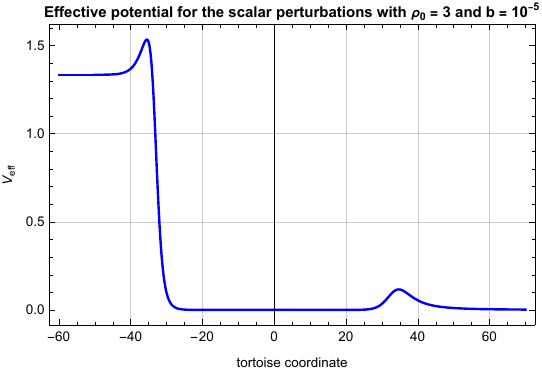}}\\
    \end{minipage}
    \caption{\footnotesize Both plots show an effective scalar potential for metric III  with $l=1$. \textbf{The left plot} is generated for the parameters $b = 10^{-5}$ and $\rho_0 = 0.1$. The inset shows the inner region, where the potential approaches a constant value and contains a finite - height wall. \textbf{The right plot} is generated for the parameters $b = 10^{-5}$ and $\rho_0=3$}
    \label{Potm3}
    \end{figure}

In terms of compact coordinate \eqref{compactcoord}  metric III  takes the following form,
\be
  && ds^2=-\left(x^2+b^2\right)\Delta^{-1}\left(\rho(x)\right)dt^2 +\dfrac{16dx^2}{\left(1-x^2\right)^4}+\dfrac{4}{\left(1-x^2\right)^2}\d\Omega^2\, , \nonumber \\
  %\quad
  && \Delta(\rho) = \dfrac{1}{2}\left(1+\dfrac{\rho}{\sqrt{\rho^2+\rho_0^2}}\right)\, ,
\ee
where $\rho(x)$ is the function as in \eqref{rhoincompactcoord}. In what follows, similar to the previous cases, the analysis will be carried out using a  new compact coordinate
\begin{equation}
    x(\sigma)=b\sinh\sigma,\quad \sigma\in\left[-\arsinh\dfrac{1}{b},\arsinh\dfrac{1}{b}\right]\, .
\end{equation}
In terms of this  new coordinate, the metric takes the form,
\begin{equation}
    ds^2=-b^2\sinh^2\sigma\Delta^{-1}\left(\rho(\sigma)\right)dt^2+\dfrac{16b^2\cosh^2\sigma d\sigma^2}{\left(1-b^2\sinh^2\sigma\right)^4}+\dfrac{4}{\left(1-b^2\sinh^2\sigma\right)^2}d\Omega^2\, ,
\end{equation}
and the effective potential for a scalar perturbation has the following form,
\begin{gather}
    V(\sigma)=\dfrac{\left(1-b^2\sinh^2\sigma\right)^2}{4}\Delta^{-1}\left(\rho(\sigma)\right)\left[l(l+1)b^2\cosh^2\sigma+\dfrac{1}{2}\left(1-b^2\sinh^2\sigma\right)\left(2b^2\sinh^2\sigma+b^2\right)-\right.\nonumber\\
    \\
    \left.-\dfrac{b^2\cosh^2\sigma b\sinh\sigma}{1-b^2\sinh^2\sigma}\Delta^{-1}\left(\rho(b\sinh\sigma)\right)\dfrac{d}{d\rho}\Delta\left(\rho(b\sinh\sigma)\right)\right]\, .\nonumber
\end{gather}
Here we use \eqref{usefull3.2}.
This potential has the following asymptotic behavior at both infinities
\begin{gather}
    V(\sigma)\sim\begin{cases}
        \dfrac{\rho_0^2\left(1+b^2\right)}{4}\left(l(l+1)+1\right),\quad \sigma\rightarrow -\arsinh\dfrac{1}{b}\\
        (1-b\sinh\sigma)^2l(l+1)\left(1+b^2\right),\quad \sigma\rightarrow \arsinh\dfrac{1}{b}
    \end{cases}
\end{gather}
At infinity in the inner region, the potential approaches a constant value, as shown in Fig.~\ref{Potm3} for two sets of parameters. As a result, a massless scalar field effectively acquires a mass that depends on the metric parameters $b$ and $\rho_0$, as well as on the angular momentum $l$. For small values of $\rho_0$, the potential reaches a large constant, creating a high barrier that acts as an effective impenetrable wall. As $\rho_0$ increases, this constant decreases, and the wall correspondingly lowers.

Since in the inner region the field effectively becomes massive, the hyperboloidal approach does not work. Therefore, in order to calculate the QNMs we use the matrix method, which is described in the section \ref{sec:matrixmethod} (see \cite{Konoplya:2004wg}, \cite{Konoplya:2006br} for more details on massive fields in the Schwarzschild and Kerr backgrounds). For this purpose we rewrite the wave equation $-\partial^2_t\phi(t,z_{\ast})+\partial^2_{z_{\ast}}\phi(t,z_{\ast}) - V_{s=0}(z_{\ast})\phi(t,z_{\ast}) = 0$ in terms of the compact coordinate $\sigma$,
\begin{gather}
    \psi''(\sigma)+p(\sigma)\psi'(\sigma)+q(\sigma)\psi(\sigma)=0\, ,
    \label{eqnmetric3compcoord}
\end{gather}
where $z_{\ast}$ is the tortoise coordinate, the prime is the derivative with respect to $\sigma$ and we used $\phi(t,z_{\ast})=\exp{(-i\omega t)}\psi(z_{\ast})$
The explicit expressions for the functions $p(\sigma)$ and $q(\sigma)$ are provided in the appendix (Eq.~\eqref{pandqmetric3}). Their asymptotic behavior at both ends of the compact domain $[-\arsinh(1/b), \arsinh(1/b)]$ is given by the following expressions:
\begin{gather}
    p(\sigma)\sim\begin{cases}
        \dfrac{1}{1+b\sinh\sigma}\sqrt{1+b^2},\quad\sigma\rightarrow -\arsinh\dfrac{1}{b}\\[1.5ex]
        -\dfrac{2}{1-b\sinh\sigma}\sqrt{1+b^2},\quad\sigma\rightarrow\arsinh\dfrac{1}{b}
    \end{cases}\nonumber
\end{gather}
\begin{gather}
    q(\sigma)\sim\begin{cases}
         \left(\dfrac{\omega^2\rho_0^2}{4}-\left(l(l+1)+1\right)\right)\dfrac{1}{\left(1+b\sinh\sigma\right)^2},\quad\sigma\rightarrow -\arsinh\dfrac{1}{b}\\[2ex]
        \dfrac{\omega^2}{\left(1-b\sinh\sigma\right)^4},\quad\sigma\rightarrow \arsinh\dfrac{1}{b}
    \end{cases}
\end{gather}
This behavior of the functions leads to the following asymptotic behavior of the field at the boundaries:
\begin{equation}
    \psi(\sigma)\sim\begin{cases}
        \left(1+b\sinh\sigma\right)^{\pm i\dfrac{\rho_0}{2\sqrt{1+b^2}}\sqrt{\omega^2-4\left(1+b^2\right)\left(l(l+1)+1\right)/\rho_0^2}},\quad\sigma\rightarrow -\arsinh\dfrac{1}{b}\\[1.5ex]
        \left(1-b\sinh\sigma\right)^{\mp i \dfrac{2+b^2}{\sqrt{1+b^2}^3}\omega}e^{\pm i \dfrac{\omega}{\sqrt{1+b^2}}},\quad\sigma\rightarrow \arsinh\dfrac{1}{b}
    \end{cases}
\end{equation}

The asymptotic conditions for QNMs should have the following form in terms of the tortoise coordinate,
\begin{equation}
    \psi(z_{\ast})\sim\begin{cases}
        e^{i\omega z_{\ast}},\quad z_{\ast}\rightarrow\infty\\[1.5ex]
        e^{-i\sqrt{\omega^2-m_{\text{eff}}^2}z_{\ast}},\quad z_{\ast}\rightarrow -\infty
    \end{cases}
    \label{scalingmetric3}
\end{equation}
where $m^2_{\text{eff}} = 4\left(1+b^2\right)\left(l(l+1)+1\right)/\rho_0^2$. Then, following the algorithm described in section \ref{sec:matrixmethod} and choosing the correct asymptotic behavior that corresponds to QNMs, one can rescale the field $\psi$
\begin{equation}
    \psi(\sigma)=\left(1+b\sinh\sigma\right)^{- i\dfrac{\rho_0}{2\sqrt{1+b^2}}\sqrt{\omega^2-4\left(1+b^2\right)\left(l(l+1)+1\right)/\rho_0^2}}\left(1-b\sinh\sigma\right)^{- i \dfrac{2+b^2}{\sqrt{1+b^2}^3}\omega}e^{ i \dfrac{\omega}{\sqrt{1+b^2}}}v(\sigma)
\end{equation}
In this case, the function $v(\sigma)$, which corresponds to QNMs, becomes regular at the boundaries $v(\sigma=\pm\arsinh 1/b)=C_{\pm}$. For computational convenience, in order to remove arbitrary constants $C_{\pm}$, we rescale the function $v(\sigma)=u(\sigma)/(1-b^2\sinh^2\sigma)$, which leads us to the final equation,
\begin{equation}
\tau(\sigma,\omega)u''(\sigma)+\lambda(\sigma,\omega)u'(\sigma)+s(\sigma,\omega)u(\sigma)=0\, .
\label{finaleqnmetric3}
\end{equation}
The matrix method is then applied to this equation. Below we present the results of our numerical analysis.

\bigskip

%\begin{itemize}
    %\item 
 \noindent $\bullet$ \textbf{Waveform.} Fig.~\ref{Potm3} shows the effective potential for a scalar field with $l=1$, for two different sets of parameters $b$ and $\rho_0$. As discussed above, the potential approaches a constant value in the inner region, which increases as $\rho_0$ decreases. In addition, in front of the finite-height wall in the inner region, there is a maximum independent of the parameters, and a separate peak appears in the physical region, reproducing the shape of the Schwarzschild potential peak. The combination of the peak and the finite-height wall gives rise to the echo effect.

Part of the signal originating from the physical region propagates toward the wormhole, is scattered by the peak in the physical region, and continues into the inner region, where it is scattered by the finite wall before returning to the physical region. Most of the signal is reflected by the wall, with the reflected fraction increasing as the wall height increases. This results in smaller echo amplitudes for lower walls compared to higher ones. Simultaneously, a portion of the signal becomes trapped between the two peaks, producing repeating, decaying bursts, as the system is dissipative and part of the energy escapes to infinity. These bursts are referred to as echo signals.

\begin{figure}[h!]
    \begin{minipage}[h!]{0.49\linewidth}
    \center{\includegraphics[width=8cm,height=6cm]{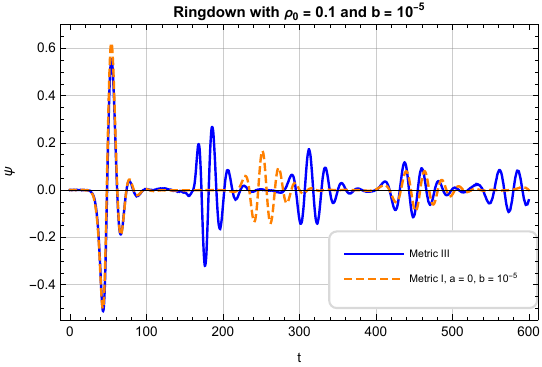}}\\
    \end{minipage}
    \begin{minipage}[h!]{0.49\linewidth}
    \center{\includegraphics[width=8cm,height=6cm]{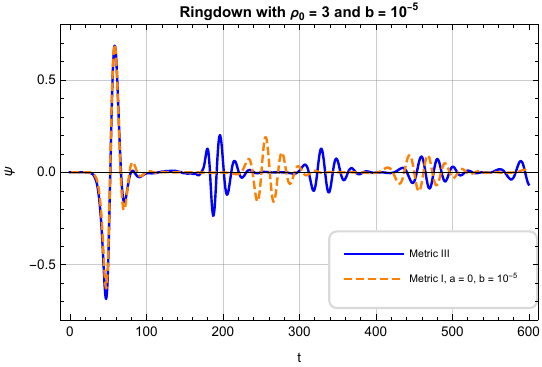}}\\
    \end{minipage}
    \caption{\footnotesize The plots show the ringdown stage for several values of the parameter $\rho_0$ as a function of time at infinity. These signals are excited by a Gaussian initial condition placed near the peak in the physical region. For comparison we show the ringdown for the symmetric version of metric I. In these plots,  we take the $b = 10^{-5}$ and $l = 1$. \textbf{The left plot} shows the ringdown for $\rho_0 = 0.1$. \textbf{The right plot} shows the rigndown for $\rho_0=3$}.
    \label{ringdown3}
    \end{figure}

Fig.~\ref{ringdown3} shows the echo signals obtained for two potentials with parameter sets $b = 10^{-5}$, $\rho_0 = 0.1$ and $b = 10^{-5}$, $\rho_0 = 3$. For comparison, the ringdown of the symmetric version of metric I is also shown. Comparing the effective potentials for metric I (Fig.~\ref{Potm1}) and metric III (Fig.~\ref{Potm3}), we observe that, for the same parameter $b = 10^{-5}$, the distance between the two peaks in metric I differs from the distance between the peak and the finite-height wall in metric III. Consequently, the echo time delay is shorter in metric III, resulting in a larger number of bursts within the interval $t < 600$ compared to metric I. As discussed in the next section, this difference impacts the quasinormal modes, since their imaginary part determines the decay rate, with $\text{Im}(\omega_n) \sim 1/L^{2l+3}$.

\bigskip

\begin{figure}
    \begin{minipage}[h!]{0.49\linewidth}
    \center{\includegraphics[width=8cm,height=6cm]{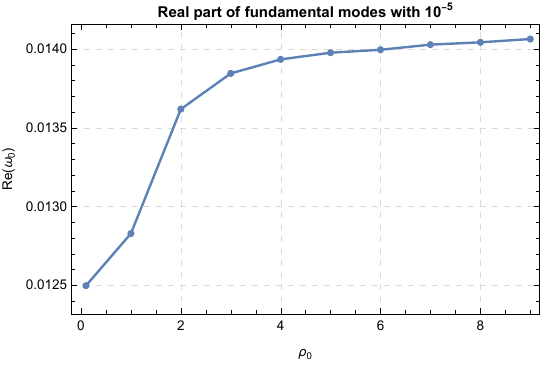}}\\
    \end{minipage}
    \begin{minipage}[h!]{0.49\linewidth}
    \center{\includegraphics[width=8cm,height=6cm]{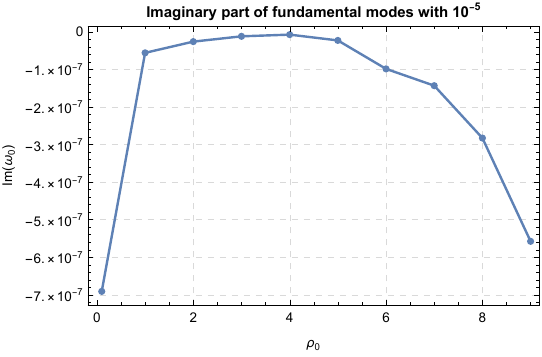}}\\
    \end{minipage}
    \begin{minipage}[h!]{0.49\linewidth}
    \center{\includegraphics[width=8cm,height=6cm]{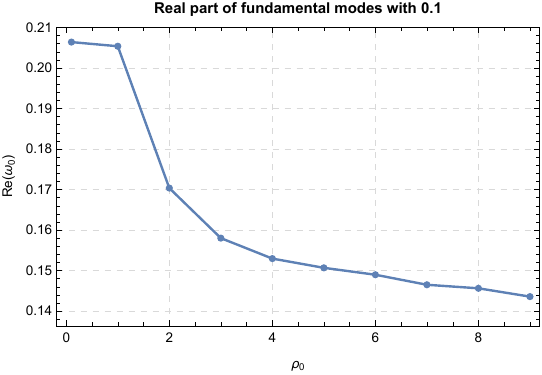}}\\
    \end{minipage}
    \begin{minipage}[h!]{0.49\linewidth}
    \center{\includegraphics[width=8cm,height=6cm]{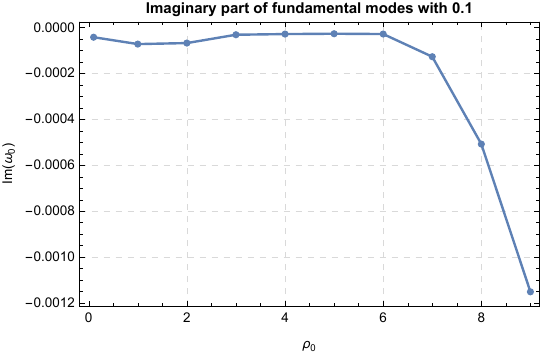}}\\
    \end{minipage}
    \caption{\footnotesize \textbf{The left plots} show the dependence of the real part of the fundamental mode on the parameter $\rho_0$, for various  values of parameter  $b=10^{-5}$, $b=0.1$ and $l=1$. \textbf{The right plots} present analogous dependencies for the imaginary parts.}
    \label{QNM31}
\end{figure}

 %   \item 
 \noindent $\bullet$ \textbf{QNMs.} Fig.~\ref{QNM31} shows how the fundamental modes depend on the parameter $\rho_0$ for two parameter sets, $b = 10^{-5}$, $l=1$ and $b=0.1$, $l=1$. The plots include values of $\rho_0$ that lie outside the allowed range (see Fig.~\ref{Const3}). Table \ref{QNM3s} lists several values of the fundamental modes for different combinations of $b$ and $\rho_0$, computed using the matrix method described in Sec.~\ref{sec:matrixmethod}. Because the effective radial potential (Fig.~\ref{Potm3}) features a trapping region, waves incident on it decay very slowly. Consequently, the imaginary part of the fundamental QNMs is extremely small, of the order of $10^{-9}$-$10^{-6}$, as shown in the table. In other words, modes trapped in the potential cavity leak out only gradually.

An important general feature is that the primary signal, just as in the symmetric case discussed in Sec.~\ref{sec:symworm}, is identical to the signal that would be produced by a single peak located in the physical region \cite{Cardoso:2016rao, Cardoso:2017njb, Cardoso:2017cqb, Mark:2017dnq}. Therefore, observing only the primary ringdown stage does not reveal the full spectrum. The QNMs corresponding to the primary signal coincide with those of the single-peak potential; these modes are not long-lived and have significantly larger imaginary parts than the true wormhole QNMs. To extract the full spectrum, one must also detect the echoes.

%\end{itemize}
\begin{table}[h!]
    \begin{center}
        \begin{tabular}{|c|p{2cm}|p{2cm}|p{2cm}|p{2cm}|p{2cm}|}
             \hline
             \multirow{1}{2em}{$b$} &  \multicolumn{5}{|c|}{$M\omega_n$}\\
             \cline{2-6}
              & $\rho_0 = 0.1$ & $\rho_0=1$ & $\rho_0=2$ & $\rho_0=3$ & $\rho_0=4$\\
             \hline\hline
             $10^{-5}$ & $0.0124974-6.911\cdot 10^{-7}i$ & $0.0128283-5.612\cdot 10^{-8}i$ & $0.013618-2.627\cdot 10^{-8}i$ & $0.013845-1.218\cdot 10^{-8}i$ & $0.0139341-7.651\cdot 10^{-9}i$\\
             \hline
             \hline
             $10^{-4}$ & $0.0177483-2.679\cdot 10^{-6}i$ & $0.0173769-1.22\cdot 10^{-6}i$ & $0.0182472-3.101\cdot 10^{-6}i$ & $0.0184891-1.481\cdot 10^{-6}i$ & $0.0185759-8.403\cdot 10^{-7}i$\\
             \hline\hline
             $10^{-3}$ & $0.041342-5.725\cdot 10^{-6}i$ & $0.0247928-4.972\cdot 10^{-6}i$ & $0.0255275-3.205\cdot 10^{-6}i$ & $0.0257127-3.041\cdot 10^{-6}i$ & $0.0257662-3.36\cdot 10^{-6}i$\\
             \hline\hline
             $10^{-2}$ & $0.0810951-6.208\cdot 10^{-4}i$ & $0.0504455-1.703\cdot 10^{-6}i$ & $0.0474537-3.591\cdot 10^{-6}i$ & $0.046608-5.534\cdot 10^{-6}i$ & $0.0463036-8.661\cdot 10^{-6}i$\\
             \hline\hline
             $0.1$ & $0.206368-4.287\cdot 10^{-4}i$ & $0.205345-7.278\cdot 10^{-4}i$ & $0.170344-6.835\cdot 10^{-4}i$ & $0.157999-3.209\cdot 10^{-4}i$ & $0.152948-2.918\cdot 10^{-4}i$\\
             \hline\hline
             $0.3$ & $0.421174-0.009431i$ & $0.41412-0.004288i$ & $0.412651-0.00342i$ & $0.38631-0.001449i$ & $0.373717-0.0004242i$\\
             \hline
        \end{tabular}
    \end{center}
    \caption{\footnotesize Scalar fundamental QNMs for  metric III with $l=1$.}
    \label{QNM3s}
\end{table}

\subsubsection{Metric IV} 
\label{sec:metricIV}

Metric IV has several distinctive features, so that its analysis is similar to that of asymptotically AdS spacetime \cite{Berti:2009kk}, \cite{Horowitz:1999jd}, \cite{Cardoso:2003cj}, \cite{Cardoso:2001bb}, \cite{Konoplya:2002zu}, \cite{Boyanov:2023qqf}. As in the earlier cases, scalar perturbations will be studied using  a compact coordinate,
\be
x(\sigma) = b\sinh \sigma,\quad \sigma\in\left[-\arsinh\dfrac{1}{b},~\arsinh\dfrac{1}{b}\right]\, ,
\label{comphyper4}
\ee
that is different from \eqref{compactcoord}, for the same reasons as discussed for the previous metrics. In terms of the new compact coordinate, the metric takes the form, 
\be
ds^2 = -b^2\cosh^2\sigma\Delta^{-n}\left(\rho(b\sinh\sigma)\right)dt^2+\dfrac{16b^2\cosh^2\sigma d\sigma^2}{\left(1-b^2\sinh^2\sigma\right)^4}+\dfrac{4}{\left(1-b^2\sinh^2\sigma\right)^2}d\Omega^2,\quad n>1
\ee
and the tortoise coordinate is now given by relation,
\be 
z_{\ast,\sigma}(\sigma) = \dfrac{4\Delta^{n/2}\left(\rho(b\sinh\sigma)\right)}{\left(1-b^2\sinh^2\sigma\right)^2}\, ,
\ee
where $\rho(x)$ is given in \eqref{rhoincompactcoord} and $_{,\sigma}$ is derivative with respect to the compact coordinate $\sigma$. Due to the asymptotic behavior of the function $\left.\Delta(\rho(b\sinh\sigma))\right|_{\sigma\rightarrow -\arsinh (1/b)}\sim (\rho_0(1+b\sinh\sigma)/2)^{2}$, the tortoise coordinate behaves as $\left.z_{\ast,\sigma}(\sigma)\right|_{\sigma\rightarrow -\arsinh (1/b)}\sim \text{const}$. When $\sigma\rightarrow\arsinh(1/b)$, the tortoise coordinate $\left.z_{\ast,\sigma}(\sigma)\right|_{\sigma\rightarrow\arsinh(1/b)}\sim 1/(1-b\sinh\sigma)^2$. Thus, the tortoise coordinate has singularity only at infinity in physical region. 

The effective potential takes the following form,
\begin{gather}
V(\sigma) = \dfrac{\left(1-b^2\sinh^2\sigma\right)^2}{4}\Delta^{-n}\left(\rho(b\sinh\sigma)\right)\left[l(l+1)b^2\cosh^2\sigma+\dfrac{1}{2}\left(1-b^2\sinh^2\sigma\right)\left(2b^2\sinh^2\sigma+b^2\right) -\right.\nonumber\\
\\
\left.-\dfrac{nb^2\cosh^2\sigma b\sinh\sigma}{1-b^2\sinh^2\sigma}\Delta^{-1}\left(\rho(b\sinh\sigma)\right)\dfrac{d}{d\rho}\Delta\left(\rho(b\sinh\sigma)\right)\right],\quad n>1\, .\nonumber
\label{potm4}
\end{gather}
Here we use \eqref{usefull3.2}. Due to the asymptotic behavior $\left.\Delta(\rho(b\sinh\sigma))\right|_{\sigma\rightarrow -\arsinh (1/b)}\sim (\rho_0(1+b\sinh\sigma)/2)^{2}$, the potential diverges at the left boundary of the compact domain, where one has $\sigma = -\arsinh(1/b)$,
\be
\left.V_{s=0}(\sigma)\right|_{\sigma\rightarrow -\arsinh(1/b)} \sim \dfrac{4^n\left(1+b^2\right)\left(l(l+1)+n\right)}{\rho_0^{2n}\left(1+b\sinh\sigma\right)^{2(n-1)}},\quad n>1
\ee
Let us consider the scalar field equation in the background of metric IV \eqref{QNMeqn} in coordinates $(t,~\sigma)$
\be
\Psi_{,\sigma\sigma}(\sigma) + p(\sigma)\Psi_{,\sigma}(\sigma)+q(\sigma)\Psi(\sigma) = 0
\ee
The asymptotic behavior of the $p(\sigma)$ and $q(\sigma)$ (the exact form of the functions can be found in the appendix \ref{sec:appCoeffHypApp}) functions at the left boundary $\sigma=-\arsinh(1/b)$ of the compact domain is as follows,
\be
\left.p(\sigma)\right|_{\sigma\rightarrow -\arsinh(1/b)}\sim\dfrac{2-n}{1+b\sinh\sigma}\sqrt{1+b^2}\nonumber\\
\\
~~~\left.q(\sigma)\right|_{\sigma\rightarrow -\arsinh(1/b)}\sim -\dfrac{l(l+1)+n}{\left(1+b\sinh\sigma\right)^2}\left(1+b^2\right)\, .\nonumber
\ee
This  leads to the following asymptotic behavior of the solution near the boundary,
\be
\left.\Psi(\sigma)\right|_{\sigma\rightarrow -\arsinh(1/b)}\sim c_+\left(1+b\sinh\sigma\right)^{k_+} + c_-\left(1+b\sinh\sigma\right)^{k_-}\, ,
\ee
where $k_{\pm} = (n-1\pm\sqrt{(1+n)^2+4l(l+1)})/2$. Since we are interested in the regular solutions, it is necessary impose condition $c_- =0$, which in turn imposes a Dirichlet boundary condition, $\Psi(\sigma=-\arsinh(1/b)) = 0$, at the left boundary \cite{Berti:2009kk}, \cite{Boyanov:2023qqf}. At the right boundary $\sigma=\arsinh(1/b)$, due to the asymptotic behavior $\left.\Delta(\rho(b\sinh\sigma))\right|_{\sigma\rightarrow\arsinh (1/b)}\sim 1$, the potential \eqref{potm4} vanishes, and the equation \eqref{QNMeqn} reduces to the free wave equation, resulting in the standard outgoing wave condition $\Psi\sim e^{i\omega z_{\ast}}$.

The next step is to construct a hyperboloidal slicing suitable for metric IV. We will use the minimal gauge method described in Sec.~\ref{sec:hyp}. Since the tortoise coordinate has a singularity only at one boundary of the compact domain, corresponding to asymptotic infinity in physical region, we adopt the in-out strategy, which requires isolating the singular part of the tortoise coordinate at this boundary,
\be
z_{\ast,\sigma}^{+,~\text{sing}}(\sigma) = \dfrac{2-b\sinh\sigma}{(1-b\sinh\sigma)^2}\Delta^{n/2}\left(\rho(b\sinh\sigma)\right)\, ,
\ee
for convenience, we include the $\Delta(\rho)$ function in the singular part. In this case, since the tortoise coordinate has a singular part only in the physical region, we will follow the strategy of \cite{Boyanov:2023qqf} and take this singular contribution as the derivative of the height function (as only this derivative is needed) $h_{,\sigma}=-~z_{\ast,\sigma}^{+,~\text{sing}}(\sigma)$. 
The functions \eqref{pfunction} used to construct the differential operator $\textbf{L}$ can be found in the appendix \ref{pfunctionmetric4}.
 It can be seen that the potential $\hat{V}$ remains divergent at the left boundary $\sigma=-\arsinh(1/b)$. Unlike, metrics I and II discussed in sections \ref{sec:m1} and \ref{sec:m2}, the hyperboloidal slicing does not automatically impose the required boundary conditions at both boundaries. The outgoing condition at the right boundary $\sigma=\arsinh(1/b)$ of the compact domain is imposed automatically, whereas the Dirichlet condition at the left boundary $\sigma=-\arsinh(1/b)$ must be imposed by hand. Due to this, and because the potential remains divergent, it is necessary to perform an additional transformation of the equation by rescaling
\be
\Psi(t,~\sigma) =  \Delta^{n/4}\left(\rho(b\sinh\sigma)\right)\eta(t,~\sigma)
\ee
and multiplying the resulting equation by a factor $ \Delta^{n/4}\left(\rho(b\sinh\sigma)\right)$. As a result, we obtain an equation analogous to the original one \eqref{L1L2hyper}
\begin{gather}
        \Delta^{n/2}\left(\rho(b\sinh\sigma)\right)\partial_{\tau}\begin{pmatrix}
            \eta(t,~\sigma)\\[4pt]
            \partial_{\tau}\eta(t,~\sigma)
        \end{pmatrix} = i\underbrace{\dfrac{1}{i}\begin{pmatrix}
            0 & \Delta^{n/2}\left(\rho(b\sinh\sigma)\right)\\[4pt]
            \mathbf{\bar{L}}_1 & \mathbf{\bar{L}}_2
        \end{pmatrix}}_{\bar{\mathbf{L}}}\begin{pmatrix}
            \eta(t,~\sigma)\\[4pt]
            \partial_{\tau}\eta(t,~\sigma)
        \end{pmatrix}\nonumber\\[6pt]
       % \\
        \mathbf{\bar{L}}_1 = \dfrac{1}{\bar{p}_{\tau\tau}(x)}\left[\partial_\sigma\left(\bar{p}_{\sigma\sigma}(\sigma)\partial_\sigma\right) - \hat{\bar{V}}(\sigma)\right],\quad \mathbf{L}_2 = \dfrac{1}{\bar{p}_{\tau\tau}(\sigma)}\left[2\bar{p}_{\tau\sigma}(\sigma)\partial_\sigma + \partial_\sigma\bar{p}_{\tau\sigma}(\sigma)\right]\nonumber
        \label{L1L2hypermetric4}
\end{gather}
In terms of the new rescaled functions (see Appendix \ref{pfunctionmetric41}), the transformed potential $\hat{\bar{V}}$ becomes free of divergences. Applying the Fourier decomposition $\eta(\tau, \sigma) = e^{-i\omega \tau} \bar{\eta}(\sigma)$ then leads to a generalized eigenvalue problem. In the numerical procedure, to implement the Dirichlet boundary condition, the rows and columns of the matrix $\mathbf{\bar{L}}$ corresponding to the right boundary, where the condition is applied, must be removed. For simplicity, all results presented below correspond to metric IV with $n=2$.

\bigskip

\begin{figure}[h!]
    \begin{minipage}[h!]{0.49\linewidth}
    \center{\includegraphics[width=8cm,height=6cm]{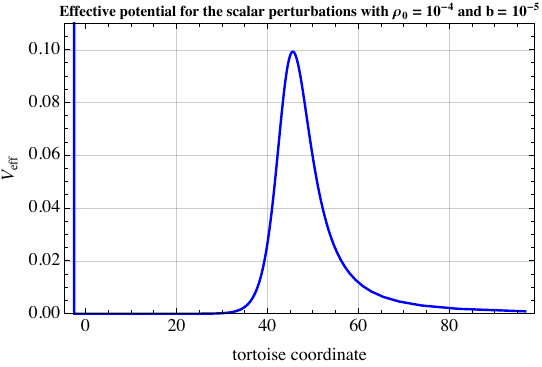}}\\
    \end{minipage}
    \begin{minipage}[h!]{0.49\linewidth}
    \center{\includegraphics[width=8cm,height=6cm]{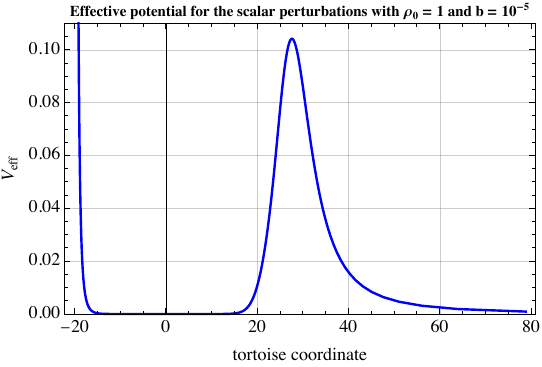}}\\
    \end{minipage}
    \caption{\footnotesize Both plots show an effective scalar potential for metric IV  with $l=1$. \textbf{The left plot} is generated for the parameters $b = 10^{-5}$ and $\rho_0 = 10^{-4}$. \textbf{The right plot} is generated for the parameters $b = 10^{-5}$ and $\rho_0=1$}
    \label{Potm4}
    \end{figure}

%\begin{itemize}
   % \item 
   \noindent $\bullet$ \textbf{Waveform}.   We begin by illustrating the waveform corresponding to a set of parameters located within the allowed region shown in Fig.~\ref{Const4}. Fig.~\ref{Potm4} also depicts the effective potential for scalar perturbations in tortoise coordinates for two different parameter sets. As noted previously, the potential exhibits an infinite wall at a finite value of the tortoise coordinate. This feature is reminiscent of the behavior observed in the SAdS black hole \cite{Horowitz:1999jd, Boyanov:2023qqf} as well as in gravastar models \cite{Cardoso:2014sna}. The combination of an infinite wall in the internal region and a potential peak in the physical region, which closely resembles the effective potential for scalar perturbations in the Schwarzschild case, gives rise to an echo effect analogous to that observed in the symmetric case discussed in Sec.~\ref{sec:symworm}.
    
\begin{figure}[h!]
    \begin{minipage}[h!]{0.49\linewidth}
    \center{\includegraphics[width=8cm,height=6cm]{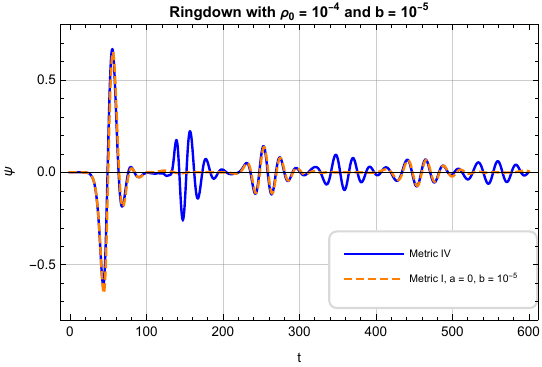}}\\
    \end{minipage}
    \begin{minipage}[h!]{0.49\linewidth}
    \center{\includegraphics[width=8cm,height=6cm]{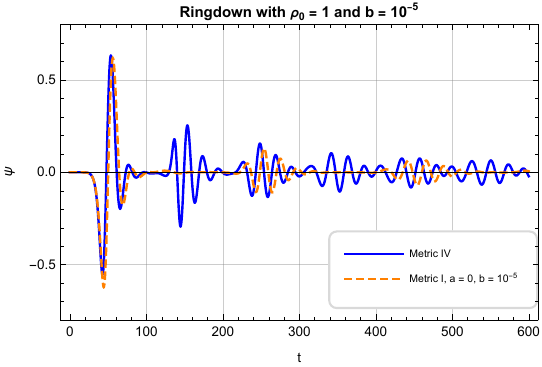}}\\
    \end{minipage}
    \caption{\footnotesize The plots show the ringdown stage for several values of the parameter $\rho_0$ as a function of time at infinity. These signals are excited by a Gaussian initial condition placed near the peak in the physical region. For comparison there are shown the ringdown for the symmetric version of metric I. In all plots, the parameters $b = 10^{-5}$ and $l = 1$ are used. \textbf{The left plot} shows the ringdown for $\rho_0 = 10^{-4}$. \textbf{The right plot} shows the rigndown for $\rho_0=1$}
    \label{ringdown4}
    \end{figure}

     Fig.~\ref{ringdown4} shows the ringdown signal generated from an initial Gaussian pulse placed near the peak in the physical region, with the observer located at infinity in the physical region. For comparison, we also show the ringdown for the symmetric version of metric I with $a = 0$ and $b = 10^{-5}$. The Fig.~\ref{ringdown4} clearly illustrates the appearance of the echo effect. This arises because the signal, falling from the outside onto the wormhole throat, first encounters the potential peak. After scattering, the entire signal is completely reflected back by the infinite wall into the physical region. Since the travel time is finite, the observer records not only the main signal but also subsequent echo signals.

By comparing the effective potentials for metric I (Fig.~\ref{Potm1}) and metric IV (Fig.~\ref{Potm4}), we observe that for the same value $b=10^{-5}$, the distance between the two peaks in metric I differs from the distance between the peak and the infinite wall in metric IV. This difference accounts for the shorter echo time delay. Specifically, for $\rho_0=10^{-4}$, the distance between the peak and the vertical wall at the zero of the tortoise coordinate is half of the peak-to-peak distance in metric I. Consequently, the echo signal in metric IV exhibits additional bursts between the echoes observed in metric I (see Fig.~\ref{ringdown4}).
    
     \begin{figure}[h!]
    \begin{minipage}[h!]{0.49\linewidth}
    \center{\includegraphics[width=8cm,height=6cm]{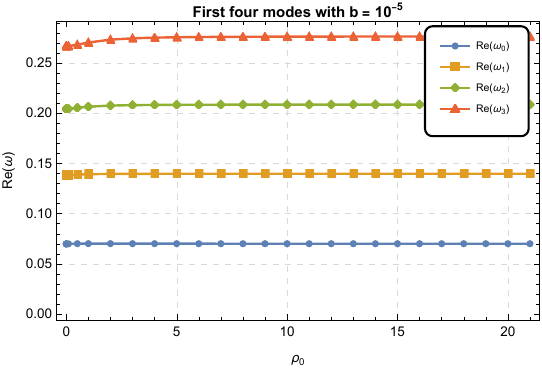}}\\
    \end{minipage}
    \begin{minipage}[h!]{0.49\linewidth}
    \center{\includegraphics[width=8cm,height=6cm]{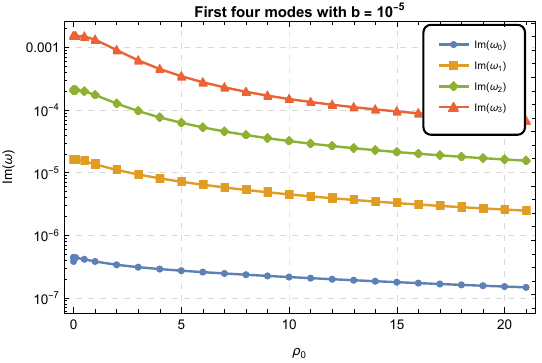}}\\
    \end{minipage}
    \caption{\footnotesize Both plots (\textbf{left} - real part, \textbf{right} - imaginary part) show the dependence of the first four QNMs on the parameter $\rho_0$ for $b=10^{-5}$ and $l=1$.}
    \label{QNM44}
    \end{figure}

\bigskip

%    \item
\noindent $\bullet$  \textbf{QNMs}.
Fig.~\ref{QNM44} shows the first four QNMs as functions of the parameter $\rho_0$ for $b=10^{-5}$. As seen in the Fig.~\ref{QNM44}, the fundamental mode has an imaginary part of order $10^{-7}-10^{-6}$, consistent with the previous observation that the signal becomes trapped between the potential peak and the wall. The small imaginary part indicates that this mode is long-lived. The overtones, in contrast, have larger imaginary parts. As noted above, the typical behavior is that the real part scales approximately as $\sim 1/L$, while the imaginary part scales as $\sim 1/L^{2l+3}$, where $L$ represents the effective size of the cavity.
    
However, as described in Sec.~\ref{sec:symworm}, the primary signal is dominated by the QNMs of the black hole rather than those of the wormhole mimicker. This is evident in the signal itself (see Fig.~\ref{ringdown4}): the initial response, produced by scattering on the potential peak, is brief and decays rapidly. Such behavior cannot be captured by a long-lived fundamental quasinormal mode. The QNMs of the full system become relevant only once the signal reaches the wall, reflects, and returns to the physical region.

\begin{figure}
    \begin{minipage}[h!]{0.49\linewidth}
    \center{\includegraphics[width=8cm,height=6cm]{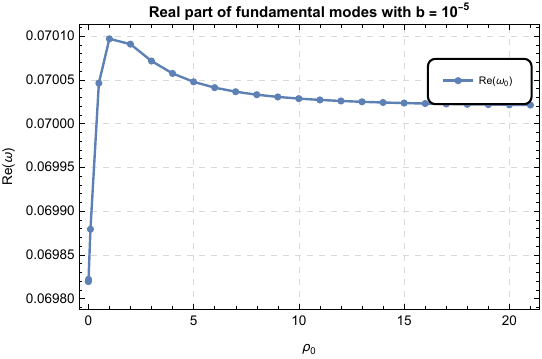}}\\
    \end{minipage}
    \begin{minipage}[h!]{0.49\linewidth}
    \center{\includegraphics[width=8cm,height=6cm]{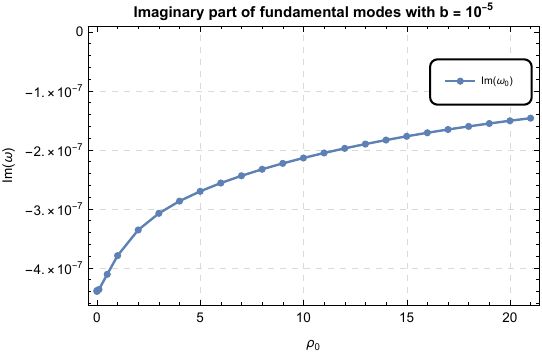}}\\
    \end{minipage}
    \begin{minipage}[h!]{0.49\linewidth}
    \center{\includegraphics[width=8cm,height=6cm]{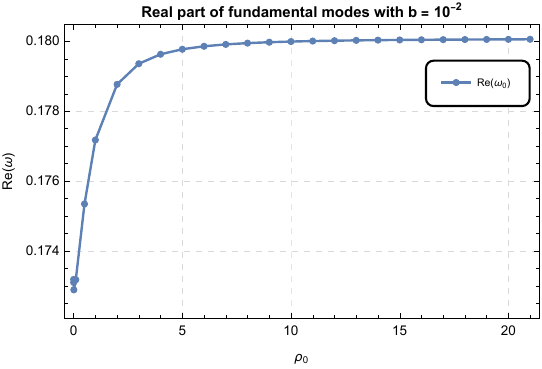}}\\
    \end{minipage}
    \begin{minipage}[h!]{0.49\linewidth}
    \center{\includegraphics[width=8cm,height=6cm]{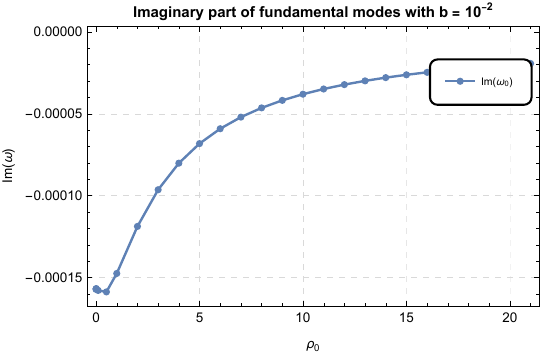}}\\
    \end{minipage}
    \caption{\footnotesize \textbf{The left plots} show the dependencies of the real parts of the fundamental mode on the parameter $\rho_0$, for the values $b=10^{-5}$, $b=10^{-2}$ and $l=1$. \textbf{The right plots} present analogous dependencies for the imaginary parts.}
    \label{QNM45}
    \end{figure}

Fig.~\ref{QNM45} shows the dependence of the fundamental modes on the parameter $\rho_0$ for two values of $b$, $b = 10^{-5}$ and $b = 10^{-2}$, with $l = 1$. The plot also includes values of $\rho_0$ that lie outside the allowed region (see Fig.~\ref{Const4}); the corresponding allowed values are listed in Table~\ref{QNM4s}. From the figure, it is clear that the real part of the fundamental mode approaches a constant as $\rho_0$ increases. Specifically, for $b = 10^{-5}$, $\text{Re}(\omega_0) \sim 0.07$, while for $b = 10^{-2}$, $\text{Re}(\omega_0) \sim 0.18$. A similar trend is observed for other values of $b$ as well as for the overtones.

%\end{itemize}
\begin{table}[h!]
    \begin{center}
        \begin{tabular}{|c|c|p{2cm}|p{2cm}|p{2cm}|p{2cm}|p{2cm}|p{2cm}|}
             \hline
             \multirow{2}{2em}{$b$} & \multirow{2}{2em}{$n$} & \multicolumn{6}{|c|}{$M\omega_n$}\\
             \cline{3-8}
             & & $\rho_0 = 10^{-4}$ & $\rho_0=10^{-3}$ & $\rho_0=10^{-1}$ & $\rho_0=0.5$ & $\rho_0=1$ & $\rho_0=2$\\
             \hline\hline
             \multirow{3}{2em}{$10^{-5}$} & $0$ & $0.06982-4.402\cdot 10^{-7}i$ & $0.06982-4.392\cdot 10^{-7}i$ & $0.06988-4.37\cdot 10^{-7}i$ & $0.070046-4.113\cdot 10^{-7}i$ & $0.070097-3.796\cdot 10^{-7}i$ & $0.07009-3.361\cdot 10^{-7}i$\\
             \cline{2-8}
             & $1$ & $0.138188-1.620\cdot 10^{-5}i$ & $0.138188-1.617\cdot 10^{-5}i$ & $0.138295-1.616\cdot 10^{-5}i$ & $0.138879-1.519\cdot 10^{-5}i$ & $0.139257-1.356\cdot 10^{-5}i$ & $0.139513-1.102\cdot 10^{-5}i$\\
             \cline{2-8}
             & $2$ & $0.204058-2.072\cdot 10^{-4}i$ & $0.204058-2.073\cdot 10^{-4}i$ & $0.204159-2.075\cdot 10^{-4}i$ & $0.20526-1.99\cdot 10^{-4}i$ & $0.206401-1.736\cdot 10^{-4}i$ & $0.207504-1.267\cdot 10^{-4}i$\\
             \hline
             \hline
             \multirow{3}{2em}{$10^{-4}$} & $0$ & $0.087556-1.671\cdot 10^{-7}i$ & $0.087556-1.672\cdot 10^{-6}i$ & $0.08765-1.667\cdot 10^{-6}i$ & $0.087962-1.575\cdot 10^{-6}i$ & $0.088086-1.445\cdot 10^{-6}i$ & $0.088117-1.258\cdot 10^{-6}i$\\
             \cline{2-8}
             & $1$ & $0.171719-7.834\cdot 10^{-5}i$ & $0.171719-7.834\cdot 10^{-5}i$ & $0.171852-7.844\cdot 10^{-5}i$ & $0.172917-7.507\cdot 10^{-5}i$ & $0.1738-6.662\cdot 10^{-5}i$ & $0.174539-5.167\cdot 10^{-5}i$\\
             \cline{2-8}
             & $2$ & $0.250308-0.00115i$ & $0.250307-0.00115i$ & $0.250417-0.001152i$ & $0.25204-0.001128i$ & $0.254392-0.001021i$ & $0.257258-0.0007173i$\\
             \hline\hline
             \multirow{3}{2em}{$10^{-3}$} & $0$ & $0.117093-9.247\cdot 10^{-6}i$ & $0.116918-1.027\cdot 10^{-5}i$ & $0.117087-1.028\cdot 10^{-5}i$ & $0.117793-9.828\cdot 10^{-6}i$ & $0.119195-8.968\cdot 10^{-6}i$ & $0.118403-7.566\cdot 10^{-6}i$\\
             \cline{2-8}
             & $1$ & $0.224433-6.571\cdot 10^{-4}i$ & $0.224371-6.554\cdot 10^{-4}i$ & $0.224235-6.538\cdot 10^{-4}i$ & $0.226168-6.496\cdot 10^{-4}i$ & $0.22854-5.899\cdot 10^{-4}i$ & $0.231136-4.362\cdot 10^{-4}i$\\
             \cline{2-8}
             & $2$ & $0.318902-0.008208i$ & $0.318822-0.008192i$ & $0.318603-0.008145i$ & $0.32069-0.008232i$ & $0.325096-0.007973i$ & $0.332873-0.006056i$\\
             \hline\hline
             \multirow{3}{2em}{$10^{-2}$} & $0$ & $0.173183-1.572\cdot 10^{-4}i$ & $0.173095-1.567\cdot 10^{-4}i$ & $0.173174-1.579\cdot 10^{-4}i$ & $0.175344-1.588\cdot 10^{-4}i$ & $0.177174-1.475\cdot 10^{-4}i$ & $0.178766-1.119\cdot 10^{-4}i$\\
             \cline{2-8}
             & $1$ & $0.310221-0.009422i$ & $0.310098-0.009395i$ & $0.310027-0.009362i$ & $0.312905-0.009619i$ & $0.318812-0.009634i$ & $0.329392-0.0079012i$\\
             \cline{2-8}
             & $2$ & $0.433883-0.05026i$ & $0.433726-0.05015i$ & $0.43364-0.04992i$ & $0.436877-0.05064i$ & $0.443433-0.0516i$ & $0.461439-0.04863i$\\
             \hline\hline
             \multirow{3}{2em}{$0.1$} & $0$ & $0.297614-0.01026i$ & $0.29761-0.01026i$ & $0.297679-0.01025i$ & $0.303922-0.01135i$ & $0.31475-0.0127i$ & $0.333275-0.0126i$\\
             \cline{2-8}
             & $1$ & $-$ & $-$ & $0.469661-0.1144i$ & $0.476239-0.1183i$ & $0.487697-0.1244i$ & $0.519252-0.1332i$\\
             % \cline{2-8}
             % & $2$ & $0.3825 - 0.04889i$ & $0.3825-0.04919i$ & $0.3824-0.04975i$ & $0.3824-0.05062i$ & $0.3821-0.05266i$ & $0.3773-0.06856i$\\
             \hline\hline
             \multirow{3}{2em}{$0.3$} & $0$ & $0.424125-0.1004i$ & $0.418028-0.0734i$ & $0.418714-0.07361i$ & $0.425014-0.07988i$ & $0.440459-0.09237i$ & $0.485085-0.1166i$\\
             \hline
        \end{tabular}
    \end{center}
    \caption{\footnotesize Scalar QNMs of metric IV with $l=1$.}
    \label{QNM4s}
    \end{table}
    
%\newpage

\section{Conclusion}

In this paper, we have studied several generalizations of the spherically symmetric Damour-Solodukhin (DS) wormhole \cite{Damour:2007ap}. By relaxing the $Z_2$ symmetry of the metric under transformation $\rho \rightarrow -\rho$, a richer spacetime structure emerges. We identified four classes of metrics: (i) a metric with two unequal peaks in the effective radia potentials, (ii) a metric in which $-g_{tt}$  tends to zero at asymptotic infinity  in the inner region, (iii) a metric with a semi-permeable wall in the potential in the inner region, and (iv) a metric with an impenetrable wall, depending on the asymptotic behavior of the functions $g(\rho)$ and $r(\rho)$ in the metric. For each class, we proposed the phenomenological representative metrics.

For these test metrics, we analyzed two observational effects: optical signatures (shadows) and gravitational wave signals (ringdown). We found that asymmetry leads to the appearance of a two-shadow effect, in which the shadow gradually reduces in size from a larger to a smaller radius over time. This arises from the non-symmetric effective potential for null geodesics. Such an effect does not occur in the symmetric case, where both peaks in the potential are of equal height, producing identical photon-sphere radii. A similar phenomenon has been previously reported in the literature \cite{Wielgus:2020uqz, Wang:2020emr}. Additionally, we derived constraints on the model parameters from observational data \cite{EventHorizonTelescope:2019ggy, EventHorizonTelescope:2022xqj}.

The obtained ringdown signals for metrics I, III, and IV confirm the echo effect discussed in earlier studies, a sequence of decaying bursts separated by time intervals, previously observed for the symmetric DS metric \cite{Bueno:2017hyj} and for other metrics containing a second potential peak or a wall \cite{Cardoso:2016oxy, Cardoso:2016rao, Hui:2019aox, Abedi:2016hgu}. No significant differences in this effect were observed compared to the symmetric case, unlike the optical properties. This can be explained by the fact that wave effects depend primarily on the presence of a potential barrier, whereas optical effects are sensitive to its height.

Asymmetry does, however, influence the amplitudes of the primary and echo signals: as shown in Fig.~\ref{Ringdown1}, the echoes can be either strongly suppressed or significantly enhanced, while the general echo structure remains unchanged. For metric II, the ringdown signal closely resembles that of the Schwarzschild black hole, consistent with the absence of a second potential peak.

We also computed the QNMs for all metrics considered. The presence of two potential peaks, or a peak together with a finite or infinite potential wall, leads to the formation of trapped, long-lived modes, resulting in a slowly decaying ringdown signal. The negative imaginary parts of these modes indicate the stability of scalar perturbations on these wormhole backgrounds. The fundamental QNM of metric II is close to that of the Schwarzschild black hole, as expected.

These results confirm the general conclusion that compact horizonless objects are difficult to distinguish observationally from black holes. When such objects are sufficiently compact, their observable properties closely resemble those of the Schwarzschild metric: their shadows are similar in size, and their ringdown signals consist of a primary burst, almost identical to that of a black hole, followed by a series of echoes (for metrics I, III, and IV) delayed by roughly twice the distance between the main potential peak (analogous to the black hole case) and the inner barrier. This distance depends logarithmically on the metric parameters, which are expected to be small, possibly arising from quantum corrections, resulting in long time delays between echoes.

Similarly, the shadow-decreasing effect observed for metrics with an asymmetric second peak or wall in the radial potential also begins with a comparable delay, since light must travel to the inner barrier and back to the observable region. Consequently, distinguishing a compact horizonless object (such as a wormhole) from a black hole requires long-time observations, either waiting for the appearance of echo signals or for the onset of shadow changes.

Metric II, however, shows a strong similarity to a black hole: it exhibits neither a second shadow nor echo bursts in its ringdown signal, yet it still describes an object with no horizon at any finite distance. This metric has a null surface at asymptotic infinity in the inner region. Whether this surface can be considered a degenerate horizon is a matter for further discussion. We are, however, reluctant to treat it as a horizon, since it remains at an infinite value of the affine parameter for any light geodesic. The spacetime is geodesically complete, meaning that no physical observer or light can reach this ``horizon''  within a finite proper time or affine parameter.

Altogether, these findings reveal the remarkable stability of classical black hole behavior under various deformations of the metric that are small in the exterior region but can be large in the inner region. They also highlight the need to identify additional observational signatures that could distinguish compact horizonless objects from true black holes.

\newpage

\section*{Appendix}
\appendix
\section{Travel time for metrics III and IV}
\label{sec:apptraveltime3and4}
The equation for the minimum of the functions $g(\rho)$ in Metrics III and IV has the form,
\be
\dfrac{g'_{\text{sch}}(\rho)}{g_{\text{sch}}(\rho)+b^2} = n\dfrac{\sqrt{\rho^2+\rho_0^2}-\rho}{\rho^2+\rho_0^2}\, .
\label{appeqn1}
\ee
It is convenient to use the compact coordinate \eqref{compactcoord}. In terms of this coordinate one has, 
\eqref{appeqn1} becomes
\be
\dfrac{x(1-x^2)^2}{2(x^2+b^2)} = n\dfrac{\sqrt{\rho^2(x)+\rho_0^2}-\rho(x)}{\rho^2(x)+\rho_0^2} \Rightarrow -
\dfrac{4}{\rho_0}b^2 + \left(2+\dfrac{16b^2}{\rho_0^2}\right)x = 0 \, ,
\label{appeqmin31}
\ee
where we used the approximation for small $x$ (since at $\rho_0=0$ the minimum is located at $x=0$). It allows us to obtain an approximate solution to equation  \eqref{appeqmin31}, 
\be
x_{\text{min}} \approx \dfrac{2b^2n\rho_0}{\rho_0^2 + 8b^2n} \Rightarrow \rho_{\text{min}} \approx \dfrac{4b^2n\rho_0}{\rho_0^2+8b^2n-4b^4n^2\rho_0^2} + \ln\dfrac{\rho_0^2+8b^2n+2b^2n\rho_0}{\rho_0^2+8b^2n-2b^2n\rho_0}\, ,
\ee
where we set $M=1$. The travel time \eqref{travel time}then can be estimated  as,
\be
\Delta t \sim \sqrt{\dfrac{8\rho_0^2}{\rho_0^2-8b^2n}}\ln\dfrac{\rho_0^2}{2b^2n\rho_0^2-8b^4n^2}\, .
\ee

\section{Numerical approach}
\label{sec:appNumApp}
The Chebyshev polynomial of the first kind of order $n$ is given by
\begin{equation}
    T_n(x)=\cos\left(n\arccos x\right),\quad x\in [-1,~1]\, .
\end{equation}
The polyniomials for  an orthogonal basis in the space $L^2([-1,~1],1/\sqrt{1-x^2}dx)$, so any function on the interval can be expanded in this basis. The expansion coefficients decrease exponentially with the order $n$, which makes Chebyshev polynomials well suited for approximations.

Throughout this work, for the discretization of equations and differential operators, we use the numerical Chebyshev collocation method, which is based on a special numerical grid constructed from the extrema of the Chebyshev polynomials $T_n(x)$.  Within the interval  $[-1,~1]$, a Chebyshev polynomial $T_n(x)$ has $n-1$ extrema, which together with the boundaries $x=\pm 1$ of the interval form a grid of $n+1$ points,
\begin{equation}
    x_i=\cos\left(\dfrac{\pi i}{n}\right),\quad i\in\{0,~1,~\dots,~n\}\, .
    \label{gridChebyshevLobatto}
\end{equation}
This grid is called the Chebyshev-Lobatto grid and it minimizes the Runge effect (oscillations near the edges of the interval when using an equidistant grid for polynomial interpolation). This grid can be stretched to an arbitrary interval $[a,~b]$ using an affine transformation,
\begin{equation}
    \Tilde{x}_i = \dfrac{1}{2}(b+a)+\dfrac{1}{2}(b-a)x_i\, .
\end{equation}
Any function $f(x)$ on this interval is then represented by a column vector $\left(f(\tilde{x}_0),~f(\tilde{x}_1),~\dots,~f(\tilde{x}_n)\right)^{T}$. 
Since we work with derivatives, the numerical differentiation is required, 
\begin{equation}
    f'(\tilde{x}_i)=\sum_{j=0}^n \mathbf{D}^{(1)}_{ij}f(\tilde{x}_j)\, ,
\end{equation}
where $\mathbf{D}^{(1)}_{ij}$ is matrix which represents the first derivative,
\begin{equation}
    \mathbf{D}^{(1)}_{ij}=\dfrac{2}{b-a}\begin{cases}
        \dfrac{2n^2+1}{6},\quad\quad\quad i=j=0\\
        -\dfrac{2n^2+1}{6},\quad\quad ~i=j=n\\
        -\dfrac{x_j}{2\left(1-x_j^2\right)},\quad 0<i=j<n\\
        \dfrac{\alpha_i}{\alpha_j}\dfrac{(-1)^{i+j}}{x_{i}-x_{j}},\quad\quad i\neq j
    \end{cases},\quad \alpha_{i}=\begin{cases}
        2,\quad i\in\{0,~n\}\\
        1,\quad i\in\{1,~\dots,~n-1\}
    \end{cases}
    \label{derivativematrix}
\end{equation}
The second derivative matrix can be constructed by matrix multiplication $\mathbf{D}^{(2)}=\mathbf{D}^{(1)}\cdot\mathbf{D}^{(1)}$. As seen from the definition \eqref{derivativematrix}, these matrices contain non-zero elements in all rows and columns, unlike the almost diagonal derivative matrix in the finite difference method. Thus, the derivatives and the scheme itself incorporate information from the entire interval rather than only from neighboring nodes, which improves the numerical method.

\section{Coefficients in the hyperboloidal approach}
\label{sec:appCoeffHypApp}
Here we list the functions \eqref{pfunction}  which are used as coefficients in the differential equation \eqref{hyperdiffeq} for the three types of metrics (I, II and IV) to which the hyperboloidal method is applied. In what follows we use \eqref{usefull3.2}.
\begin{itemize}
    \item \textbf{Metric I} 
    \begin{gather}
    p_{\tau\tau}(\sigma)=4-\left(b\sinh\sigma - a\right)^2,\quad p_{\tau\sigma}(\sigma)=\dfrac{1}{2}\left(b\sinh\sigma - a\right)\left(-3+\left(b\sinh\sigma - a\right)^2\right)\nonumber\\
    \nonumber\\
    p_{\tau}(\sigma)=\dfrac{3}{2}\left(-1+\left(b\sinh\sigma - a\right)^2\right)b\cosh\sigma,\quad p_{\sigma\sigma}(\sigma)=\dfrac{\left(1-\left(b\sinh\sigma - a\right)^2\right)^2}{4}\\
    \nonumber\\
    p_{\sigma}(\sigma)=\left(b\sinh\sigma - a\right)\left(-1+\left(b\sinh\sigma - a\right)^2\right)b\cosh\sigma\nonumber
    \label{pfunctionmetric1}
    \end{gather}
    
    \item \textbf{Metric II}
    \begin{gather}
    p^{\text{in-out}}_{\tau\tau}(\sigma) = \dfrac{2-b\sinh\sigma}{\left(1-b\sinh\sigma\right)^2}\left[4-(2-b\sinh\sigma)(1+b\sinh\sigma)^2\sqrt{\Delta\left(\rho(b\sinh\sigma)\right)}\right]\nonumber\\
    \nonumber\\
    p^{\text{in-out}}_{\tau x}(\sigma) = 1 - \dfrac{1}{2}(2-b\sinh\sigma)(1+b\sinh\sigma)^2\sqrt{\Delta\left(\rho(b\sinh\sigma)\right)}\nonumber\\
    \nonumber\\
    p^{\text{in-out}}_{xx}(\sigma) = \dfrac{\left(1-b^2\sinh^2\sigma\right)^2}{4}\sqrt{\Delta\left(\rho(b\sinh\sigma)\right)}\\
    \nonumber\\
    p^{\text{in-out}}_{\tau}(\sigma) = -b\cosh\sigma\sqrt{\Delta\left(\rho(b\sinh\sigma)\right)}\left[\dfrac{3}{2}\left(1-b^2\sinh^2\sigma\right)^2+\dfrac{2-b\sinh\sigma}{\left(1-b\sinh\sigma\right)^2}\cdot\right.\nonumber\\
    \nonumber\\
    \left.\cdot\Delta^{-1}\left(\rho(b\sinh\sigma)\right)\dfrac{d}{d\rho}\Delta\left(\rho(b\sinh\sigma)\right)\right]\nonumber\\
    \nonumber\\
    p^{\text{in-out}}_{x}(\sigma) = b\cosh\sigma\sqrt{\Delta\left(\rho(b\sinh\sigma)\right)}\left[-\left(1-b^2\sinh^2\sigma\right)b\sinh\sigma+\dfrac{1}{2}\Delta^{-1}\left(\rho(b\sinh\sigma)\right)\dfrac{d}{d\rho}\Delta\left(\rho(b\sinh\sigma)\right)\right]\nonumber
    \label{pfunctionmetric2}
    % \dfrac{\sqrt{\rho^2(b\sinh\sigma)+\rho_0^2}-\rho(b\sinh\sigma)}{\rho^2(b\sinh\sigma)+\rho^2_0}
    \end{gather}
    \item \textbf{Metric IV} 
    \begin{gather} 
    p_{\tau\tau}(\sigma)=\Delta^{n/2}\left(\rho(b\sinh\sigma)\right)\dfrac{12+12~b\sinh\sigma+3~b^2\sinh^2\sigma-2~b^3\sinh^3\sigma-b^4\sinh^4\sigma}{4(1+b\sinh\sigma)^2}\nonumber\\
    \nonumber\\
    p_{\tau \sigma}(\sigma) = \dfrac{1}{4}\left(1+b\sinh\sigma\right)^2\left(2-b\sinh\sigma\right)\nonumber\\
    \nonumber\\
    p_{\sigma\sigma}(\sigma) = \dfrac{\left(1-b^2\sinh^2\sigma\right)^2}{4}\Delta^{-n/2}\left(\rho(b\sinh\sigma)\right),\quad p_{\tau}(\sigma) = \dfrac{3}{4}b\cosh\sigma\left(1-b^2\sinh^2\sigma\right)\\
    \nonumber\\
    p_{\sigma}(\sigma) = -b\cosh\sigma\Delta^{-n/2}\left(\rho(b\sinh\sigma)\right)\left[b\sinh\sigma\left(1-b^2\sinh^2\sigma\right)+\dfrac{n}{2}\Delta^{-1}\left(\rho(b\sinh\sigma)\right)\dfrac{d}{d\rho}\Delta\left(\rho(b\sinh\sigma)\right)\right]\nonumber\\
    \nonumber\\
    \hat{V}(\sigma) = \Delta^{-n/2}\left(\rho(b\sinh\sigma)\right)\left[l(l+1)b^2\cosh^2\sigma+\dfrac{1}{2}\left(1-b^2\sinh^2\sigma\right)\left(2b^2\sinh^2\sigma+b^2\right) -\right.\nonumber\\
    \nonumber\\
    \left.-\dfrac{nb^2\cosh^2\sigma b\sinh\sigma}{1-b^2\sinh^2\sigma}\Delta^{-1}\left(\rho(b\sinh\sigma)\right)\dfrac{d}{d\rho}\Delta\left(\rho(b\sinh\sigma)\right)\right]\nonumber
    \label{pfunctionmetric4}
    \end{gather}

    Rescaling the function $\Psi(t,~\sigma) =  \Delta^{n/4}\left(\rho(b\sinh\sigma)\right)\eta(t,~\sigma)$ and multiplying the equation by $\Delta^{n/4}\left(\rho(b\sinh\sigma)\right)$ yields new coefficients
    \begin{gather}
    \bar{p}_{\tau\tau}(\sigma) = \Delta\left(\rho(b\sinh\sigma)\right)\dfrac{12+12~b\sinh\sigma+3~b^2\sinh^2\sigma-2~b^3\sinh^3\sigma-b^4\sinh^4\sigma}{4(1+b\sinh\sigma)^2}\nonumber\\
    \nonumber\\
    \bar{p}_{\sigma\sigma}(\sigma)=\dfrac{\left(1-b^2\sinh^2\sigma\right)^2}{4},\quad \bar{p}_{\sigma}(\sigma) =-b^2\cosh\sigma\sinh\sigma\left(1-b^2\sinh^2\sigma\right)\nonumber\\
    \\
    \bar{p}_{\tau\sigma}(\sigma)=\dfrac{1}{4}\left(1+b\sinh\sigma\right)^2\left(2-b\sinh\sigma\right)\Delta^{n/2}\left(\rho(b\sinh\sigma)\right)\nonumber\\
    \nonumber\\
    \bar{p}_{\tau}(\sigma)=b\cosh\sigma\Delta^{n/2}\left(\rho(b\sinh\sigma)\right)\left[\dfrac{3}{4}\left(1-b^2\sinh^2\sigma\right)+\dfrac{n}{2}\dfrac{2-b\sinh\sigma}{\left(1-b\sinh\sigma\right)^2}\Delta^{-1}\left(\rho(b\sinh\sigma)\right)\dfrac{d}{d\rho}\Delta\left(\rho(b\sinh\sigma)\right)\right]\nonumber
    \end{gather}
    \begin{gather}
    \hat{\bar{V}}(\sigma) = l(l+1)b^2\cosh^2\sigma+\dfrac{1}{2}\left(1-b^2\sinh^2\sigma\right)\left(2b^2\sinh^2\sigma+b^2\right) - \nonumber\\
    % \dfrac{n}{4}b\sinh\sigma\Delta^{-1}\left(\rho(b\sinh\sigma)\right)\dfrac{d}{d\rho}\Delta\left(\rho(b\sinh\sigma)\right)-\nonumber\\
    \nonumber\\
    -\left(\dfrac{n}{4}b\sinh\sigma+\dfrac{nb^2\cosh^2\sigma b\sinh\sigma}{1-b^2\sinh^2\sigma}\right)\Delta^{-1}\left(\rho(b\sinh\sigma)\right)\dfrac{d}{d\rho}\Delta\left(\rho(b\sinh\sigma)\right) - \dfrac{4b^2\cosh^2\sigma}{\left(1-b^2\sinh^2\sigma\right)^2}\cdot\nonumber\\
    \nonumber\\
    \cdot\left[\dfrac{n}{4}\dfrac{d}{d\rho}\left(\Delta^{-1}\left(\rho(b\sinh\sigma)\right)\dfrac{d}{d\rho}\Delta\left(\rho(b\sinh\sigma)\right)\right)-\dfrac{n^2}{16}\left(\Delta^{-1}\left(\rho(b\sinh\sigma)\right)\dfrac{d}{d\rho}\Delta\left(\rho(b\sinh\sigma)\right)\right)^2\right]\nonumber
    \label{pfunctionmetric41}
    \end{gather}
\end{itemize}

\section{Coefficients for metric III in matrix method}
\label{sec:appCoeffMatrixMethod}
Since metric III is treated with a different approach -  the matrix method - in this section we list the coefficients of the original equation \eqref{eqnmetric3compcoord} written using a compact coordinate. According to the method described in Section \ref{sec:matrixmethod}, these coefficients will be used to construct new coefficients for the equation \eqref{finaleqnmetric3} obtained by substituting the ansatz \eqref{scalingmetric3} and performing further the transformation $v(\sigma)=u(\sigma)/(1-b^2\sinh^2\sigma)$. In this work, we do not explicitly write out the resulting coefficients, as they are rather cumbersome and not very informative. We find that (we use \eqref{usefull3.2}), 
\begin{gather}
    p(\sigma) = -\dfrac{4b\cosh\sigma}{\left(1-b^2\sinh^2\sigma\right)^2}\left[\left(1-b^2\sinh^2\sigma\right)b\sinh\sigma+\dfrac{1}{2}\Delta^{-1}\left(\rho(b\sinh\sigma)\right)\dfrac{d}{d\rho}\Delta\left(\rho(b\sinh\sigma)\right)\right]\nonumber\\
    \nonumber\\
    q(\sigma) = \dfrac{16\omega^2}{\left(1-b^2\sinh^2\sigma\right)^4}\Delta\left(\rho(\sigma)\right) - \dfrac{4}{\left(1-b^2\sinh^2\sigma\right)^2}\left[l(l+1)b^2\cosh^2\sigma\dfrac{1}{2}\left(1-b^2\sinh^2\sigma\right)\cdot\right.\nonumber\\
    \\
    \left.\cdot\left(2b^2\sinh^2\sigma+b^2\right)-\dfrac{b^2\cosh^2\sigma b\sinh^2\sigma}{1-b^2\sinh^2\sigma}\Delta^{-1}\left(\rho(b\sinh\sigma)\right)\dfrac{d}{d\rho}\Delta\left(\rho(b\sinh\sigma)\right)\right]\nonumber
    \label{pandqmetric3}
\end{gather}

\section{Convergence tests}
\label{Convergence tests}
\begin{figure}[h!]
    \begin{minipage}[h!]{0.49\linewidth}
    \center{\includegraphics[width=6cm,height=4cm]{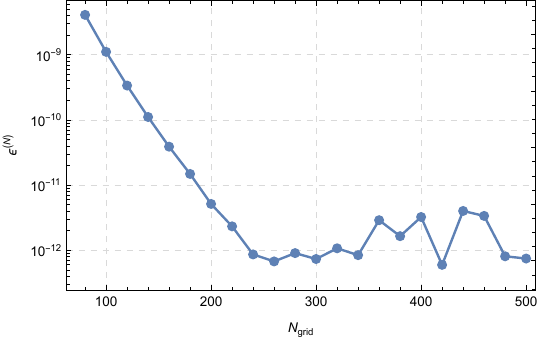}}\\
    \end{minipage}
    \hfill
    \begin{minipage}[h!]{0.49\linewidth}
    \center{\includegraphics[width=6cm,height=4cm]{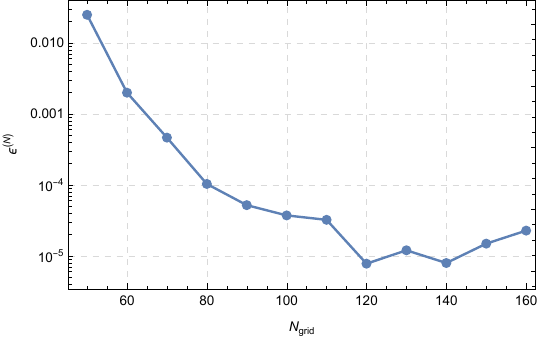}}\\
    \end{minipage}
    \begin{minipage}[h!]{0.49\linewidth}
    \center{\includegraphics[width=6cm,height=4cm]{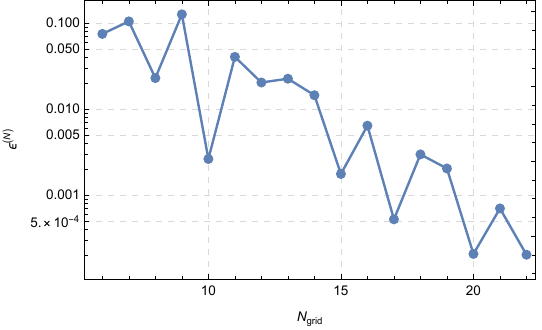}}\\
    \end{minipage}
    \begin{minipage}[h!]{0.49\linewidth}
    \center{\includegraphics[width=6cm,height=4cm]{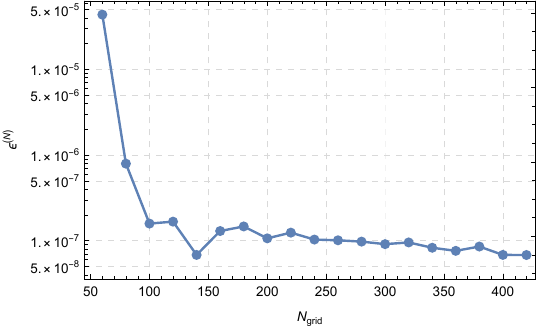}}\\
    \end{minipage}
    \caption{\footnotesize The plots show $\epsilon^{(N)}$ as a function of a grid size. \textbf{The left top plot} shows the convergence test for the metric I with the parameters $b=10^{-5}$ and $a=0.03$. \textbf{The right top plot} shows the convergence test for the metric II with the parameters $b=10^{-5}$ and $\rho_0=10^{-5}$. \textbf{The left bottom plot} shows the convergence test for the metric III with the parameters $b=10^{-5}$ and $\rho_0=0.1$. \textbf{The right bottom plot} shows the convergence test for the metric IV with the parameters $b=10^{-5}$ and $\rho_0=10^{-4}$.}
    \label{ConvergTest}
    \end{figure}

    To confirm the validity of our numerical computation of the  QNMs, we consider the following ratio as a convergence test,
    \begin{equation}
        \epsilon^{(N)} = \dfrac{|\omega_0^{(N)}-\omega_0^{(N-1)}|}{|\omega_0^{(N-1)}|}~,
    \end{equation}
    where $\omega_0^{(N)}$ is the fundamental QNM computed using a grid of size $N$. Fig.~\ref{ConvergTest} shows the convergence test for the metrics which we consider in the paper.

\section{Double-delta potential}
\label{app:doubledelta}
In this section we consider a simple but very instructive exactly solvable problem for finding QNMs: the double-delta potential. We study the Schrödinger equation $-\partial_x^2\psi(x)+V_{\text{DD}}(x)\psi(x)=\omega^2\psi(x)$ with a potential
\begin{equation}
    V_{\text{DD}}(x)=V_1\delta(x-a)+V_2\delta(x+a)\, ,
\end{equation}
where $V_1$ and $V_2$ are positive  and are not required to be equal. Since in the region between the delta peaks the Schrödinger equation becomes free, the solution can be written in a simple form,
\begin{equation}
    \psi(x)=\begin{cases}
        \psi_{\text{I}}(x) = A_{\text{in}}e^{-i\omega x}+A_{\text{out}}e^{i\omega x},\quad a\leq x\\
        \psi_{\text{II}}(x) = B_{\text{in}}e^{-i\omega x}+B_{\text{out}}e^{i\omega x},\quad -a\leq x< a\\
        \psi_{\text{III}}(x) = e^{-i\omega x},\quad x <- a\\
    \end{cases}
\end{equation}
The coefficients $A_{\text{in}}$, $A_{\text{out}}$, $B_{\text{in}}$ and $B_{\text{in}}$ must be determined. The presence of delta peaks requires the derivatives of the wave function $\psi(x)$ to have discontinuities at the peak locations. This leads to a system of four equations that are solved to determine the coefficients,
\begin{equation}
    \begin{cases}
        \psi_{\text{I}}(a)=\psi_{\text{II}}(a)\\
        -\psi'_{\text{I}}(a)+\psi'_{\text{II}}(a)+V_2\psi_{\text{I}}(a)=0\\
        \psi_{\text{II}}(-a)=\psi_{\text{III}}(-a)\\
        -\psi'_{\text{II}}(-a)+\psi'_{\text{III}}(-a)+V_1\psi_{\text{II}}(-a)=0
    \end{cases}
    \label{qnmdeltaeqn}
\end{equation}
Because QNMs are solutions satisfying the specific boundary conditions $\psi(x\rightarrow\pm\infty)\rightarrow e^{\pm i\omega x}$, the condition $A_{\text{out}}=0$ gives the equation for finding the QNMs,
\begin{equation}
    A_{\text{out}} = \left(1-\dfrac{V_1}{2i\omega}\right)\left(1-\dfrac{V_2}{2i\omega}\right)+\dfrac{V_1V_2}{4\omega^2}e^{4i\omega a}=0\, .
\end{equation}

If the potential contains only one delta peak (say $V_1=0$ and $V_2=V$), then there is only a single QNM $\omega_{\text{QNM}}=-iV/2$, which is purely imaginary. In the case when both potentials are present, one obtains a transcendental equation \eqref{qnmdeltaeqn} for the modes. Assuming large separation $L=2a$ and using an ansatz $\omega_{n} = \sum_{k=0}^{\infty}\omega^{(k)}/L^{k+1}$, one finds 
\begin{equation}
    \omega_n = \dfrac{\pi n}{L}\left(1-\dfrac{1}{L}\left(\dfrac{1}{V_1}+\dfrac{1}{V_2}\right)+\dfrac{1}{L^2}\left(\dfrac{1}{V_1}+\dfrac{1}{V_2}\right)^2+\dots\right)-i\dfrac{\pi^2n^2}{L^3}\left(\dfrac{1}{V_1^2}+\dfrac{1}{V_2^2}\right) +\dots\, .
\end{equation}
Thus, the imaginary part appears only at third order in $L$, consistent with the behavior of QNMs in cavities. At first order, the mode reproduces the normal modes of a box $\pi n/L$. Further examples of systems with cavities can be found in the literature \cite{Bueno:2017hyj}, \cite{Boonserm:2010px}.


\newpage
\begin{thebibliography}{999}

%\cite{Schwarzschild:1916uq}
\bibitem{Schwarzschild:1916uq}
K.~Schwarzschild,
%``On the gravitational field of a mass point according to Einstein's theory,''
Sitzungsber. Preuss. Akad. Wiss. Berlin (Math. Phys. ) \textbf{1916} (1916), 189-196
\href{https://arxiv.org/abs/physics/9905030}{[arXiv:physics/9905030 [physics]]}.

%\cite{Narayan:2005ie}
\bibitem{Narayan:2005ie}
R.~Narayan,
%``Black holes in astrophysics,''
New J. Phys. \textbf{7} (2005), 199
%doi:10.1088/1367-2630/7/1/199
\href{https://arxiv.org/abs/gr-qc/0506078}{[arXiv:gr-qc/0506078 [gr-qc]]}.

%\cite{LIGOScientific:2016sjg}
\bibitem{LIGOScientific:2016sjg}
B.~P.~Abbott \textit{et al.} [LIGO Scientific and Virgo],
%``GW151226: Observation of Gravitational Waves from a 22-Solar-Mass Binary Black Hole Coalescence,''
Phys. Rev. Lett. \textbf{116} (2016) no.24, 241103
%doi:10.1103/PhysRevLett.116.241103
\href{https://arxiv.org/abs/1606.04855}{[arXiv:1606.04855 [gr-qc]]}.

%\cite{LIGOScientific:2016aoc}
\bibitem{LIGOScientific:2016aoc}
B.~P.~Abbott \textit{et al.} [LIGO Scientific and Virgo],
%``Observation of Gravitational Waves from a Binary Black Hole Merger,''
Phys. Rev. Lett. \textbf{116} (2016) no.6, 061102
%doi:10.1103/PhysRevLett.116.061102
\href{https://arxiv.org/abs/1602.03837}{[arXiv:1602.03837 [gr-qc]]}.
%13330 citations counted in INSPIRE as of 29 Jul 2025

%\cite{Blanchet:2013haa}
\bibitem{Blanchet:2013haa}
L.~Blanchet,
%``Post-Newtonian Theory for Gravitational Waves,''
Living Rev. Rel. \textbf{17} (2014), 2
%doi:10.12942/lrr-2014-2
\href{https://arxiv.org/abs/1310.1528}{[arXiv:1310.1528 [gr-qc]]}.

%\cite{Buonanno:1998gg}
\bibitem{Buonanno:1998gg}
A.~Buonanno and T.~Damour,
%``Effective one-body approach to general relativistic two-body dynamics,''
Phys. Rev. D \textbf{59} (1999), 084006
% doi:10.1103/PhysRevD.59.084006
\href{https://arxiv.org/abs/gr-qc/9811091}{[arXiv:gr-qc/9811091 [gr-qc]]}.

%\cite{Blanchet:2009sd}
\bibitem{Blanchet:2009sd}
L.~Blanchet, S.~L.~Detweiler, A.~Le Tiec and B.~F.~Whiting,
%``Post-Newtonian and Numerical Calculations of the Gravitational Self-Force for Circular Orbits in the Schwarzschild Geometry,''
Phys. Rev. D \textbf{81} (2010), 064004
%doi:10.1103/PhysRevD.81.064004
\href{https://arxiv.org/abs/0910.0207}{[arXiv:0910.0207 [gr-qc]]}.
%\cite{Campanelli:2005dd}

\bibitem{Campanelli:2005dd}
M.~Campanelli, C.~O.~Lousto, P.~Marronetti and Y.~Zlochower,
%``Accurate evolutions of orbiting black-hole binaries without excision,''
Phys. Rev. Lett. \textbf{96} (2006), 111101
%doi:10.1103/PhysRevLett.96.111101
\href{https://arxiv.org/abs/gr-qc/0511048}{[arXiv:gr-qc/0511048 [gr-qc]]}.

%\cite{Vishveshwara:1970zz}
\bibitem{Vishveshwara:1970zz}
C.~V.~Vishveshwara,
%``Scattering of Gravitational Radiation by a Schwarzschild Black-hole,''
Nature \textbf{227} (1970), 936-938
% \href{https://www.nature.com/articles/227936a0}{Nature \textbf{227} (1970), 936-938}
%doi:10.1038/227936a0

%\cite{Nollert:1999ji}
\bibitem{Nollert:1999ji}
H.~P.~Nollert,
%``TOPICAL REVIEW: Quasinormal modes: the characteristic `sound' of black holes and neutron stars,''
Class. Quant. Grav. \textbf{16} (1999), R159-R216
%doi:10.1088/0264-9381/16/12/201

%\cite{Dreyer:2003bv}
\bibitem{Dreyer:2003bv}
O.~Dreyer, B.~J.~Kelly, B.~Krishnan, L.~S.~Finn, D.~Garrison and R.~Lopez-Aleman,
%``Black hole spectroscopy: Testing general relativity through gravitational wave observations,''
Class. Quant. Grav. \textbf{21} (2004), 787-804
%doi:10.1088/0264-9381/21/4/003
\href{https://arxiv.org/abs/gr-qc/0309007}{[arXiv:gr-qc/0309007 [gr-qc]]}.

%\cite{Berti:2009kk}
\bibitem{Berti:2009kk}
E.~Berti, V.~Cardoso and A.~O.~Starinets,
%``Quasinormal modes of black holes and black branes,''
Class. Quant. Grav. \textbf{26} (2009), 163001
%doi:10.1088/0264-9381/26/16/163001
\href{https://arxiv.org/abs/0905.2975}{[arXiv:0905.2975 [gr-qc]]}.
%2097 citations counted in INSPIRE as of 29 Jul 2025

%\cite{Konoplya:2011qq}
\bibitem{Konoplya:2011qq}
R.~A.~Konoplya and A.~Zhidenko,
%``Quasinormal modes of black holes: From astrophysics to string theory,''
Rev. Mod. Phys. \textbf{83} (2011), 793-836
% doi:10.1103/RevModPhys.83.793
\href{https://arxiv.org/abs/1102.4014}{[arXiv:1102.4014 [gr-qc]]}.

%\cite{Berti:2015itd} 
\bibitem{Berti:2015itd} E.~Berti, E.~Barausse, V.~Cardoso, L.~Gualtieri, P.~Pani, U.~Sperhake, L.~C.~Stein, N.~Wex, K.~Yagi and T.~Baker, \textit{et al.}
%``Testing General Relativity with Present and Future Astrophysical Observations,'' 
Class. Quant. Grav. \textbf{32} (2015), 243001 
%doi:10.1088/0264-9381/32/24/243001 
\href{https://arxiv.org/abs/1501.07274}{[arXiv:1501.07274 [gr-qc]]}.

%\cite{Konoplya:2016pmh}
\bibitem{Konoplya:2016pmh}
R.~Konoplya and A.~Zhidenko,
%``Detection of gravitational waves from black holes: Is there a window for alternative theories?,''
Phys. Lett. B \textbf{756} (2016), 350-353
% doi:10.1016/j.physletb.2016.03.044
\href{https://arxiv.org/pdf/1602.04738}{[arXiv:1602.04738 [gr-qc]]}.

%\cite{Berti:2005ys}
\bibitem{Berti:2005ys}
E.~Berti, V.~Cardoso and C.~M.~Will,
%``On gravitational-wave spectroscopy of massive black holes with the space interferometer LISA,''
Phys. Rev. D \textbf{73} (2006), 064030
%doi:10.1103/PhysRevD.73.064030
\href{https://arxiv.org/abs/gr-qc/0512160}{[arXiv:gr-qc/0512160 [gr-qc]]}.

%\cite{EventHorizonTelescope:2019ggy}
\bibitem{EventHorizonTelescope:2019ggy}
K.~Akiyama \textit{et al.} [Event Horizon Telescope],
%``First M87 Event Horizon Telescope Results. VI. The Shadow and Mass of the Central Black Hole,''
Astrophys. J. Lett. \textbf{875}, no.1, L6 (2019)
%doi:10.3847/2041-8213/ab1141
\href{https://arxiv.org/abs/1906.11243}{[arXiv:1906.11243 [astro-ph.GA]]}.
%1251 citations counted in INSPIRE as of 26 Aug 2024

%\cite{EventHorizonTelescope:2022xqj}
\bibitem{EventHorizonTelescope:2022xqj}
K.~Akiyama \textit{et al.} [Event Horizon Telescope],
%``First Sagittarius A* Event Horizon Telescope Results. VI. Testing the Black Hole Metric,''
Astrophys. J. Lett. \textbf{930}, no.2, L17 (2022)
%doi:10.3847/2041-8213/ac6756
\href{https://arxiv.org/abs/2311.09484}{[arXiv:2311.09484 [astro-ph.HE]]}.

%\cite{Cardoso:2008bp}
\bibitem{Cardoso:2008bp}
V.~Cardoso, A.~S.~Miranda, E.~Berti, H.~Witek and V.~T.~Zanchin,
%``Geodesic stability, Lyapunov exponents and quasinormal modes,''
Phys. Rev. D \textbf{79} (2009) no.6, 064016
% doi:10.1103/PhysRevD.79.064016
\href{https://arxiv.org/abs/0812.1806}{[arXiv:0812.1806 [hep-th]]}.

%\cite{Konoplya:2017wot}
\bibitem{Konoplya:2017wot}
R.~A.~Konoplya and Z.~Stuchl{\'\i}k,
%``Are eikonal quasinormal modes linked to the unstable circular null geodesics?,''
Phys. Lett. B \textbf{771} (2017), 597-602
% doi:10.1016/j.physletb.2017.06.015
\href{https://arxiv.org/abs/1705.05928}{[arXiv:1705.05928 [gr-qc]]}.

%\cite{Abramowicz:2002vt}
\bibitem{Abramowicz:2002vt}
M.~A.~Abramowicz, W.~Kluzniak and J.~P.~Lasota,
%``No observational proof of the black hole event-horizon,''
Astron. Astrophys. \textbf{396} (2002), L31-L34
%doi:10.1051/0004-6361:20021645
\href{https://arxiv.org/abs/astro-ph/0207270}{[arXiv:astro-ph/0207270 [astro-ph]]}.

%\cite{Mazur:2001fv}
\bibitem{Mazur:2001fv}
P.~O.~Mazur and E.~Mottola,
%``Gravitational Condensate Stars: An Alternative to Black Holes,''
Universe \textbf{9} (2023) no.2, 88
%doi:10.3390/universe9020088
\href{https://arxiv.org/abs/gr-qc/0109035}{[arXiv:gr-qc/0109035 [gr-qc]]}.

%\cite{Schunck:2003kk}
\bibitem{Schunck:2003kk}
F.~E.~Schunck and E.~W.~Mielke,
%``General relativistic boson stars,''
Class. Quant. Grav. \textbf{20} (2003), R301-R356
%doi:10.1088/0264-9381/20/20/201
\href{https://arxiv.org/abs/0801.0307}{[arXiv:0801.0307 [astro-ph]]}.

%\cite{Herdeiro:2021lwl}
\bibitem{Herdeiro:2021lwl}
C.~A.~R.~Herdeiro, A.~M.~Pombo, E.~Radu, P.~Cunha, V.P. and N.~Sanchis-Gual,
%``The imitation game: Proca stars that can mimic the Schwarzschild shadow,''
JCAP \textbf{04} (2021), 051
% doi:10.1088/1475-7516/2021/04/051
\href{https://arxiv.org/abs/2102.01703}{[arXiv:2102.01703 [gr-qc]]}.

%\cite{Abramowicz:1997qk}
\bibitem{Abramowicz:1997qk}
M.~A.~Abramowicz, M.~Bruni, S.~Sonego, N.~Andersson and P.~Ghosh,
%``Gravitational waves from ultracompact stars: The optical geometry view of trapped modes,''
Class. Quant. Grav. \textbf{14} (1997), L189-L194
%doi:10.1088/0264-9381/14/12/002
%29 citations counted in INSPIRE as of 04 Aug 2025

%\cite{Morris:1988tu}
\bibitem{Morris:1988tu}
M.~S.~Morris, K.~S.~Thorne and U.~Yurtsever,
%``Wormholes, Time Machines, and the Weak Energy Condition,''
Phys. Rev. Lett. \textbf{61} (1988), 1446-1449
%doi:10.1103/PhysRevLett.61.1446

%\cite{Lemos:2008cv}
\bibitem{Lemos:2008cv}
J.~P.~S.~Lemos and O.~B.~Zaslavskii,
%``Black hole mimickers: Regular versus singular behavior,''
Phys. Rev. D \textbf{78} (2008), 024040
% doi:10.1103/PhysRevD.78.024040
\href{https://arxiv.org/abs/0806.0845}{[arXiv:0806.0845 [gr-qc]]}.

%\cite{Damour:2007ap}
\bibitem{Damour:2007ap}
T.~Damour and S.~N.~Solodukhin,
%``Wormholes as black hole foils,''
Phys. Rev. D \textbf{76} (2007), 024016
%doi:10.1103/PhysRevD.76.024016
\href{https://arxiv.org/abs/0704.2667}{[arXiv:0704.2667 [gr-qc]]}.

%\cite{Mathur:2005zp}
\bibitem{Mathur:2005zp}
S.~D.~Mathur,
%``The Fuzzball proposal for black holes: An Elementary review,''
Fortsch. Phys. \textbf{53} (2005), 793-827
%doi:10.1002/prop.200410203
\href{https://arxiv.org/abs/hep-th/0502050}{[arXiv:hep-th/0502050 [hep-th]]}.

%\cite{Cardoso:2019rvt} 
\bibitem{Cardoso:2019rvt} V.~Cardoso and P.~Pani, 
%``Testing the nature of dark compact objects: a status report,'' 
Living Rev. Rel. \textbf{22} (2019) no.1, 4 
%doi:10.1007/s41114-019-0020-4 
\href{https://arxiv.org/abs/1904.05363}{[arXiv:1904.05363 [gr-qc]]}.

%\cite{Cardoso:2014sna}
\bibitem{Cardoso:2014sna}
V.~Cardoso, L.~C.~B.~Crispino, C.~F.~B.~Macedo, H.~Okawa and P.~Pani,
%``Light rings as observational evidence for event horizons: long-lived modes, ergoregions and nonlinear instabilities of ultracompact objects,''
Phys. Rev. D \textbf{90} (2014) no.4, 044069
%doi:10.1103/PhysRevD.90.044069
\href{https://arxiv.org/abs/1406.5510}{[arXiv:1406.5510 [gr-qc]]}.

%\cite{Maggio:2017ivp}
\bibitem{Maggio:2017ivp}
E.~Maggio, P.~Pani and V.~Ferrari,
%``Exotic Compact Objects and How to Quench their Ergoregion Instability,''
Phys. Rev. D \textbf{96} (2017) no.10, 104047
%doi:10.1103/PhysRevD.96.104047
\href{https://arxiv.org/abs/1703.03696}{[arXiv:1703.03696 [gr-qc]]}.

%\cite{Pani:2008bzt}
\bibitem{Pani:2008bzt}
P.~Pani, V.~Cardoso, M.~Cadoni and M.~Cavaglia,
%``Ergoregion instability of black hole mimickers,''
PoS \textbf{BHGRS} (2008), 027
%doi:10.22323/1.075.0027
\href{https://arxiv.org/abs/0901.0850}{[arXiv:0901.0850 [gr-qc]]}.

%\cite{Claudel:2000yi}
\bibitem{Claudel:2000yi}
C.~M.~Claudel, K.~S.~Virbhadra and G.~F.~R.~Ellis,
%``The Geometry of photon surfaces,''
J. Math. Phys. \textbf{42} (2001), 818-838
%doi:10.1063/1.1308507
\href{https://arxiv.org/abs/gr-qc/0005050}{[arXiv:gr-qc/0005050 [gr-qc]]}.

%\cite{Teo:2020sey}
\bibitem{Teo:2020sey}
E.~Teo,
%``Spherical orbits around a Kerr black hole,''
Gen. Rel. Grav. \textbf{53} (2021) no.1, 10
%doi:10.1007/s10714-020-02782-z
\href{https://arxiv.org/abs/2007.04022}{[arXiv:2007.04022 [gr-qc]]}.

%\cite{Perlick:2021aok}
\bibitem{Perlick:2021aok}
V.~Perlick and O.~Y.~Tsupko,
%``Calculating black hole shadows: Review of analytical studies,''
Phys. Rept. \textbf{947} (2022), 1-39
%doi:10.1016/j.physrep.2021.10.004
\href{https://arxiv.org/abs/2105.07101}{[arXiv:2105.07101 [gr-qc]]}.

%\cite{Vagnozzi:2022moj}
\bibitem{Vagnozzi:2022moj}
S.~Vagnozzi, R.~Roy, Y.~D.~Tsai, L.~Visinelli, M.~Afrin, A.~Allahyari, P.~Bambhaniya, D.~Dey, S.~G.~Ghosh and P.~S.~Joshi, \textit{et al.}
%``Horizon-scale tests of gravity theories and fundamental physics from the Event Horizon Telescope image of Sagittarius A,''
Class. Quant. Grav. \textbf{40} (2023) no.16, 165007
%doi:10.1088/1361-6382/acd97b
\href{https://arxiv.org/abs/2205.07787}{[arXiv:2205.07787 [gr-qc]]}.

%\cite{Tsukamoto:2024pid}
\bibitem{Tsukamoto:2024pid}
N.~Tsukamoto,
%``Circular light orbits of a general, static, and spherical symmetrical wormhole with $Z_2$ symmetry,''
Eur. Phys. J. C \textbf{84} (2024) no.12, 1325
%doi:10.1140/epjc/s10052-024-13696-4
\href{https://arxiv.org/abs/2401.07846}{[arXiv:2401.07846 [gr-qc]]}.

%\cite{Shaikh:2018kfv}
\bibitem{Shaikh:2018kfv}
R.~Shaikh,
%``Shadows of rotating wormholes,''
Phys. Rev. D \textbf{98} (2018) no.2, 024044
%doi:10.1103/PhysRevD.98.024044
\href{https://arxiv.org/abs/1803.11422}{[arXiv:1803.11422 [gr-qc]]}.

%\cite{Wang:2020jek}
\bibitem{Wang:2020jek}
J.~Wang,
%``Multiple rings in the shadow of extremely compact objects,''
Int. J. Mod. Phys. D \textbf{30} (2021) no.15, 2150112
%doi:10.1142/S0218271821501121
\href{https://arxiv.org/abs/2012.10237}{[arXiv:2012.10237 [gr-qc]]}.

%\cite{Rosa:2022tfv}
\bibitem{Rosa:2022tfv}
J.~L.~Rosa and D.~Rubiera-Garcia,
%``Shadows of boson and Proca stars with thin accretion disks,''
Phys. Rev. D \textbf{106} (2022) no.8, 084004
d%oi:10.1103/PhysRevD.106.084004
\href{https://arxiv.org/abs/2204.12949}{[arXiv:2204.12949 [gr-qc]]}.

%\cite{Cunha:2017qtt}
\bibitem{Cunha:2017qtt}
P.~V.~P.~Cunha, E.~Berti and C.~A.~R.~Herdeiro,
%``Light-Ring Stability for Ultracompact Objects,''
Phys. Rev. Lett. \textbf{119} (2017) no.25, 251102
% doi:10.1103/PhysRevLett.119.251102
\href{https://arxiv.org/abs/1708.04211}{[arXiv:1708.04211 [gr-qc]]}.

%\cite{Keir:2014oka}
\bibitem{Keir:2014oka}
J.~Keir,
%``Slowly decaying waves on spherically symmetric spacetimes and ultracompact neutron stars,''
Class. Quant. Grav. \textbf{33} (2016) no.13, 135009
% doi:10.1088/0264-9381/33/13/135009
\href{https://arxiv.org/abs/1404.7036}{[arXiv:1404.7036 [gr-qc]]}.

%\cite{Xavier:2024iwr}
\bibitem{Xavier:2024iwr}
S.~V.~M.~C.~B.~Xavier, C.~A.~R.~Herdeiro and L.~C.~B.~Crispino,
%``Traversable wormholes and light rings,''
Phys. Rev. D \textbf{109} (2024) no.12, 124065
% doi:10.1103/PhysRevD.109.124065
\href{https://arxiv.org/abs/2404.02208}{[arXiv:2404.02208 [gr-qc]]}.

%\cite{Benomio:2024lev}
\bibitem{Benomio:2024lev}
G.~Benomio, A.~C{\'a}rdenas-Avenda{\~n}o, F.~Pretorius and A.~Sullivan,
%``Turbulence for spacetimes with stable trapping,''
Phys. Rev. D \textbf{111} (2025) no.10, 104037
\href{https://arxiv.org/abs/2411.17445}{[arXiv:2411.17445 [gr-qc]]}.

%\cite{Guo:2022ghl}
\bibitem{Guo:2022ghl}
G.~Guo, Y.~Lu, P.~Wang, H.~Wu and H.~Yang,
%``Black holes with multiple photon spheres,''
Phys. Rev. D \textbf{107} (2023) no.12, 124037
%doi:10.1103/PhysRevD.107.124037
\href{https://arxiv.org/pdf/2212.12901}{[arXiv:2212.12901 [gr-qc]]}.

%\cite{Wang:2025hzu}
\bibitem{Wang:2025hzu}
X.~J.~Wang, Y.~Meng, X.~M.~Kuang and K.~Liao,
%``Exploring black holes with multiple photon spheres by interferometric signatures,''
Phys. Rev. D \textbf{112} (2025) no.12, 124016
% doi:10.1103/t36v-kwj7
\href{https://arxiv.org/abs/2508.02355}{[arXiv:2508.02355 [gr-qc]]}.

%\cite{Sakai:2014pga}
\bibitem{Sakai:2014pga}
N.~Sakai, H.~Saida and T.~Tamaki,
%``Gravastar Shadows,''
Phys. Rev. D \textbf{90} (2014) no.10, 104013
%doi:10.1103/PhysRevD.90.104013
\href{https://arxiv.org/abs/1408.6929}{[arXiv:1408.6929 [gr-qc]]}.

%\cite{Wielgus:2020uqz}
\bibitem{Wielgus:2020uqz}
M.~Wielgus, J.~Horak, F.~Vincent and M.~Abramowicz,
%``Reflection-asymmetric wormholes and their double shadows,''
Phys. Rev. D \textbf{102} (2020) no.8, 084044
%doi:10.1103/PhysRevD.102.084044
\href{https://arxiv.org/abs/2008.10130}{[arXiv:2008.10130 [gr-qc]]}.

%\cite{Wang:2020emr}
\bibitem{Wang:2020emr}
X.~Wang, P.~C.~Li, C.~Y.~Zhang and M.~Guo,
%``Novel shadows from the asymmetric thin-shell wormhole,''
Phys. Lett. B \textbf{811} (2020), 135930
%doi:10.1016/j.physletb.2020.135930
\href{https://arxiv.org/abs/2007.03327}{[arXiv:2007.03327 [gr-qc]]}.
%61 citations counted in INSPIRE as of 24 Jul 2025

%\cite{Guerrero:2022qkh} 
\bibitem{Guerrero:2022qkh} 
M.~Guerrero, G.~J.~Olmo, D.~Rubiera-Garcia and D.~G{\'o}mez S{\'a}ez-Chill{\'o}n, 
%``Light ring images of double photon spheres in black hole and wormhole spacetimes,''
Phys. Rev. D \textbf{105} (2022) no.8, 084057
%doi:10.1103/PhysRevD.105.084057
\href{https://arxiv.org/abs/2202.03809}{[arXiv:2202.03809 [gr-qc]]}.

%\cite{Guerrero:2021pxt}
\bibitem{Guerrero:2021pxt}
M.~Guerrero, G.~J.~Olmo and D.~Rubiera-Garcia,
%``Double shadows of reflection-asymmetric wormholes supported by positive energy thin-shells,''
JCAP \textbf{04} (2021), 066
%doi:10.1088/1475-7516/2021/04/066
\href{https://arxiv.org/abs/2102.00840}{[arXiv:2102.00840 [gr-qc]]}.
%37 citations counted in INSPIRE as of 24 Jul 2025

%\cite{Bueno:2017hyj}
\bibitem{Bueno:2017hyj}
P.~Bueno, P.~A.~Cano, F.~Goelen, T.~Hertog and B.~Vercnocke,
%``Echoes of Kerr-like wormholes,''
Phys. Rev. D \textbf{97} (2018) no.2, 024040
%doi:10.1103/PhysRevD.97.024040
\href{https://arxiv.org/abs/1711.00391}{[arXiv:1711.00391 [gr-qc]]}.
%232 citations counted in INSPIRE as of 03 Aug 2025

%\cite{Guo:2021enm}
\bibitem{Guo:2021enm}
G.~Guo, P.~Wang, H.~Wu and H.~Yang,
%``Quasinormal modes of black holes with multiple photon spheres,''
JHEP \textbf{06} (2022), 060
%doi:10.1007/JHEP06(2022)060
\href{https://arxiv.org/abs/2112.14133}{[arXiv:2112.14133 [gr-qc]]}.

%\cite{Guo:2022umh}
\bibitem{Guo:2022umh}
G.~Guo, P.~Wang, H.~Wu and H.~Yang,
%``Echoes from hairy black holes,''
JHEP \textbf{06} (2022), 073
%doi:10.1007/JHEP06(2022)073
\href{https://arxiv.org/abs/2204.00982}{[arXiv:2204.00982 [gr-qc]]}.

%\cite{Shen:2025yiy}
\bibitem{Shen:2025yiy}
S.~F.~Shen, G.~R.~Li, R.~G.~Daghigh, J.~C.~Morey, M.~D.~Green, W.~L.~Qian and R.~H.~Yue,
%``Asymptotic quasinormal modes, echoes, and black hole spectral instability: a brief review,''
\href{https://arxiv.org/abs/2507.11663}{[arXiv:2507.11663 [gr-qc]]}.

%\cite{Cardoso:2016oxy} 
\bibitem{Cardoso:2016oxy} V.~Cardoso, S.~Hopper, C.~F.~B.~Macedo, C.~Palenzuela and P.~Pani, %``Gravitational-wave signatures of exotic compact objects and of quantum corrections at the horizon scale,'' 
Phys. Rev. D \textbf{94} (2016) no.8, 084031
%doi:10.1103/PhysRevD.94.084031
\href{https://arxiv.org/abs/1608.08637}{[arXiv:1608.08637 [gr-qc]]}.

%\cite{Cardoso:2016rao}
\bibitem{Cardoso:2016rao}
V.~Cardoso, E.~Franzin and P.~Pani,
%``Is the gravitational-wave ringdown a probe of the event horizon?,''
Phys. Rev. Lett. \textbf{116} (2016) no.17, 171101
[erratum: Phys. Rev. Lett. \textbf{117} (2016) no.8, 089902]
%doi:10.1103/PhysRevLett.116.171101
\href{https://arxiv.org/abs/1602.07309}{[arXiv:1602.07309 [gr-qc]]}.
%804 citations counted in INSPIRE as of 04 Aug 2025

%\cite{Hui:2019aox}
\bibitem{Hui:2019aox}
L.~Hui, D.~Kabat and S.~S.~C.~Wong,
%``Quasinormal modes, echoes and the causal structure of the Green's function,''
JCAP \textbf{12} (2019), 020
%doi:10.1088/1475-7516/2019/12/020
\href{https://arxiv.org/abs/1909.10382}{[arXiv:1909.10382 [gr-qc]]}.

%\cite{Abedi:2016hgu}
\bibitem{Abedi:2016hgu}
J.~Abedi, H.~Dykaar and N.~Afshordi,
%``Echoes from the Abyss: Tentative evidence for Planck-scale structure at black hole horizons,''
Phys. Rev. D \textbf{96} (2017) no.8, 082004
%doi:10.1103/PhysRevD.96.082004
\href{https://arxiv.org/abs/1612.00266}{[arXiv:1612.00266 [gr-qc]]}.

%\cite{Harko:2013yb}
\bibitem{Harko:2013yb}
T.~Harko, F.~S.~N.~Lobo, M.~K.~Mak and S.~V.~Sushkov,
%``Modified-gravity wormholes without exotic matter,''
Phys. Rev. D \textbf{87} (2013) no.6, 067504
%doi:10.1103/PhysRevD.87.067504
\href{https://arxiv.org/abs/1301.6878}{[arXiv:1301.6878 [gr-qc]]}.

%\cite{Moraes:2017dbs}
\bibitem{Moraes:2017dbs}
P.~H.~R.~S.~Moraes and P.~K.~Sahoo,
%``Nonexotic matter wormholes in a trace of the energy-momentum tensor squared gravity,''
Phys. Rev. D \textbf{97} (2018) no.2, 024007
%doi:10.1103/PhysRevD.97.024007
\href{https://arxiv.org/abs/1709.00027}{[arXiv:1709.00027 [gr-qc]]}.

%\cite{Svitek:2016nvm}
\bibitem{Svitek:2016nvm}
O.~Svitek and T.~Tahamtan,
%``Nonsymmetric dynamical thin-shell wormhole in Robinson{\textendash}Trautman class,''
Eur. Phys. J. C \textbf{78} (2018) no.2, 167
%doi:10.1140/epjc/s10052-018-5628-0
\href{https://arxiv.org/abs/1606.01501}{[arXiv:1606.01501 [gr-qc]]}

%\cite{Garattini:2019ivd}
\bibitem{Garattini:2019ivd}
R.~Garattini,
%``Casimir Wormholes,''
Eur. Phys. J. C \textbf{79} (2019) no.11, 951
%doi:10.1140/epjc/s10052-019-7468-y
\href{https://arxiv.org/abs/1907.03623}{[arXiv:1907.03623 [gr-qc]]}.

%\cite{Berthiere:2017tms}
\bibitem{Berthiere:2017tms}
C.~Berthiere, D.~Sarkar and S.~N.~Solodukhin,
%``The fate of black hole horizons in semiclassical gravity,''
Phys. Lett. B \textbf{786} (2018), 21-27
%doi:10.1016/j.physletb.2018.09.027
\href{https://arxiv.org/abs/1712.09914}{[arXiv:1712.09914 [hep-th]]}.

%\cite{Potaux:2021yan} \cite{Potaux:2022uxa} \cite{Potaux:2023fwm}
\bibitem{Potaux:2021yan}
Y.~Potaux, D.~Sarkar and S.~N.~Solodukhin,
%``Quantum states and their back-reacted geometries in 2D dilaton gravity,''
Phys. Rev. D \textbf{105} (2022) no.2, 025015
%doi:10.1103/PhysRevD.105.025015
\href{https://arxiv.org/abs/2112.03855}{[arXiv:2112.03855 [hep-th]]}.

%\cite{Potaux:2022uxa}
\bibitem{Potaux:2022uxa}
Y.~Potaux, D.~Sarkar and S.~N.~Solodukhin,
%``Spacetime Structure, Asymptotic Radiation, and Information Recovery for a Quantum Hybrid State,''
Phys. Rev. Lett. \textbf{130} (2023) no.26, 261501
%doi:10.1103/PhysRevLett.130.261501
\href{https://arxiv.org/abs/2212.13208}{[arXiv:2212.13208 [hep-th]]}.

%\cite{Potaux:2023fwm}
\bibitem{Potaux:2023fwm}
Y.~Potaux, D.~Sarkar and S.~N.~Solodukhin,
%``Hybrid quantum states in 2D dilaton gravity,''
Phys. Rev. D \textbf{108} (2023) no.12, 125012
%doi:10.1103/PhysRevD.108.125012
\href{https://arxiv.org/abs/2310.18745}{[arXiv:2310.18745 [hep-th]]}.

%\cite{Johannsen:2010ru}
\bibitem{Johannsen:2010ru}
T.~Johannsen and D.~Psaltis,
%``Testing the No-Hair Theorem with Observations in the Electromagnetic Spectrum: II. Black-Hole Images,''
Astrophys. J. \textbf{718} (2010), 446-454
\href{https://arxiv.org/abs/1005.1931}{[arXiv:1005.1931 [astro-ph.HE]]}.

%\cite{EventHorizonTelescope:2020qrl}
\bibitem{EventHorizonTelescope:2020qrl}
D.~Psaltis \textit{et al.} [Event Horizon Telescope],
%``Gravitational Test Beyond the First Post-Newtonian Order with the Shadow of the M87 Black Hole,''
Phys. Rev. Lett. \textbf{125} (2020) no.14, 141104
\href{https://arxiv.org/abs/2010.01055}{[arXiv:2010.01055 [gr-qc]]}.

%\cite{Fragione:2020khu}
\bibitem{Fragione:2020khu}
G.~Fragione and A.~Loeb,
%``An upper limit on the spin of SgrA$^*$ based on stellar orbits in its vicinity,''
Astrophys. J. Lett. \textbf{901} (2020) no.2, L32
doi:10.3847/2041-8213/abb9b4
\href{https://arxiv.org/abs/2008.11734}{[arXiv:2008.11734 [astro-ph.GA]]}.

% %\cite{Fragione:2020khu}
% \bibitem{Fragione:2020khu}
% G.~Fragione and A.~Loeb,
% %``An upper limit on the spin of SgrA$^*$ based on stellar orbits in its vicinity,''
% Astrophys. J. Lett. \textbf{901} (2020) no.2, L32
% doi:10.3847/2041-8213/abb9b4
% \href{https://arxiv.org/abs/2008.11734}{[arXiv:2008.11734 [astro-ph.GA]]}.

% %\cite{Fragione:2022oau}
% \bibitem{Fragione:2022oau}
% G.~Fragione and A.~Loeb,
% %``Implication of Spin Constraints by the Event Horizon Telescope on Stellar Orbits in the Galactic Center,''
% Astrophys. J. Lett. \textbf{932} (2022) no.2, L17
% \href{https://arxiv.org/abs/2205.12274}{[arXiv:2205.12274 [astro-ph.GA]]}.

%\cite{Shibata:1999hn}
\bibitem{Shibata:1999hn}
M.~Shibata,
%``Fully general relativistic simulation of coalescing binary neutron stars: Preparatory tests,''
Phys. Rev. D \textbf{60} (1999), 104052
%doi:10.1103/PhysRevD.60.104052
\href{https://arxiv.org/abs/gr-qc/9908027}{[arXiv:gr-qc/9908027 [gr-qc]]}.
%133 citations counted in INSPIRE as of 29 Jul 2025

%\cite{Shibata:2006bs}
\bibitem{Shibata:2006bs}
M.~Shibata and K.~Uryu,
%``Merger of black hole-neutron star binaries in full general relativity,''
Class. Quant. Grav. \textbf{24} (2007), S125-S138
%doi:10.1088/0264-9381/24/12/S09
\href{https://arxiv.org/abs/astro-ph/0611522}{[arXiv:astro-ph/0611522 [astro-ph]]}.
%84 citations counted in INSPIRE as of 29 Jul 2025

%\cite{Pretorius:2005gq}
\bibitem{Pretorius:2005gq}
F.~Pretorius,
%``Evolution of binary black hole spacetimes,''
Phys. Rev. Lett. \textbf{95} (2005), 121101
%doi:10.1103/PhysRevLett.95.121101
\href{https://arxiv.org/abs/gr-qc/0507014}{[arXiv:gr-qc/0507014 [gr-qc]]}.
%1396 citations counted in INSPIRE as of 29 Jul 2025

%\cite{Aylott:2009ya}
\bibitem{Aylott:2009ya}
B.~Aylott, J.~G.~Baker, W.~D.~Boggs, M.~Boyle, P.~R.~Brady, D.~A.~Brown, B.~Brugmann, L.~T.~Buchman, A.~Buonanno and L.~Cadonati, \textit{et al.}
%``Testing gravitational-wave searches with numerical relativity waveforms: Results from the first Numerical INJection Analysis (NINJA) project,''
Class. Quant. Grav. \textbf{26} (2009), 165008
%doi:10.1088/0264-9381/26/16/165008
\href{https://arxiv.org/pdf/0901.4399}{[arXiv:0901.4399 [gr-qc]]}.
%170 citations counted in INSPIRE as of 29 Jul 2025

%\cite{Decanini:2015yba}
\bibitem{Decanini:2015yba}
Y.~Decanini, A.~Folacci and M.~Ould El Hadj,
%``Waveforms produced by a scalar point particle plunging into a Schwarzschild black hole: Excitation of quasinormal modes and quasibound states,''
Phys. Rev. D \textbf{92} (2015) no.2, 024057
%doi:10.1103/PhysRevD.92.024057
\href{https://arxiv.org/abs/1506.09133}{[arXiv:1506.09133 [gr-qc]]}.
%8 citations counted in INSPIRE as of 29 Jul 2025

%\cite{Dolan:2006vj}
\bibitem{Dolan:2006vj}
S.~Dolan, C.~Doran and A.~Lasenby,
%``Fermion scattering by a Schwarzschild black hole,''
Phys. Rev. D \textbf{74} (2006), 064005
%doi:10.1103/PhysRevD.74.064005
\href{https://arxiv.org/abs/gr-qc/0605031}{[arXiv:gr-qc/0605031 [gr-qc]]}.
%100 citations counted in INSPIRE as of 29 Jul 2025

%\cite{Jing:2005pk}
\bibitem{Jing:2005pk}
J.~l.~Jing and Q.~y.~Pan,
%``Dirac quasinormal frequencies of the Kerr-Newman black hole,''
Nucl. Phys. B \textbf{728} (2005), 109-120
%doi:10.1016/j.nuclphysb.2005.08.038
\href{https://arxiv.org/abs/gr-qc/0506098}{[arXiv:gr-qc/0506098 [gr-qc]]}.
%37 citations counted in INSPIRE as of 29 Jul 2025

%\cite{Folacci:2020ekl}
\bibitem{Folacci:2020ekl}
A.~Folacci and M.~Ould El Hadj,
%``Electromagnetic radiation generated by a charged particle falling radially into a Schwarzschild black hole: A complex angular momentum description,''
Phys. Rev. D \textbf{102} (2020) no.2, 024026
%doi:10.1103/PhysRevD.102.024026
\href{https://arxiv.org/abs/2004.07813}{[arXiv:2004.07813 [gr-qc]]}.
%6 citations counted in INSPIRE as of 29 Jul 2025

%\cite{Ruffini:1972pw}
\bibitem{Ruffini:1972pw}
R.~Ruffini, J.~Tiomno and C.~V.~Vishveshwara,
%``Electromagnetic field of a particle moving in a spherically symmetric black-hole background,''
Lett. Nuovo Cim. \textbf{3S2} (1972), 211-215
%doi:10.1007/BF02772872
%35 citations counted in INSPIRE as of 29 Jul 2025

%\cite{Ching:1998mxl}
\bibitem{Ching:1998mxl}
E.~S.~C.~Ching, P.~T.~Leung, A.~Maassen van den Brink, W.~M.~Suen, S.~S.~Tong and K.~Young,
%``Quasinormal-mode expansion for waves in open systems,''
Rev. Mod. Phys. \textbf{70} (1998), 1545
%doi:10.1103/RevModPhys.70.1545
\href{https://arxiv.org/abs/gr-qc/9904017}{[arXiv:gr-qc/9904017 [gr-qc]]}.
%67 citations counted in INSPIRE as of 29 Jul 2025

%\cite{Nollert:1998ys}
\bibitem{Nollert:1998ys}
H.~P.~Nollert and R.~H.~Price,
%``Quantifying excitations of quasinormal mode systems,''
J. Math. Phys. \textbf{40} (1999), 980-1010
%doi:10.1063/1.532698
\href{https://arxiv.org/abs/gr-qc/9810074}{[arXiv:gr-qc/9810074 [gr-qc]]}.
%139 citations counted in INSPIRE as of 29 Jul 2025

%\cite{Nollert:1992ifk}
\bibitem{Nollert:1992ifk}
H.~P.~Nollert and B.~G.~Schmidt,
%``Quasinormal modes of Schwarzschild black holes: Defined and calculated via Laplace transformation,''
Phys. Rev. D \textbf{45} (1992) no.8, 2617
%doi:10.1103/PhysRevD.45.2617
%116 citations counted in INSPIRE as of 29 Jul 2025

%\cite{Andersson:1996cm}
\bibitem{Andersson:1996cm}
N.~Andersson,
%``Evolving test fields in a black hole geometry,''
Phys. Rev. D \textbf{55} (1997), 468-479
%doi:10.1103/PhysRevD.55.468
\href{https://arxiv.org/abs/gr-qc/9607064}{[arXiv:gr-qc/9607064 [gr-qc]]}.
%153 citations counted in INSPIRE as of 29 Jul 2025

%\cite{Andersson:1995zk}
\bibitem{Andersson:1995zk}
N.~Andersson,
%``Excitation of Schwarzschild black hole quasinormal modes,''
Phys. Rev. D \textbf{51} (1995), 353-363
%doi:10.1103/PhysRevD.51.353
%76 citations counted in INSPIRE as of 29 Jul 2025

%\cite{Schultz}
\bibitem{Schultz}
B.~F.~Schultz and C.~M.~Will,
%``Quasi-normal frequencies: Key analytic results,''
Astrophys.~J. \textbf{L291},~33~(1985),
%doi:10.1086/184453

%\cite{Konoplya:2004ip}
\bibitem{Konoplya:2004ip}
R.~A.~Konoplya,
%``Quasinormal modes of the Schwarzschild black hole and higher order WKB approach,''
J. Phys. Stud. \textbf{8} (2004), 93-100

%\cite{Leaver:1985ax}
\bibitem{Leaver:1985ax}
E.~W.~Leaver,
%``An Analytic representation for the quasi normal modes of Kerr black holes,''
Proc. Roy. Soc. Lond. A \textbf{402} (1985), 285-298
%doi:10.1098/rspa.1985.0119

%\cite{Lin:2019mmf}
\bibitem{Lin:2019mmf}
K.~Lin and W.~L.~Qian,
%``On matrix method for black hole quasinormal modes,''
Chin. Phys. C \textbf{43} (2019) no.3, 035105
%doi:10.1088/1674-1137/43/3/035105
\href{https://arxiv.org/abs/1902.08352}{[arXiv:1902.08352 [gr-qc]]}.
%39 citations counted in INSPIRE as of 31 Jul 2025

%\cite{Motl:2003cd}
\bibitem{Motl:2003cd}
L.~Motl and A.~Neitzke,
%``Asymptotic black hole quasinormal frequencies,''
Adv. Theor. Math. Phys. \textbf{7} (2003) no.2, 307-330
%doi:10.4310/ATMP.2003.v7.n2.a4
\href{https://arxiv.org/abs/hep-th/0301173}{[arXiv:hep-th/0301173 [hep-th]]}.

%\cite{Zenginoglu:2007jw}
\bibitem{Zenginoglu:2007jw}
A.~Zenginoglu,
%``Hyperboloidal foliations and scri-fixing,''
Class. Quant. Grav. \textbf{25} (2008), 145002
% doi:10.1088/0264-9381/25/14/145002
\href{https://arxiv.org/abs/0712.4333}{[arXiv:0712.4333 [gr-qc]]}.

%\cite{Zenginoglu:2011jz}
\bibitem{Zenginoglu:2011jz}
A.~Zenginoglu,
%``A Geometric framework for black hole perturbations,''
Phys. Rev. D \textbf{83} (2011), 127502
%doi:10.1103/PhysRevD.83.127502
\href{https://arxiv.org/abs/1102.2451}{[arXiv:1102.2451 [gr-qc]]}.
%93 citations counted in INSPIRE as of 31 Jul 2025

%\cite{Jaramillo:2020tuu}
\bibitem{Jaramillo:2020tuu}
J.~L.~Jaramillo, R.~Panosso Macedo and L.~Al Sheikh,
%``Pseudospectrum and Black Hole Quasinormal Mode Instability,''
Phys. Rev. X \textbf{11} (2021) no.3, 031003
%doi:10.1103/PhysRevX.11.031003
\href{https://arxiv.org/abs/2004.06434}{[arXiv:2004.06434 [gr-qc]]}.
%200 citations counted in INSPIRE as of 31 Jul 2025

%\cite{PanossoMacedo:2023qzp}
\bibitem{PanossoMacedo:2023qzp}
R.~Panosso Macedo,
%``Hyperboloidal approach for static spherically symmetric spacetimes: a didactical introductionand applications in black-hole physics,''
Phil. Trans. Roy. Soc. Lond. A \textbf{382} (2024) no.2267, 20230046
%doi:10.1098/rsta.2023.0046
\href{https://arxiv.org/abs/2307.15735}{[arXiv:2307.15735 [gr-qc]]}.
%35 citations counted in INSPIRE as of 31 Jul 2025

%\cite{Lin:2023rkd}
\bibitem{Lin:2023rkd}
K.~Lin,
%``Quasinormal modes by improved matrix method and weighted residual method,''
Phys. Scripta \textbf{99} (2024) no.6, 065030
%doi:10.1088/1402-4896/ad46c5
\href{https://arxiv.org/abs/2306.07782}{[arXiv:2306.07782 [gr-qc]]}.
%4 citations counted in INSPIRE as of 31 Jul 2025

%\cite{Lin:2016sch}
\bibitem{Lin:2016sch}
K.~Lin and W.~L.~Qian,
%``A Matrix Method for Quasinormal Modes: Schwarzschild Black Holes in Asymptotically Flat and (Anti-) de Sitter Spacetimes,''
Class. Quant. Grav. \textbf{34} (2017) no.9, 095004
%doi:10.1088/1361-6382/aa6643
\href{https://arxiv.org/abs/1610.08135}{[arXiv:1610.08135 [gr-qc]]}.
%72 citations counted in INSPIRE as of 31 Jul 2025

%\cite{Barausse:2014tra}
\bibitem{Barausse:2014tra}
E.~Barausse, V.~Cardoso and P.~Pani,
%``Can environmental effects spoil precision gravitational-wave astrophysics?,''
Phys. Rev. D \textbf{89} (2014) no.10, 104059
%doi:10.1103/PhysRevD.89.104059
\href{https://arxiv.org/abs/1404.7149}{[arXiv:1404.7149 [gr-qc]]}.

%\cite{Volkel:2018hwb}
\bibitem{Volkel:2018hwb}
S.~H.~V{\"o}lkel and K.~D.~Kokkotas,
%``Wormhole Potentials and Throats from Quasi-Normal Modes,''
Class. Quant. Grav. \textbf{35} (2018) no.10, 105018
%doi:10.1088/1361-6382/aabce6
\href{https://arxiv.org/abs/1802.08525}{[arXiv:1802.08525 [gr-qc]]}.
%75 citations counted in INSPIRE as of 04 Aug 2025

%\cite{Konoplya:2004wg}
\bibitem{Konoplya:2004wg}
R.~A.~Konoplya and A.~V.~Zhidenko,
%``Decay of massive scalar field in a Schwarzschild background,''
Phys. Lett. B \textbf{609} (2005), 377-384
% doi:10.1016/j.physletb.2005.01.078
\href{https://arxiv.org/abs/gr-qc/0411059}{[arXiv:gr-qc/0411059 [gr-qc]]}.

%\cite{Konoplya:2006br}
\bibitem{Konoplya:2006br}
R.~A.~Konoplya and A.~Zhidenko,
%``Stability and quasinormal modes of the massive scalar field around Kerr black holes,''
Phys. Rev. D \textbf{73} (2006), 124040
% doi:10.1103/PhysRevD.73.124040
\href{https://arxiv.org/abs/gr-qc/0605013}{[arXiv:gr-qc/0605013 [gr-qc]]}.

%\cite{Cardoso:2017njb}
\bibitem{Cardoso:2017njb}
V.~Cardoso and P.~Pani,
``The observational evidence for horizons: from echoes to precision gravitational-wave physics,''
\href{https://arxiv.org/abs/1707.03021}{[arXiv:1707.03021 [gr-qc]]}.
%97 citations counted in INSPIRE as of 04 Aug 2025

%\cite{Cardoso:2017cqb}
\bibitem{Cardoso:2017cqb}
V.~Cardoso and P.~Pani,
%``Tests for the existence of black holes through gravitational wave echoes,''
Nature Astron. \textbf{1} (2017) no.9, 586-591
%doi:10.1038/s41550-017-0225-y
\href{https://arxiv.org/pdf/1709.01525}{[arXiv:1709.01525 [gr-qc]]}.
%407 citations counted in INSPIRE as of 04 Aug 2025

%\cite{Mark:2017dnq}
\bibitem{Mark:2017dnq}
Z.~Mark, A.~Zimmerman, S.~M.~Du and Y.~Chen,
%``A recipe for echoes from exotic compact objects,''
Phys. Rev. D \textbf{96} (2017) no.8, 084002
%doi:10.1103/PhysRevD.96.084002
\href{https://arxiv.org/abs/1706.06155}{[arXiv:1706.06155 [gr-qc]]}.
%232 citations counted in INSPIRE as of 04 Aug 2025

%\cite{Berti:2006wq}
\bibitem{Berti:2006wq}
E.~Berti and V.~Cardoso,
%``Quasinormal ringing of Kerr black holes. I. The Excitation factors,''
Phys. Rev. D \textbf{74} (2006), 104020
%doi:10.1103/PhysRevD.74.104020
\href{https://arxiv.org/abs/gr-qc/0605118}{[arXiv:gr-qc/0605118 [gr-qc]]}.

%\cite{Horowitz:1999jd}
\bibitem{Horowitz:1999jd}
G.~T.~Horowitz and V.~E.~Hubeny,
%``Quasinormal modes of AdS black holes and the approach to thermal equilibrium,''
Phys. Rev. D \textbf{62} (2000), 024027
%doi:10.1103/PhysRevD.62.024027
\href{https://arxiv.org/abs/hep-th/9909056}{[arXiv:hep-th/9909056 [hep-th]]}.
%978 citations counted in INSPIRE as of 09 Aug 2025

%\cite{Cardoso:2003cj}
\bibitem{Cardoso:2003cj}
V.~Cardoso, R.~Konoplya and J.~P.~S.~Lemos,
%``Quasinormal frequencies of Schwarzschild black holes in anti-de Sitter space-times: A Complete study on the asymptotic behavior,''
Phys. Rev. D \textbf{68} (2003), 044024
% doi:10.1103/PhysRevD.68.044024
\href{https://arxiv.org/abs/gr-qc/0305037}{[arXiv:gr-qc/0305037 [gr-qc]]}.

%\cite{Cardoso:2001bb}
\bibitem{Cardoso:2001bb}
V.~Cardoso and J.~P.~S.~Lemos,
%``Quasinormal modes of Schwarzschild anti-de Sitter black holes: Electromagnetic and gravitational perturbations,''
Phys. Rev. D \textbf{64} (2001), 084017
%doi:10.1103/PhysRevD.64.084017
\href{https://arxiv.org/abs/gr-qc/0105103}{[arXiv:gr-qc/0105103 [gr-qc]]}.
%379 citations counted in INSPIRE as of 09 Aug 2025

%\cite{Konoplya:2002zu}
\bibitem{Konoplya:2002zu}
R.~A.~Konoplya,
%``On quasinormal modes of small Schwarzschild-anti-de Sitter black hole,''
Phys. Rev. D \textbf{66} (2002), 044009
% doi:10.1103/PhysRevD.66.044009
\href{https://arxiv.org/abs/hep-th/0205142}{[arXiv:hep-th/0205142 [hep-th]]}.

%\cite{Boyanov:2023qqf}
\bibitem{Boyanov:2023qqf}
V.~Boyanov, V.~Cardoso, K.~Destounis, J.~L.~Jaramillo and R.~Panosso Macedo,
%``Structural aspects of the anti{\textendash}de Sitter black hole pseudospectrum,''
Phys. Rev. D \textbf{109} (2024) no.6, 064068
%doi:10.1103/PhysRevD.109.064068
\href{https://arxiv.org/abs/2312.11998}{[arXiv:2312.11998 [gr-qc]]}.
%41 citations counted in INSPIRE as of 09 Aug 2025

%\cite{Boonserm:2010px}
\bibitem{Boonserm:2010px}
P.~Boonserm and M.~Visser,
%``Quasi-normal frequencies: Key analytic results,''
JHEP \textbf{03} (2011), 073
%doi:10.1007/JHEP03(2011)073
\href{https://arxiv.org/abs/1005.4483}{[arXiv:1005.4483 [math-ph]]}.

%\cite{Correia:2018apm}
\bibitem{Correia:2018apm}
M.~R.~Correia and V.~Cardoso,
%``Characterization of echoes: A Dyson-series representation of individual pulses,''
Phys. Rev. D \textbf{97} (2018) no.8, 084030
%doi:10.1103/PhysRevD.97.084030
\href{https://arxiv.org/abs/1802.07735}{[arXiv:1802.07735 [gr-qc]]}.



\end{thebibliography}
\end{document}